\documentclass[twocolumn,showkeys,showpacs,preprintnumbers,amsmath,amssymb,superscriptaddress,nofootinbib]{revtex4-2}

\usepackage{mathrsfs,amsmath}
\usepackage[scr=boondox]{mathalfa}
\usepackage{comment}
\usepackage{graphicx, float}
\usepackage{dcolumn,booktabs}
\usepackage{bm}
\usepackage{makecell}
\usepackage{epstopdf}
\usepackage{hyperref,url}
\usepackage[usenames, dvipsnames]{color} 
\usepackage[table, dvipsnames]{xcolor} 
\usepackage{colortbl}
\usepackage[caption=false]{subfig}

\makeatletter
\newcommand{\hwidth}[1]{%
	\noalign{\hrule \@height #1}%
}
\makeatother

\begin{document}
	
	\preprint{APS/123-QED}
	
	\title{Fundamental and higher-order excited modes of radial oscillation of \protect\\ neutron stars for various types of cold nucleonic and hyperonic matter}
	
	\author{D\'{a}niel Barta}
	\email{barta.daniel@wigner.hu}
	\affiliation{%
		Wigner Research Centre for Physics,
		H-1525 Budapest 114, P.O. Box 49, Hungary
	}%
	
	\date{\today}
	
	\begin{abstract}
		This research paper complements our earlier qualitative study of the effect of viscosity and thermal conductivity on the radial oscillation and relaxation of non-rotating neutron stars. The fundamental and first two lowest-frequency excited modes of radial oscillation have been computed in the high nuclear density regime for a set of seven realistic equations of state (EoS) as functions of central energy density. Various types of zero-temperature EoS of cold nucleonic and hybrid nucleon--hyperon--quark matter models are used in the inner core to determine the internal structure in and around the hydrostatic equilibrium states and investigate the influence of each EoS on the dynamical behavior of non-rotating neutron stars. We confirm the principal results of earlier, related studies that suggest an underlying correlation between the frequency spectrum of the fundamental oscillation mode and the variation of the adiabatic index over the high nuclear-density regime. We provide valuable information to impose further constraints on the plausible set of realistic EoS models, in addition to the practical applications for the rapidly evolving field of asteroseismology of compact objects.
	\end{abstract}
	
	
	\maketitle
	
	\section{\label{sec:intro}Introduction}
	Neutron stars (NSs) provide us with unique insights into the physics of the extremely dense and cold nuclear matter, which cannot be reached in terrestrial experiments. The central density of NSs is expected to reach up to several times of the nuclear saturation density ($n_{0} \simeq 0.16 \text{ fm}^{-3}$). The constraints on the attributes of matter at supernuclear densities (see, e.g. \cite{Chirenti2015} and references therein) and relatively low temperatures (compared to particle collisions producing comparable energy densities) rely heavily on observations of macroscopic equilibrium parameters of NSs. Such macroscopic parameters as the total gravitational mass $(M)$ and the circumferential radius $(R)$ of NSs in binary systems are determined by high-precision timing observations of radio and X-ray pulsars. \cite{Ozel2016,Lattimer2001,Read2009} The recent discovery of high-mass NSs of masses up to around $M \approx 2.01 M_{\odot}$ (e.g., PSR J1614-2230 has yielded a mass of $1.927 \pm 0.017 M_{\odot}$ \cite{Fonseca2016} and the mass of PSR J0348+0432 has been measured at around $2.01 \pm 0.04 M_{\odot}$ \cite{Antoniadis2013}) has ruled out several equations of state (EoS) models (see \ref{fig:MR}). Due to the tremendous advances in the measurements, precise masses for $\sim 35$ currently known NSs range from $1.17$ to $2.01$ $M_{\odot}$. Also more than a dozen radii are known in the range of $9.9$ to $11.2$ kms, but current estimates for radii are still dominated by systematic errors. \cite{Miller2016} More strict constraints on the plausible EoS models will be imposed by yielding masses and radii of a few stars to $\sim 5\%$ precision from the recently lunched NICER mission \cite{Gendreau2012} and from the upcoming LOFT mission \cite{Watts2016}. Constraints on the EoS, inferred from GW events GW170817 \cite{Abbott2018} and GW190425 \cite{Abbott2020} that are associated with binary NS mergers, are important confirmations of the great potential that multi-messenger observations offer.
	
	X-ray and $\gamma$-ray burst phenomena are clearly explosive events that perturb their sources \cite{Lindblom1992} and have been generally associated with neutron stars (see, e.g. \cite{Ramaty1981} and references therein). In a manner analogous to terrestrial seismology, observational methods and techniques in asteroseismology are using the frequency of seismic waves rippling throughout stars with the aim of probing their internal structure and thermodynamic properties. \cite{Sagun2020} The frequencies appearing in the spectrum of NS oscillations, which sensitively rely on accurate stellar models, are matched to the observed frequencies. The period of stellar oscillations for non-relativistic stars are in the range of minutes, whilst for NSs the periods are much shorter; they typically range from about 0.2 to 0.9 milliseconds (cf. eq. \eqref{time-scale}). \cite{Detweiler1983} These oscillations occur when a star is perturbed away from its dynamical equilibrium and a restoring force tries to return it back to that equilibrium state. Among the various types of oscillation modes, in this paper we focus on the three lowest-frequency modes (fundamental and the next two higher modes) of radial oscillations where the pressure provides the dominant restoring force that produces oscillations. In this mode the Lagrangian displacement of fluid elements in the star is purely radial and the spherical symmetry is preserved. \cite[p.~55]{Phelan2008}
	
	As the radial modes are the simplest oscillation modes of NSs, they have been comprehensively investigated over the past half-century. In his pioneering papers \cite{Chandrasekhar1964}, Chandrasekhar introduced a variational method to impose a sufficient criterion for the dynamical stability of radial and non-radial stellar oscillations. The identification of stable modes has been in the focus of interest ever since and methods for obtaining spectra of oscillation modes have been thoroughly investigated by various authors (e.g \cite{Harrison1965,Zeldovich1971,Chanmugam1977,Glass1983,Vath1992} and references therein). The first extensive analysis of radial modes for stellar models with various zero-temperature EoSs was presented by Glass \& Lindblom in Ref. \cite{Glass1983}. Although Glass \& Lindblom used appropriate equations, the numerical results for the eigenfrequencies were essentially flawed, as V\"{a}th \& Chanmugam later pointed out in Ref. \cite{Vath1992}. The correct eigenfrequencies were computed for six EoSs by Ref. \cite{Vath1992} and verified by the two different numerical schemes of Ref. \cite{Kokkotas2001}. In addition to having the earlier studies re-examined, Kokkotas \& Ruoff also included further six zero-temperature EoSs in Ref. \cite{Kokkotas2001}.
		 
	Our present paper, together with its accompanying paper \cite{Barta2019}, is devoted to study the radial oscillations of NS models. In the earlier paper \cite{Barta2019}, we proved that the fundamental equation for radial pulsation expressed by a set of effective variables which involve dissipative terms can be recast in a self-adjoint form. In contrast to the common non-dissipative case, the associated Sturm--Liouville eigenvalue problem (SL-EVP) is generalized for a discrete set of eigenfunctions with complex eigenvalues which correspond to the squared frequencies of the oscillation modes and the imaginary part corresponds to the damped solution. The imaginary part directly determines the minimum period of observable pulsars, hence it is imperative to make an accurate estimation on the relative timescale of thermal conductivity. In addition to providing a formulation of the dynamical equations that govern the radial oscillations of neutron stars through a perturbation scheme, an order-of-magnitude estimation was given for the timescale of energy dissipation, rather than a precise one, by an analytical approximation method without relying on explicit numerical computations. In the present paper, the SL-EVP for radial-oscillation modes is converted to a system of finite difference equations where we implement a second-order accurate differencing scheme so the resulting system of finite-difference equations emerges as a tridiagonal matrix eigenvalue problem. In a manner similar to the approach of Kokkotas \& Ruoff \cite{Kokkotas2001}, we compute the four lowest-frequency radial-oscillation modes of neutron stars constructed from various representative EoS models considered by \"{O}zel \& Freire \cite{Ozel2016}. The rapidly growing list of observational data combined with advances in EoS modeling has motivated us to investigate neutron stars in a wide range of total gravitational mass and for various types of representative EoS models, listed in Table \ref{tab:EoS}, based on nucleonic and hybrid nucleon--hyperon--quark matter models. The algorithm yields zero-frequency modes at the maxima and minima of the mass curves, as emphasized by \cite{Vath1992,Kokkotas2001}, while the adiabatic index of the equilibrium state characterizes the stiffness of the EoS at a given density. 
	
	Although proto-NSs \cite{Gondek1997} and strange stars \cite{Vath1992,Gondek1999} were also studied with finite-temperature EoS, we resort to the study of zero-temperature EoSs only in this paper. Some of our equations in this paper involve dissipative terms which will be applicable in a following research, where advanced theoretical data on the heat transfer of dense matter shall be implemented in order to accurately compute the rate at which viscosity and thermal conductivity drain the energy deposited in oscillation modes. Properties of the bulk and the shear viscosities, and of the thermal conductivity in NSs have been studied more in detail by a number of recent of works \cite{Shternin2008,Shternin2013,Shternin2017,Tolos2016}. The zero-temperature approximation is evidently no longer sufficient in case one investigates damped oscillations, due to internal friction or viscous effects, as a result of the finite temperature inside the star. The density and temperature dependence of transport coefficients in the crust and in the core have been described for different EoSs by \cite{Shternin2008}. Thermal conductivity and shear viscosity of nuclear matter arising from nucleon--nucleon interaction in non-superfluid neutron-star cores were considered by \cite{Shternin2013}, whereas those arising from the collisions among phonons in superfluid neutron stars were considered by \cite{Tolos2016}. The conversion of kinetic energy into heat and effects of viscosity on stellar pulsations in general have been addressed first by \cite{Kopal1964,Higgins1968,Mihalas1983}. And the effects of radial heat flow on the dynamical instability and on the thermal evolution of NSs have been extensively studied by Herrera and his colleagues in Refs. \cite{Herrera1989,Chan1993,Herrera1997}.

	\section{Realistic equation of state and thermodynamic properties of matter} \label{sec:EoS}
	Internal structure and macroscopic properties of NSs are strongly correlated with the EoS of dense matter, even though the exact EoS remains exceedingly uncertain especially at high densities. Although, the latest discovery of high-mass NSs PSR J1614-2230 \cite{Fonseca2016} and PSR J0348+0432 \cite{Antoniadis2013} has ruled out several  realistic models of EoS, suggesting that the maximal mass for NSs has to be larger than $M \sim 2 M_{\odot}$ for a given EoS, but the number of candidate models with maximal mass below this limit is still considerably large. In NS cores, the temperature of matter is far below the Fermi energy of its constituent particles and its particular thermodynamic state at $T \simeq 0$ is accurately described by the isentropic one-parameter equation of state
	\begin{equation} \label{EoS-eq}
	p = p(\rho),\, \epsilon = \epsilon(\rho),
	\end{equation}
	relating the pressure $p$ and energy density $\epsilon$ to the rest-mass density $\rho$ which exceeds nuclear density \cite{Lackey2014}
	\begin{equation} \label{nucdens-eq}
	\rho_{\text{nuc}} \simeq 2.3 \times 10^{14} \text{ g/cm}^3.
	\end{equation}
	In fact, densities in the cores are expected to be as large as $\rho \sim 5 - 10 \rho_{0}$, where the nuclear matter at saturation (i.e. at the minimum of the energy per nucleon) has the density $\rho_{0} \simeq 2.8 \times 10^{14} \text{ g/cm}^3$ or $n_{0} \simeq 0.16 \text{ fm}^{-3}$, where the baryon-number density is related to the baryon-mass density as $n_{B} = \rho_{B}/m_{u}$ and $m_{u} = 931.494$ MeV is the atomic mass unit. Given that neutrons geometrically overlap at $\rho \sim 4 \rho_{0}$ and with increasing overlap between nucleons, transitions to non-nucleonic phase are expected. \cite{Ozel2016} It is possible for ultra-dense matter to contain hyperon, pion or kaon condensates. \cite{Lackey2006} Some of the possibilities considered to date also include free quarks or color superconducting phases. \cite{Alford2005} 
	
	\subsection{Realistic nuclear-theory-based EoS models} \label{sec:realistic-EoS}
\begin{figure}
		\centering
		\includegraphics[scale=0.60]{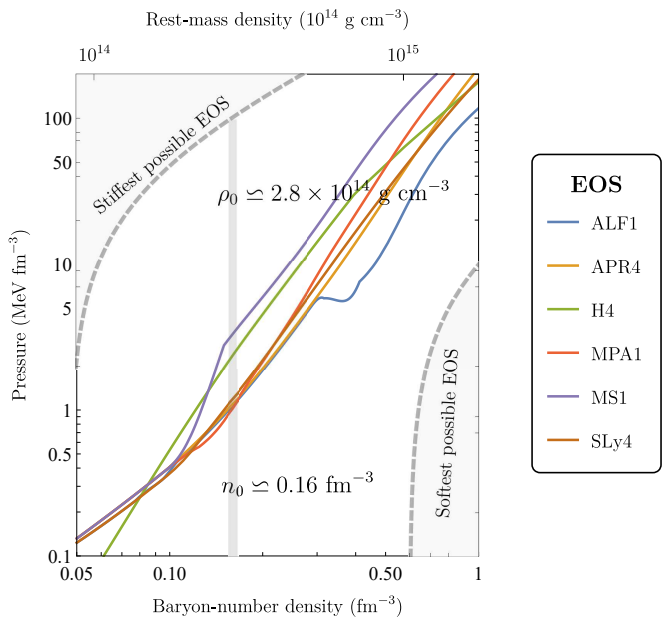}
		\caption{Pressure as a function of baryon-number density in the crust for some nucleonic and hybrid nucleon--hyperon--quark matter models based on different microphysics. The nuclear saturation density $n_{0} \simeq 0.16 \text{ fm}^{-3}$ is denoted by a shaded gray line. The range of pressures at $n_{0}$ is approximately a factor of $1$ to $3$ $\text{MeV fm}^{-3}$. The pressure and number density series were reproduced from Ref. \cite{Ozel2016}.}
		\label{fig:EoS}
	\end{figure}

	With the intention of covering a wide range of potential types of representative EoS models and generation methods, here we consider four EoSs of cold nucleonic matter (pure $npe\mu$ matter, i.e. the hypothetical components composed of neutrons, protons, electrons, and muons) and we follow the widespread naming convention of Refs. \cite{Ozel2016,Lattimer2001,Read2009}:
	\begin{itemize}
		\item \textsf{APR4} was derived by a variational method with modern nuclear potentials \cite{Akmal1998}; 
		\item \textsf{MPA1} was derived by a relativistic Brueckner--Hartree--Fock theory \cite{Muther1987}; 
		\item \textsf{MS1} was derived by a relativistic mean-field theory \cite{Mueller1996}; 
		\item \textsf{SLy4} was derived by a potential method \cite{Douchin2001};
	\end{itemize}
	and we also include three EoSs of `exotic' non-nucleonic matter (i.e. the hypothetical components composed of hybrid nucleon--hyperon--quark matter):
	\begin{itemize}
		\item \textsf{SQM1} is a hybrid EoS which describes relativistic non-interacting gas mixed with strange quark matter \cite{Prakash1995};		
		\item \textsf{H4} is a stiffer hyperon-based EoS derived by a relativistic mean-field theory \cite{Lackey2006}; 
		\item \textsf{ALF1} is a hybrid EoS which describes a APR4 nuclear matter for a low density and a color--flavor-locked quark matter for a high density with the transition density $3\rho_{0}$, where $\rho_{0} \simeq 2.8 \times 10^{14} \text{ g/cm}^3$ \cite{Alford2005}. 
	\end{itemize}
	\begin{table}[ht]
		\centering
		\begin{tabular}[t]{|@{\extracolsep{\fill}} c @{\extracolsep{\fill}} | c @{\extracolsep{\fill}} c @{\extracolsep{\fill}} c @{\extracolsep{\fill}} c @{\extracolsep{\fill}} c @{\extracolsep{\fill}}|}
			\Xhline{2\arrayrulewidth}
			\rowcolor[gray]{0.96}
			\textbf{ EoS } & \textbf{Constituents } &  $\bm{M_{\text{\textbf{max}}}/M_{\odot}}$ \textbf{ } & $\bm{R}$\textbf{/km } & \textbf{ } $\bm{r_{s}/R}$ \textbf{ } & \textbf{ Ref. } \\
			\Xhline{2\arrayrulewidth}
			APR4 & Nucleons & 2.22 & 9.94 & 0.66 & \cite{Muther1987} \\
			
			MPA1 & Nucleons & 2.47 & 11.25 & 0.65 & \cite{Muther1987} \\
			
			MS1 & Nucleons & 2.78 & 13.27 & 0.62 & \cite{Mueller1996} \\
			
			SLy4 & Nucleons & 2.06 & 9.9 & 0.61 & \cite{Douchin2001} \\
			
			SQM1 & Nuc., s quarks & 1.56 & 8.54 & 0.54 & \cite{Prakash1995} \\
			
			H4 & Nuc., $\Sigma,\, \Lambda$ & 2.04 & 11.81 & 0.51 & \cite{Lackey2006} \\
			
			ALF1 & u, d, s quarks & 1.50 & 9.02 & 0.49 & \cite{Alford2005} \\
			
			\Xhline{2.3\arrayrulewidth}
		\end{tabular}
		\caption{Nucleonic and hybrid nucleon--hyperon--quark matter models of NS cores. The upper four models correspond to matter with nucleonic degrees of freedom only, whereas the lower three models involves non-nucleonic states of nuclear matter, such as kaon condensates or hyperons. We present some characteristic parameters of the maximal-mass-mass configuration for each EoS (marked in Fig. \ref{fig:MR} by the symbols $\diamond$): $M_{\text{max}}$ is the maximal mass in units of $M_{\odot}$, $R$ is the circumferential radius in units of km, $r_{s}/R$ is the compactness, respectively.} \label{tab:EoS}
	\end{table}%
	Due to the fact that the realistic EoSs are given in tabulated forms only, one has to interpolate between the given values of pressure and energy density to obtain stellar models with continuous functions of radius $r$. \cite[p.~24]{Ruoff2000} We have simultaneously interpolated the pressure and its first and second derivatives as functions of energy density. We used different degrees of spline interpolants, including linear, quadratic and cubic spline curves, to find the smoothest possible interpolating function at the tabulated set of points. The following densities are determined at the boundary of two neighboring pieces for the zero-temperature EoSs that are describe in Sec \ref{sec:polytropic-EoS} by piecewise-polytropes with $n = 3$ pieces: $\rho_{1} \simeq 10^{14} \text{ g/cm}^3,\ \rho_{2} \simeq 5.012 \times 10^{14} \text{ g/cm}^3,\ \rho_{3} \simeq 10^{15} \text{ g/cm}^3.$  ALF1 has the lowest pressure among the above considered types of EoSs and thus, making it the softest one. APR4, MPA1 and Sly4 have also relatively small pressure as in the case of ALF1 for a low-density region $\rho_{1} \leq \rho \leq \rho_{3}$, but for $\rho_{2} \lesssim \rho \leq \rho_{3}$, the pressure is higher than that for ALF1. Thus, for $\rho < \rho_{3}$, which NSs of canonical mass $1.3 \text{ -- }1.4 M_{\odot}$ have, these EoSs are soft as far as the canonical NSs are concerned. It is worthy of note that for a relatively small value of $p_{2}$, the adiabatic index, as illustrated in Fig. \ref{fig:gamma}, is as large as $\Gamma_{2} \sim 3$, owing to the fact that the maximal mass of NS has to be $M_{\text{max}} \lesssim 2M_{\odot}$ for a given EoS. Thus, an EoS that is soft at $\rho = \rho_{2}$ has to be in general stiff for $\rho \gtrsim \rho_{3}$. However, MPA1 has pressure that even exceeds that of H4 for a high-density region $\rho \gtrsim \rho_{3}$. By contrast, H4 and MS1 have pressure higher than the rest for $\rho \lesssim \rho_{3}$, although the EoS becomes softer for a high-density region $\rho \gtrsim \rho_{3}$. In particular, MS1 has extremely high pressure among many other types of EoSs for $\rho \lesssim \rho_{3}$, and on that account, it is the stiffest EoS. All the distinguishing features mentioned above are reflected in Fig. \ref{fig:EoS} which displays the pressure in NS as a function of the baryon-number density or of rest-mass density. Table \ref{tab:EoS} lists the representative EoSs, the constituent particles, the maximal mass and circumferential radius of NSs of total mass $1.4 M_{\odot}$, respectively, in descending order of compactness.
	
	\subsection{One-piece and piecewise-polytropic EoSs for heating and cooling processes} \label{sec:polytropic-EoS}
\begin{figure}
		\centering
		\includegraphics[scale=0.60]{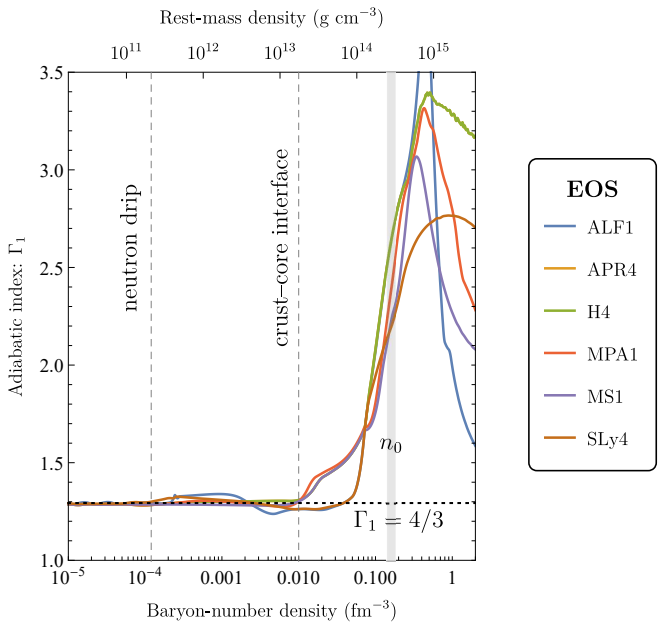}
		\caption{The effective adiabatic index $\Gamma_{1}$ as a function of baryon-number density (or rest-mass density) for the set of representative EoS models considered in Table \ref{tab:EoS}. The dotted horizontal line corresponds to $\Gamma_{1} = 4/3$. The average value of the exponent $\Gamma_{1} = d \log p / d \log n_{B} \simeq 2$ holds for nucleonic EoS models in the vicinity of nuclear saturation density $n_{0} \simeq 0.16 \text{ fm}^{-3}$, denoted by a shaded gray line. The dashed vertical lines correspond to the neutron drip and crust-core interface points, $\rho_{\text{ND}} \simeq 4 \times 10^{11} \text{ g/cm}^3,\ \rho_{\text{CC}} \simeq 1.75 \times 10^{13} \text{ g/cm}^3$, respectively.}
		\label{fig:gamma}
	\end{figure}

	Although the tabulated, nuclear-theory-based EoS models listed in Table (\ref{tab:EoS}) are more realistic, they cannot be given in analytic terms over the whole range of energy densities inside NSs. A very common closed-form EoS is the polytropic one,
	\begin{equation} \label{polytropic-EoS}
	p = K\rho^{\Gamma},
	\end{equation}
	which describes a non-interacting, degenerate matter. In general, the `polytropic constant' $K = K(s)$ depends on the entropy, however, the degenerated matter dynamics in zero-temperature approximation can be modeled as an adiabatic flow with a constant $K$. The internal energy is given by the first law of thermodinamics for adiabatic process $(\delta Q = 0)$, which can be integrated to obtain
	\begin{equation} \label{internal-energy}
	\epsilon = \rho + \frac{1}{\Gamma-1}K\rho^{\Gamma} = \rho + \frac{1}{\Gamma-1}p,
	\end{equation}
	where we have imposed that $\lim\limits_{\rho \to 0}\epsilon/\rho = 1$. \cite{Tooper1964} With the adiabatic assumption, eqs. (\ref{polytropic-EoS}--\ref{internal-energy}) represent a barotropic fluid where the pressure is just a function of $\rho$. The adiabatic index $\Gamma_{1}$, defined by 
	\begin{equation} \label{adiabatic-index}
	\Gamma_{1} = \frac{d \log p}{d \log \rho} = \frac{\epsilon+p}{p}\frac{dp}{d\epsilon},
	\end{equation}
	is an important dimensionless parameter characterizing the stiffness of the EoS \eqref{polytropic-EoS} at a given density. \cite[p.~190]{Bona2009} For instance, a non-relativistic degenerate Fermi gas is reasonably well described by a polytropic EoS that scales as $p \propto \rho^{5/3}$, and for highly relativistic degenerate Fermi gases, $p \propto \rho^{4/3}$. Generally, $\Gamma_{1}$ and the speed of sound $c_{s}$ depend on the dynamical regime determined by $\rho$ (or $\epsilon$) as presented in Fig. \ref{fig:gamma} for the representative EoSs listed in Table \ref{tab:EoS}. The non-monotonic variation of $\Gamma_{1}$ with baryon-number density (or rest-mass density) is revealed by Figs. \ref{fig:gamma} and \ref{fig:seed-of-sound}, respectively. As a sudden increase of $\Gamma_{1}$ comes about at $\rho \simeq 1.3 \times 10^{13} \text{ g/cm}^3$ for the stiffer MPA1 and MS1, but for most EoSs, it only occurs at $\rho \simeq 10^{14} \text{ g/cm}^3$. An abrupt change occurs in the stiffness of the matter when the Fermi energy of free neutrons exceeds the rest-mass energy of the neutrons. It is quite apparent that the matter undergoes phase transitions at this point which is associated with the `neutron drip point' and only slightly depends on the particular EoS model. \cite{Kokkotas2001} The decay rate of the adiabatic index after its maximum also varies from one EoS to another, cf. Ref. \cite{Ibanez2018}. The following two conditions must be satisfied by these EoSs: 
	\begin{enumerate}
		\item The thermodynamic stability requires the EoS be monotonic ($dp/d\rho \geq 0$ and $dp/d\epsilon \geq 0$), and therefore $\Gamma_{1}$ must be positive.
		\item The causality requires the speed of sound $c_{s}$ be less than the speed of light which is expressed by
		\begin{equation} \label{speed-of-sound}
		c_{s}^{2} = \frac{dp}{d\epsilon}\bigg\rvert_{s} \leq 1.
		\end{equation}
	\end{enumerate}
	The constraint of eqs. (\ref{adiabatic-index}--\ref{speed-of-sound}) on the pressure-averaged adiabatic index, defined by
	\begin{equation}
	\bar{\Gamma}_{1} = \frac{\int_{0}^{R}\Gamma_{1}p(4\pi r^{2})dr}{\int_{0}^{R}p(4\pi r^{2})dr},
	\end{equation}
	demand that
	\begin{equation}
	\bar{\Gamma}_{1} = \frac{\epsilon+p}{p}c_{s}^{2} \geq \frac{4}{3}
	\end{equation}
	within a dynamically stable star of radius $R$. \cite{Ivanov2017} Expressly, $\bar{\Gamma}_{1} < 4/3$ is a sufficient condition to ensure the dynamical stability of spherical stars in Newtonian gravity (as we have outlined in Appx. \ref{sec:dynamical-stability}). However, in the stronger gravity of General Relativity (GR), even models with the stiffest EoSs become unstable for some value $R/M > 9/8$. The more rigorous constraint on $\bar{\Gamma}_{1}$ for a star is to be stable against radial perturbation
	\begin{equation}
	\bar{\Gamma}_{1} < \frac{4}{3} + K\frac{4M}{R},
	\end{equation}
	where $K$ is positive and of order of unity. For dynamical oscillations of neutron stars, the adiabatic index $\Gamma_{1}$ does not coincide with the polytropic one $\Gamma$, and the critical point (see Sec. \ref{sec:dynamical-stability-fundamental-mode}) implies a secular instability whose growth time is long compared to the dynamical timescale of stellar oscillations \eqref{time-scale}. \cite{Ivanov2017}
	
	It has been demonstrated by Ref. \cite{Read2009} that a piecewise-polytropic EoS with three pieces $(n = 3)$ above the nuclear density approximately reproduces most properties of the representative EoSs listed in Table \ref{tab:EoS}. These models of nuclear-theory-based zero-temperature EoSs at high density are modeled with a small number of parameters and the expression for pressure \eqref{polytropic-EoS} is written in a parameterized form as
	\begin{equation} \label{piecewise-eq}
	p(\rho) = K_{i}\rho^{\Gamma_{i}} \quad \text{for} \quad \rho_{i} \leq \rho < \rho_{i+1} \quad (0 \leq i \leq n),
	\end{equation}
	where $n$ is the number of the pieces used to parameterize a EoS at high density, $\rho_{i}$ is the rest-mass density at the boundary of two neighboring $(i-1)$-th and $i$-th pieces, $K_{i}$ is the polytropic constant for the $i$-th piece, and $\Gamma_{i}$ is the adiabatic index for the $i$-th piece. Here, $\rho_{0} = 0$, $\rho_{1}$ denotes a nuclear density evaluated in eq. \eqref{nucdens-eq}, and $\rho_{n+1} \to \infty$. Other parameters $(\rho_{i};\, K_{i};\, \Gamma_{i})$ are determined by fitting with an EoS. Requiring the continuity of the pressure at each $\rho_{i},\, 2n$ free parameters, say $(K_{i};\, \Gamma_{i})$, determine the EoS completely. \cite{Hotokezaka2012}
	
	One of the most serious drawbacks of the polytropic EoS is that, although \eqref{polytropic-EoS} is a good approximation for a `cold' star, there are extremely energetic processes, like the merger of stars or accretion from a disk, which can increase enormously the temperature and a simple polytrope will not provide a physical description. A more realistic EoS in closed form can be obtained by a combination of the polytripic EoS to describe the cold part and an ideal EoS for the thermal one, allowing for fluid heating due to shocks. The hybrid EoS is given by
	\begin{equation}
	p = K\rho^{\Gamma} + (\Gamma_{\text{th}} -1)\rho\epsilon_{\text{th}}
	\end{equation}
	with an adiabatic thermal index $\Gamma_{\text{th}}$ that can be different from the adiabatic cold index $\Gamma$. The internal energy can be split into a thermal and a cold part,
	\begin{equation}
	\epsilon = \epsilon_{\text{th}} + \epsilon_{\text{cold}}.
	\end{equation}
	The total internal energy density $\epsilon$ can be obtained from the evolution of the conserved quantities, whilst the cold part is described by \eqref{internal-energy}, leading to the explicit expression
	\begin{equation}
	p = K\frac{\Gamma-\Gamma_{\text{th}}}{\Gamma -1}\rho^{\Gamma} + (\Gamma_{\text{th}} -1)\rho\epsilon_{\text{th}}.
	\end{equation}
	It is possible to extend this approach by using a collection of continous piecewise-polytropes \eqref{piecewise-eq} in hybrid EoS models, which in turn allows an accurate match with any type of realistic nuclear-theory-based EoS at high density.

	\subsection{Stress--energy tensor for dissipative fluids}
	\begin{table*}[ht]
		\centering
		\begin{tabular}[t]{|@{\extracolsep{\fill}} c @{\extracolsep{\fill}} c @{\extracolsep{\fill}} c @{\extracolsep{\fill}} c @{\extracolsep{\fill}}|}
			\Xhline{2\arrayrulewidth}
			\rowcolor[gray]{0.96}
			\textbf{ Process } & \textbf{ Reaction } & \begin{tabular}{@{}c@{}}\textbf{ Heat flux } \\ $\text{[erg }\text{cm}^{-3}\text{s}^{-1}]$ \end{tabular} & \begin{tabular}{@{}c@{}}\textbf{ Local luminosity } \\ $[\text{erg }\text{s}^{-1}]$ \end{tabular} \\
			\Xhline{2\arrayrulewidth}
			Direct Urca & \begin{tabular}{@{}c@{}}$n \rightarrow p + e + \bar{\nu}_{e}$ \\ $p + e \rightarrow n + \nu_{e}$ \end{tabular} & $Q \sim 3 \times 10^{27}T_{9}^{8}$ & $L_{\nu} \sim 10^{46}T_{9}^{8}$ \\		
			
			Modified Urca & \begin{tabular}{@{}c@{}}$n + N \rightarrow p + e + N + \bar{\nu}_{e}$ \\ \textbf{ } $p + e + N \rightarrow n + e + N + \nu_{e}$ \textbf{ } \end{tabular} & $Q \sim 10^{20-22}T_{9}^{8}$ & $L_{\nu} \sim 10^{38-40}T_{9}^{8}$ \\
			
			Bresmmstrahlung & $N + N \rightarrow N + N + \nu + \bar{\nu}$ & $Q \sim 10^{18-20}T_{9}^{8}$ & $L_{\nu} \sim 10^{36-38}T_{9}^{8}$ \\
			\Xhline{2.3\arrayrulewidth}
		\end{tabular}
		\caption{Possible mechanisms of neutron star cooling by various neutrino-emission processes due to nucleon--nucleon collisions assumed to take part in the core. The modified Urca process has the neutron and the proton branch, each including a direct and an inverse, where $N = n$ or $p$, respectively. \cite{Lim2017}} \label{tab:emission}
	\end{table*}%

	Neutrino emission processes are supposed to be the main sources of energy loss in the stellar core in the later stages of stellar evolution. For the reason, the equations of relativistic fluid dynamics to describe energy--momentum conservation are written as
	\begin{equation} \label{energy-conservation}
	\partial_{\beta}T^{\alpha\beta} = -Q_{\nu}u^{\alpha},
	\end{equation}
	where $Q_{\nu}$ is the total neutrino emissivity of all processes outlined in Table \ref{tab:emission}, and the stress--energy tensor for dissipative fluids can be written as the sum of three components
	\begin{equation} \label{energy-momentum-tensor}
	T^{\alpha\beta} = T^{\alpha\beta}_{\text{pf}} + T^{\alpha\beta}_{\text{visc}} + T^{\alpha\beta}_{\text{heat}},
	\end{equation}
	where  
	\begin{equation} \label{energy-momentum-tensor2}
	\begin{array}{l}
	T^{\alpha\beta}_{\text{pf}} = (\epsilon + p)u^{\alpha}u^{\beta} + p h^{\alpha\beta}, \\[10pt]
	T^{\alpha\beta}_{\text{heat}} = Q^{\alpha}u^{\beta} + Q^{\beta}u^{\alpha} \\[10pt]
	T^{\alpha\beta}_{\text{visc}} = -\zeta\Theta h^{\alpha\beta} - 2\eta\sigma^{\alpha\beta}
	\end{array}
	\end{equation}
	are the perfect fluid, heat flux and viscosity stress--energy tensors, respectively. \cite{Gusakov2005} Note that despite its causality and stability problems, the above description of stress--energy tensor has been widely used in Eckart's theory of relativistic irreversible thermodynamics \cite{Eckart1940}. As in eq. \eqref{EoS-eq}, the variables $p$ and $\epsilon$ represent the isotropic pressure and energy density measured by a comoving observer with velocity $u^{\alpha}$, which satisfies $u^{\alpha}u_{\alpha} = 1$ with $u^{0} > 0$, and
	\begin{equation} \label{projection}
	h^{\alpha\beta} = g^{\alpha\beta} + u^{\alpha}u^{\beta}
	\end{equation}  
	is the standard projection tensor onto 3-space normal to flow. The symmetric trace-free spatial shear tensor is defined as
	\begin{equation} \label{shear-tensor}
	\sigma^{\alpha\beta} = \frac{1}{2}\left(u^{\alpha}_{;\mu}h^{\mu\beta} + u^{\beta}_{;\mu}h^{\mu\alpha}\right) - \frac{1}{3}\Theta h^{\alpha\beta}
	\end{equation}
	and expansion scalar (or dilatation rate)
	\begin{equation} \label{dilatation-rate}
	\Theta = u^{\alpha}_{\; ;\alpha}
	\end{equation}
	is associated with the convergence (or divergence) of the fluid world lines. The heat flux density in eqs. (\ref{energy-conservation}--\ref{energy-momentum-tensor2}) written as
	\begin{equation} \label{heat-flux}
	Q^{\alpha} = -\kappa h^{\alpha\beta}\left(T_{,\beta}-Ta_{\beta}\right)
	\end{equation}
	is a spacelike vector, $Q^{\alpha}u_{\alpha} = 0$, that describes the flow of thermal energy per unit of area along spatial cordinate axis $x^{\alpha}$ per unit of time. The first term in eq. \eqref{heat-flux} corresponds to the non-relativistic Fourier's law of heat conduction, the second term takes into account the relativistic effect of isothermal heat flux due to the inertia of energy with $a_{\beta} = u^{\gamma}u_{\beta;\gamma}$ being the acceleration of fluid.  The negative sign indicates that heat flows from higher to lower temperature regions. 
	
	In eqs. \eqref{energy-momentum-tensor2} and \eqref{heat-flux}, $\eta,\, \zeta$, and $\kappa$ are collectively called transport coefficients (or dissipation coefficients). The bulk viscosity coefficient $\zeta$ defines the resistance of the medium to gradual uniform compression or expansion; and $\kappa$ is non-negative and accounts for the thermal conductivity, respectively. \cite{Rezzola2013} The shear (also called as `common' or `dynamic') viscosity coefficient $\eta$ describes the fluid's resistance to gradual shear deformation which is assumed to be equal to the electron shear viscosity $\eta_{e}$ in the stellar core; and we take it from \cite{Gusakov2005} in this paper.  Shternin and Yakovlev (2008) \cite{Shternin2008} concluded that the shear viscosity of neutrons and of protons can be neglected for the reason that it depends strongly on the nuclear interaction model and the many-body theory.

	\section{Equilibrium stellar models} \label{sec:equilibrium-stellar-model}
	\begin{figure}
	\centering
	\includegraphics[scale=0.7]{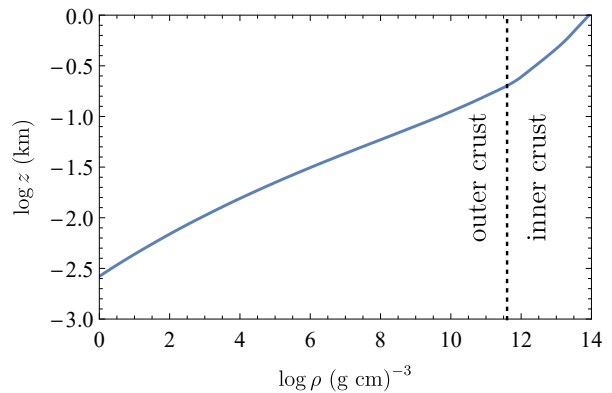}
	\caption{The internal structure of the crust in neutron stars of total mass $M = 1.4M_{\odot}$ with SLy4 EoS in their cores is characterized by the proper depth $z$ below the surface as a function of rest-mass density $\rho$. A point of inflection in $z$ marks the border of the outer and inner crust which is indicated by a dashed line.}
	\label{fig:Depth}
	\end{figure}
\begin{figure}
	\centering
	\includegraphics[scale=0.6]{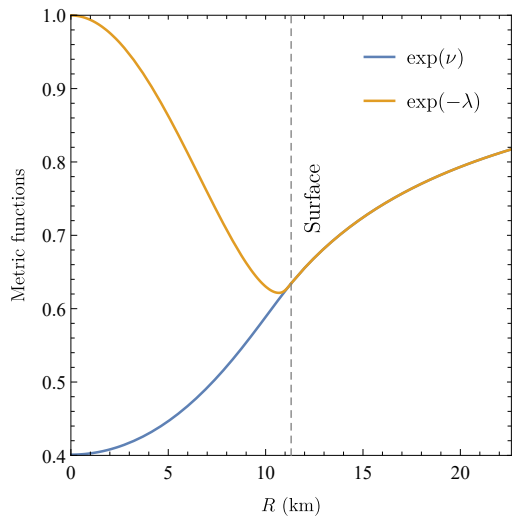}
	\caption{The radial profile of metric functions of neutron stars of total mass $M = 1.4M_{\odot}$ with SLy4 EoS in their cores. The metric function $\exp(\nu)$ steadily increases from the center towards the assimptotic region. $\exp(-\lambda)$ is flat near the center and reaches a minimum near the surface, where it joins its counterpart.}
	\label{fig:metric}
	\end{figure}

	
	We consider a static spherically symmetric star, described by the metric
	\begin{equation} \label{line-element}
	ds^{2} = e^{\nu}dt^{2} - e^{\lambda}dr^{2} - r^{2}d\Omega^{2},
	\end{equation}
	where $t$ and $r$ are the time and radial coordinates, $d\Omega$ is a solid angle element in a spherical frame with the origin at the stellar center and 
	\begin{equation} \label{metric-func}
	\nu = \nu(t,r), \quad \lambda = \lambda(t,r)
	\end{equation}
	are the metric functions. The later function is often replaced by the expression
	\begin{equation} \label{mass-func}
	e^{\lambda} = \left(1-2m/r\right)^{-1},
	\end{equation}
	where the gravitational mass contained within the radius $r$ is given as
	\begin{equation} \label{mass-func2}
	m(r) = 4 \pi \int_{0}^{r}\rho(r')r'^{2}dr'.
	\end{equation}
	Outside the star the metric reduces to the Schwarzschild form:
	\begin{equation} \label{exterior-line-element}
	e^{\lambda} = (1 - r_{s}/r)^{-1}.
	\end{equation}
	The characteristic scale factor, defined by 
	\begin{equation} \label{Schwarzschild-radius}
	r_{s} = 2GM/c^{2},
	\end{equation}
	is referred to as Schwarzschild radius or `gravitational radius' and the total gravitational mass of the star, defined by $M = m(R)$, which acts as the source of the gravitational field outside the star $(r > R)$. The crust corresponds to the layer $r_{cc} < r < R$, where $r_{cc}$ determines the crust-core interface. The proper depth below the stellar surface is defined by
	\begin{equation}
	z(r) = \int_{r}^{R}e^{\lambda(r')}dr',
	\end{equation}
	as the proper radial distance between the stellar surface $R$ and a given layer of radius $r$. The proper depth and the metric functions of NSs of total mass $M = 1.4M_{\odot}$ with SLy4 EoS in their cores are shown in Figs. \ref{fig:Depth}--\ref{fig:metric} as function of radius in units of km. (Cf. Figure 37 in Ref. \cite{Chamel2008} and Figure 4.1. in Ref \cite{Camenzind2007}.)
	
	As in Ref. \cite{Barta2019}, the physical variables for energy density $\epsilon$ and for isotropic pressure $p$ are replaced by the corresponding effective variables
	\begin{equation} \label{Effective-variables-eq}
	\bar{\epsilon} = \epsilon + (T_{\text{visc}} + T_{\text{heat}})^{0}_{\;0}, \quad \bar{p} = p - (T_{\text{visc}} + T_{\text{heat}})^{1}_{\;1}
	\end{equation}
	that incorporate time-dependent dissipative contributions of the stress--energy tensor \eqref{energy-momentum-tensor2}. From now on, we will only use these effective variables and for the sake of simplicity, we shall not put any bar over them. The structure of compact stars in hydrostatic equilibrium is then constrained by the following set of equations:
	\begin{subequations} \label{Einstein-eq}
		\begin{align}
		& \displaystyle \frac{dm}{dr} = \displaystyle 4\pi r^{2}\epsilon   \label{Einstein-eq:a} \\
		& \displaystyle\frac{dp}{dr} = \displaystyle -\frac{(\epsilon + p)(m + 4\pi r^3 p)}{r(r-2m)} + \frac{2\ddot{\lambda} + \dot{\lambda}(\dot{\lambda}-\dot{\nu})}{16\pi}e^{-\nu}   \label{Einstein-eq:b} \\
		& \displaystyle\frac{d\nu}{dr} = \displaystyle -\frac{2}{\epsilon + p}\frac{dp}{dr},   \label{Einstein-eq:c}
		\end{align}
	\end{subequations}
	\begin{figure}
	\centering
	\includegraphics[scale=0.6]{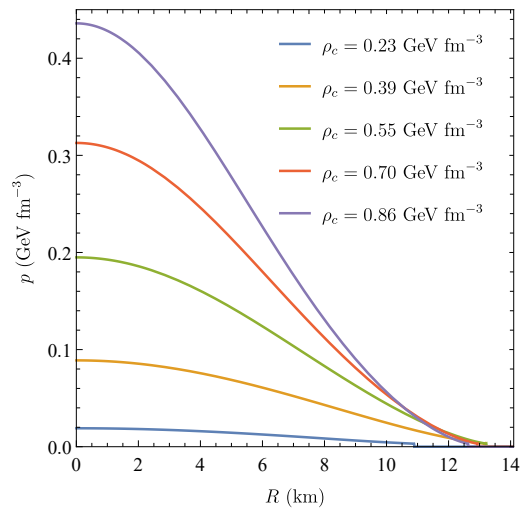}
	\caption{Radial profile of pressure in units of $\text{ GeV fm}^{-3}$ for different central densities in neutron stars with MS1 EoS in their cores. The blue line represents the pressure for central density $\rho_{c} = 0.23 \text{ GeV fm}^{-3}$ associated with the minimum mass $M = 1.17$ $M_{\odot}$ allowed for neutron stars. \cite{Suwa2018} The purple line represents the pressure for central density $\rho_{c} = 0.86 \text{ GeV fm}^{-3}$ associated with the maximal mass $M = 2.78$ $M_{\odot}$ allowed for EoS MS1. The other three lines represent the pressure for values of central density equidistant from the aforementioned extremal central densities.}
	\label{fig:PR}
	\end{figure}
	\begin{figure*}
	\begin{minipage}{.48\linewidth}
		\centering
		\subfloat[Total gravitational mass--radius relations.]{\label{fig:MR}
			\includegraphics[width=1.0\linewidth]{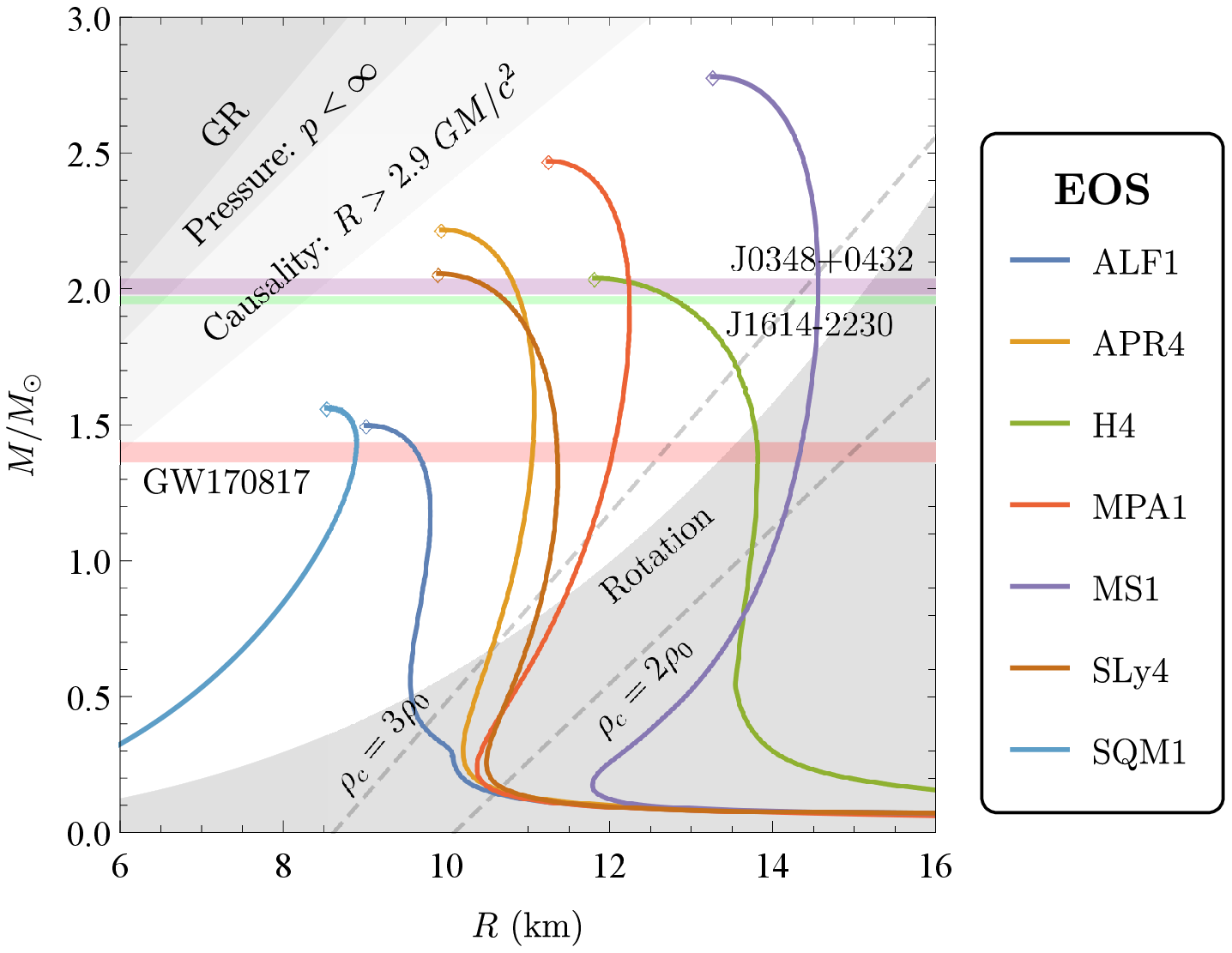}}
	\end{minipage}
	\begin{minipage}{.48\linewidth}
		\centering
		\subfloat[Total gravitational mass--central energy density relations.]{\label{fig:ME}
			\includegraphics[width=1.0\linewidth]{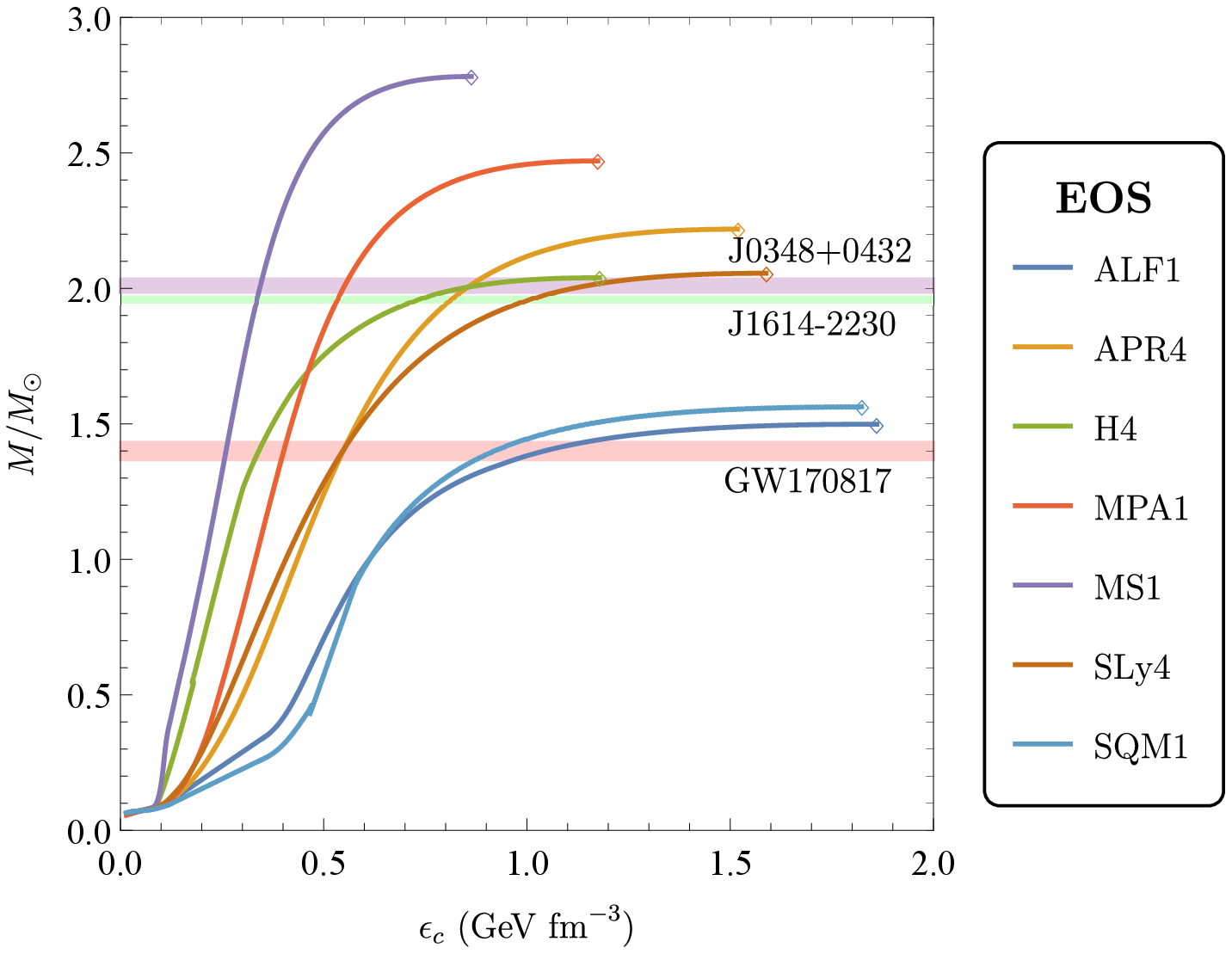}}
	\end{minipage}\par\medskip
	\caption{$M(R)$ and $M(\epsilon_{c})$ curves for each nucleonic EoS (APR4, MPA1, MS1, SLy4) are shown as light-colored curves, the blue curves refer to self-bound quark stars (ALF1, SQM1), and the green curve to a strange star model (H4), corresponding to the realistic EoS models displayed on the $p$--$n$ plane in Fig. \ref{fig:EoS}. The symbols $\diamond$ mark the maximal-mass configuration of the respective family of stellar models. Most EoSs involving non-nucleonic matter, such as kaon condensates or hyperons, tend to predict an upper limit around $2.01 M_{\odot}$ for the maximal mass of neutron stars. The purple and the green bands indicate the rapidly rotating NSs in millisecond pulsars, cataloged as PSR J1614-2230 \cite{Fonseca2016} and J0348+0432 \cite{Antoniadis2013}, with the highest-known mass of $1.97 \pm 0.04$ $M_{\odot}$ and $2.01 \pm 0.04$ $M_{\odot}$, respectively. The light red band shows the interval of total binary NS masses inferred from gravitational-wave signal GW170817. The dashed gray lines refer to stars whose central density $\rho_{c}$ is double or triple of nuclear saturation density $\rho_{0}$, respectively. The upper left areas of different shades of grayscale refer to regions of the $M$--$R$ plane excluded by GR constraint for $R>2GM/c^2$, by finite pressure for $R > 2.25GM/c^2$, and by causality for $R>2.9GM/c^2$. The lower shaded area indicates the region bounded by the realistic mass-shedding limit $R/\text{10 km} < \mathcal{C}^{2/3}(M/M_{\odot})^{1/3}(f_{K}/\text{1 kHz})$ for the highest-known Keplerian frequency, $f_{K} \backsimeq 716$ Hz, for the uniformly rotating NS PSR J1748-2446ad. The deviation of $\mathcal{C}$ from its Newtonian value of $1.838$ depends, in GR, (as computed by \cite{Haensel2009}) on the NS interior mass distribution. For a hadronic EoS, $\mathcal{C} = 1.08$, whilst for a strange star with a crust, $\mathcal{C} \backsimeq 1.15$.}
	\end{figure*}
	where overdots denote partial differentiation with respect to $t$. \cite{Vath1992} The mass-radius equation \eqref{Einstein-eq:a} and the Tolman--Oppenheimer--Volkoff (TOV) equation \eqref{Einstein-eq:b} constraint the stellar structure in static gravitational equilibrium, whereas eq. \eqref{Einstein-eq:c} determines the metric function $\nu(r)$. Supplemented with a given EoS \eqref{EoS-eq}, the set of equations \eqref{Einstein-eq} can be integrated from the center of the star with initial condition $m(0) = 0$ and an arbitrary value for the central density $\rho_{c} = \rho(0)$, until the pressure $p(r)$ will vanish at some radius $R$ which corresponds to the star's surface.\footnote{\label{footnote1} It is clear from eq. \eqref{Einstein-eq:b} that $p(r)$ is strictly decreasing with increasing $r$. However, for some of the representative EoSs, $p$ may not drop to zero by the time $dp/dr$ becomes negligible. Therefore the numerical integration has to be stopped when the decrease of $dp/dr$ comes to a halt and the value of $p$ has fallen sufficiently low.} Even though in Sec. \ref{sec:realistic-EoS} we interpolate between the tabulated values of pressure and energy density such a way that produce the smoothest possible EoS, the stellar models obtained from the integration of eqs. \eqref{Einstein-eq} are not as smooth as those of polytropic ones. This hindrance is especially evident in the vicinity of the stellar surface, where the pressure is zero, but it gets immensely increased as one moves into the stellar interior. (See footnote \ref{footnote1} and Fig. \ref{fig:PR}.) The underlying reason is that $\rho(r)$ and other matter functions contain unwanted bumps and sharp bends as the matter undergoes phase transitions for increasing pressure. \cite[p.~24]{Ruoff2000} As stated in Sec. \ref{sec:polytropic-EoS}, these sharp phase transitions from hadronic state into another (e.g. hyperons, pion or kaon condensates) change the state and the stiffness of matter quite abruptly, together with the adiabatic index $\Gamma_{1}$. The stiffening of EoS plays a crucial role in setting the maximal mass: the maximal mass increases as the EoS gets stiffer at high densities. \cite{Ozel2016} The extra pressure from the repulsive force raises the maximal mass. Considering that among the EoS models present in this paper, MS1 allows the largest maximal mass to pile up, Fig. \ref{fig:PR} displays the radial profile of isotropic pressure for different central densities of NSs with MS1 EoS in their core.
	
	For each particular EoS there is a unique family of equilibrium configurations which is parameterized by $\rho_{c}$. Figs. \ref{fig:MR}--\ref{fig:ME} show the corresponding mass--radius and mass--central energy density relations, respectively, which can be interpreted as sequences of stellar models $M = M(R)$ and $M = M(\rho_{c})$. $M(R)$ curves demonstrate that low-mass configurations, having low central densities, mainly below  $\rho_{0}$, depend dominantly on the low-density part of the EoS, whereas the maximal mass is hardly affected (see also \cite{Lindblom1992}). On the other hand the Harrison--Zel'dovich--Novikov criterion \cite{Harrison1965,Zeldovich1971} for static stability of compact stars implies that $\partial M(\rho_{c})/\partial\rho_{c} \geq 0$. As mass piles up with increasing $\rho_{c}$, the $M$--$R$ relation probes the higher density part of the EoS, in the range $\rho_{0} \lesssim \rho_{c} \lesssim 3 \rho_{0}$. Lattimer \& Prakash \cite{Lattimer2001} showed that across various EoS models, the scaling of radius is approximately $R \propto p^{1/4}$ at a central density $\rho_{c} \backsimeq 1.5\rho_{0}$, so that a radius measurement is probing the EoS, just above nuclear density \eqref{nucdens-eq}. The $M(R)$ curves that have attained sufficient mass exhibit vertical segments with radii varying from $10$ to $16$ km. The vertical segments can be understood from the scaling of hadronic EoS models which is typically close to $p \propto \rho^{2}$ for higher densities (as demonstrated by $\Gamma_{1} \backsimeq 2$ in Fig. \ref{fig:gamma}). In this case, the radius is nearly independent of mass. Those models of hadronic EoS with extreme softening (due to a kaon or pion condensate, high abundances of hyperons, or a low-density quark-hadron phase transition) do not have pronounced vertical segments. \cite{Lattimer2012}

	\section{Adiabatic radial perturbations and speed of their propagation} \label{sec:perturbations}
	To move beyond the state of equilibrium, we now introduce small, adiabatic perturbations of eqs. \eqref{Einstein-eq} in such a way that does not violate the spherical symmetry of the star.\footnote{Note that the inclusion of transverse perturbations results in the appearance of non-radial oscillation modes. Although non-radial modes of NSs (such as g-modes and r-modes, being in the sensitivity bands of aLIGO and aVirgo detectors) are of great importance for being accompanied by the emission of gravitational waves, in order to make the problem more easily tractable, we specialize to purely radial modes in this paper.} The metric functions from the line element \eqref{line-element} can then be recast as
	\begin{equation}
	\nu(r,t) = \nu_{0}(r) + \delta\nu(r,t); \quad \lambda(r,t) = \lambda_{0}(r) + \delta\lambda(r,t),
	\end{equation}
	where a small parameter $\delta \ll 1$ denotes the ratio between the scale of variation of the metric functions $(\nu_{0},\, \lambda_{0})$ corresponding to the unperturbed equilibrium stellar model and $(\delta\nu,\, \delta\lambda)$ are metric perturbations due to the radial pulsations. To obtain the equations that govern the radial pulsations, let us consider a fluid element displaced radially outward from an initial position at $r$ to $r + \delta r$. The fluid element expands (or, if displaced inward, contracts) with the radial velocity
	\begin{equation} \label{radial-velocity}
	\delta u_{r} = \frac{\partial}{\partial t}\delta r - \frac{(\delta T_{\text{visc}})^{1}_{\;0} + (\delta T_{\text{heat}})^{1}_{\;0}}{\bar{p}_{0} + \bar{\epsilon}_{0}}
	\end{equation}
	as its pressure adjusting with the group velocity \eqref{group-velocity} across the fluid to the external pressure, given by
	\begin{equation}
	\delta p = \frac{dp}{dr}\delta r.
	\end{equation}
	Note that the term involving $\delta T_{\text{visc}}$ and $\delta T_{\text{heat}}$ represents the radial component of linear momentum density and is present only if viscosity or heat transfer occurs across the fluid elements, cf. \eqref{t1components}. We assume that the motion of the fluid element, moving with a four-velocity defined by
	\begin{equation} \label{four-velocity}
	u_{\mu} = (-e^{\nu_{0}/2},\, e^{\lambda_{0}-\nu_{0}/2}\delta u_{r},\, 0,\, 0),
	\end{equation}
	is faster than the time for heat to flow out of the fluid element. The perturbation is thus adiabatic, and from eqs. (\ref{adiabatic-index}--\ref{speed-of-sound}) the change in the energy density of the fluid element due to the perturbation is
	\begin{equation}
	\delta\epsilon = \frac{d\epsilon}{dp}\bigg\rvert_{s}\delta p = \frac{d\epsilon}{dp}\bigg\rvert_{s}\frac{dp}{dr}\delta r = \frac{\epsilon + p}{\Gamma p}\frac{dp}{dr}\delta r,
	\end{equation}
	where the adiabatic index of the perturbation $\Gamma$ is assumed to be equal to the adiabatic index of the equilibrium configuration $\Gamma_{1}$ in eq. \eqref{adiabatic-index}.
	\begin{figure}
		\centering
		\includegraphics[scale=0.60]{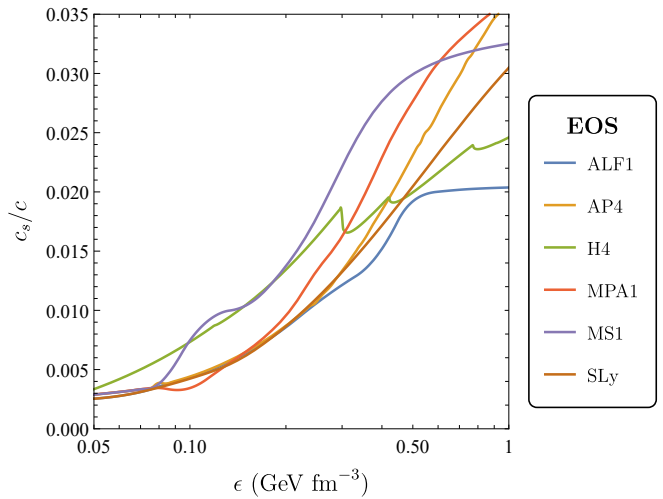}
		\caption{The local speed of sound relative to the speed of light is displayed for realistic EoS models in the same range as in Fig. \ref{fig:EoS}. It depends on the dynamical regime determined by the local energy density in compact stars.}
		\label{fig:seed-of-sound}
	\end{figure}
	
	Let us consider the radial displacement of the fluid element $\delta r$ which satisfies a linear wave equation being expanded in an asymptotic series
	\begin{equation} \label{time-dependence}
	\delta r = \exp[i\theta(r,t)]\sum_{n=0}^{\infty}X_{n}(r,t),
	\end{equation}
	where the angular frequency and the wavenumber (or spatial frequency), defined as
	\begin{equation} \label{four-wavevector}
	\omega = -\frac{\partial\theta}{\partial t} \quad \text{and} \quad k = \frac{\partial\theta}{\partial r},
	\end{equation}
	are the coordinate components of the four-wavevector, respectively. \cite{Glass1983b} The higher-order terms $X_{n}$ in the series \eqref{time-dependence} become successively smaller in the small parameter $\delta \ll 1$ where $\partial X_{n}/\partial t = \partial X_{n}/\partial r = 0$. The equation for $X_{0}^{2} = X_{0}X_{0}^{*}$ can be obtained from the Lagrangian
	\begin{equation}
	L = G(\omega,k)X_{0}^{2},
	\end{equation}
	where $G(\omega,k) = 0$ is the desired dispersion relation. The Lagrangian for the pulsation equation \eqref{pulsation-equation} is
	\begin{equation}
	\begin{array}{ll}
	\mkern-8mu L = & \mkern-4mu \displaystyle e^{\nu_{0}/2+3\lambda_{0}/2}\frac{p+\epsilon}{r^{2}}\Bigg(\omega^{2}(\delta r)^{2} - 8\pi e^{\nu_{0}}p(\delta r)^{2} \\
	& \mkern-45mu \displaystyle + \frac{e^{\nu_{0}-\lambda_{0}}}{p+\epsilon}\left[\left(\frac{dp}{dr}\right)^{2}\frac{(\delta r)^{2}}{p+\epsilon} - \frac{4}{r}\left(\frac{dp}{dr}\right)(\delta r)^{2} - \Gamma p\left(\frac{d}{dr}\delta r\right)^2 \right]\Bigg),
	\end{array}
	\end{equation}
	where one can take the first-order approximation of the radial displacement \eqref{time-dependence}:
	\begin{equation} \label{time-dependence2}
	\delta r \sim X_{0}\cos(kr - \omega t)
	\end{equation}
	and regard the $\sin^{2}$ and $\cos^{2}$ terms as time averaged to 1/2. \cite{Glass1983b} As reported by Ref. \cite[p.~392--399]{Whitham1974}, the dispersion relation can be identified from the Lagrangian as
	\begin{equation}
	\begin{array}{ll}
	\mkern-8mu G(\omega,k) = & \mkern-4mu \displaystyle e^{\nu_{0}/2+3\lambda_{0}/2}\frac{p+\epsilon}{r^{2}}\Bigg[\omega^{2} - c_{s}^{2}k^{2}e^{\nu_{0}-\lambda_{0}} - 8\pi e^{\nu_{0}}p \\
	& \mkern-5mu \displaystyle  + \frac{e^{\nu_{0}-\lambda_{0}}}{(p+\epsilon)^{2}}\left(\frac{dp}{dr}\right)^{2} - \frac{4}{r}\frac{e^{\nu_{0}-\lambda_{0}}}{p+\epsilon}\left(\frac{dp}{dr}\right) \Bigg] = 0.
	\end{array}
	\end{equation}
	
	Now, let us assume, instead of \eqref{time-dependence} and \eqref{time-dependence2}, that the radial displacement possesses a simple harmonic time-dependence of the form 
	\begin{equation} \label{time-dependence3}
	\delta r(r, t) = X(r)e^{i\omega t},
	\end{equation}
	where $X(r)$ is the radially-dependent amplitude and $\omega$ is a complex characteristic eigenfrequency which consists of the so called `damped frequency' and `relaxation time',
	\begin{equation} \label{damping-freq}
	\begin{array}{l}
	\omega_{d} \equiv \operatorname{Re}(\omega) = \omega_{n}\sqrt{1 - \zeta_{d}^2}, \\[10pt]
	1/\tau_{d} \equiv \operatorname{Im}(\omega) = -\omega_{n}\zeta_{d},
	\end{array}
	\end{equation}
	respectively. The aforementioned quantities are related to the natural frequency $\omega_{n}$ of the undamped system and the damping ratio $0 \leq \zeta_{d} \leq 1$ which describes the rate at which the normal oscillation modes decay and critically determines the dynamical behavior through the energy-dissipation rate \eqref{energy-dissipation}. \cite{Barta2019} Being bilinear in the adiabatic perturbations, the total-energy density $\epsilon$ has a time dependence $\exp[-2\operatorname{Im}(\omega)t]$. \cite{Cutler1987} Subsequently, the time derivative of the total-energy density implies that
	\begin{equation}
	\frac{d\epsilon}{dt} = -2\operatorname{Im}(\omega)\epsilon,
	\end{equation}
	which together with the energy-dissipation rate
	\begin{equation}  \label{energy-dissipation}
	- \frac{d\epsilon}{dt} = 2\eta\delta\sigma^{\mu\nu}\delta\sigma^{*}_{\mu\nu} + \zeta(\delta\Theta)^{2} + \frac{\kappa}{T}\nabla_{\mu}\delta T \nabla^{\mu} \delta T^{*},
	\end{equation}
	directly determines the characteristic damping time of perturbations as
	\begin{equation}
	\tau_{d} = -1/\omega_{n}\zeta_{d} = -2\epsilon/\dot{\epsilon}.
	\end{equation}
	
	In order to discuss causality, eqs. \eqref{four-wavevector} must be transformed to a local Lorentz frame. With respect to $(e^{-\nu_{0}/2}\partial_{t},\, e^{-\lambda_{0}/2}\partial_{r})$ the transformed quantities are
	\begin{equation}
	\tilde{\omega} = e^{-\nu_{0}/2}\omega, \quad \tilde{k} = e^{-\lambda_{0}/2}k.
	\end{equation}
	The transformed dispersion relation is 
	\begin{equation} \label{dispersion-relation}
	\tilde{\omega}^{2} = c_{s}^{2}\tilde{k}^{2} - \mu^{2}
	\end{equation}
	where
	\begin{equation}
	\mu^{2}(r) = -8\pi p + 4\tilde{\omega}_{1}^{2} + r^{2}\tilde{\omega}_{1}^{4}
	\end{equation}
	and
	\begin{equation}
	\tilde{\omega}_{1}^{2} =  \left(\frac{m}{r^{3}} + \frac{4\pi}{p}\right)\left(1 - \frac{2m}{r}\right)^{-1}.
	\end{equation}
	The group velocity with respect to local light cones is
	\begin{equation} \label{group-velocity}
	c_{g} = \frac{d\tilde{\omega}}{d\tilde{k}} = c_{s}\left(1 - \frac{\mu^{2}}{c_{s}^{2}\tilde{k}^{2}}\right)^{-1/2},
	\end{equation}
	which is always greater than the local speed of sound and therefore $c_{s} < 1$ is merely a weak condition (the very least required by causal wave propagation). \cite{Glass1983} In line with Jeans' criterion for general relativistic spherical stars, for all sufficiently long wavelengths such that
	\begin{equation} \label{Jeans-criterion}
	\tilde{k} < \mu/c_{s}
	\end{equation}
	a dynamical instability sets in. In the Newtonian limit, $\mu = 2(Gm/r^{3})^{1/2}$ gives the spherical version of Jeans' classic result for an infinite homogenous medium. \cite{Chandrasekhar1984} Note that there is a local, purely general relativistic instability at any point in the sphere where $2m(r)$ approaches $r$. This supports the well-known result that configurations at equilibrium must have $2m < r$ at all interior points.

	\section{Linear pulsation equation and associated eigenvalue problem} \label{sec:pulsation}
	Chandrasekhar \cite[eq.~(61)]{Chandrasekhar1964} showed that, with the assumption of harmonic time-dependence \eqref{time-dependence3}, the fundamental equation for radial pulsation can be cast in the form
	\begin{equation} \label{pulsation-equation}
	\begin{array}{l}
	\displaystyle c_{s}^{2}X'' + \left[(c_{s}^{2})' - Z + 4\pi r \Gamma pe^{\lambda} - \nu'/2\right]X' \\[5pt]
	\displaystyle + \big[(\nu^{2})'/2 - 2r^{-3}me^{\lambda} - Z' - 4\pi(\rho + p)Zre^{\lambda} \\[5pt]
	+ \omega^{2}e^{\lambda-\nu}\big]X = i(p + \epsilon)^{-1}(\omega \mathcal{S}_{1} - \omega^{-1} \mathcal{S}_{2})
	\end{array}
	\end{equation}
	where the functions of equilibrium stellar configuration $\{p(r),\, \rho(r),\, m(r),\, \nu(r),\, \lambda(r)\}$ in eq. \eqref{Einstein-eq}, the effective adiabatic index $\Gamma(r)$ and sound speed $c_{s}$ in eqs. (\ref{adiabatic-index}--\ref{speed-of-sound}) and 
	\begin{equation}
	Z(r) = c_{s}^{2}\left(\frac{\nu'}{2} - \frac{2}{r}\right)
	\end{equation}
	are computed for specific stellar models. \cite{Kokkotas2001} 
	
\begin{figure}
		\centering
		\includegraphics[scale=0.6]{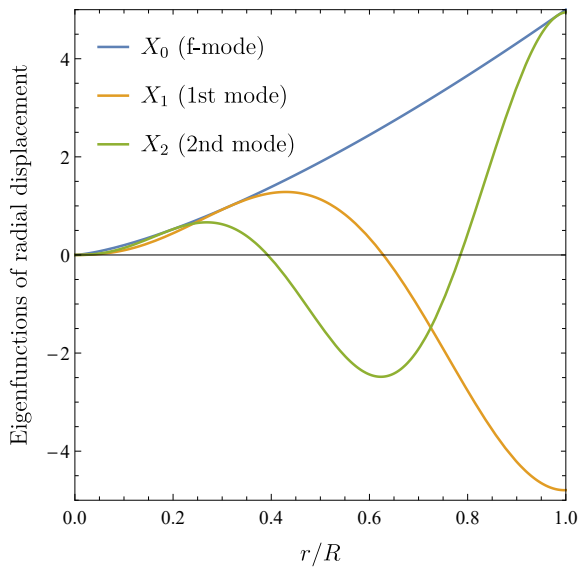}
		\caption{Eigenfunctions functions of radial displacement for the first three lowest-frequency oscillation modes $(X_{0},\, X_{1},\, X_{2})$ as a function of the fractional radius $r/R$ obtained for SLy4 EoS at a central density $\rho_{c} = 0.547 \text{ GeV fm}^{-3}$. The displacement amplitude has been renormalized such that $X_{0} = 1$. The total mass and circumferential radius of this particular neutron-star configuration, which corresponds to Figs. \ref{fig:Depth}--\ref{fig:metric}, are $M = 1.4M_{\odot}$ and $R = 11.36$ km.}
		\label{fig:Eigenfunctions}
	\end{figure}

	The fundamental equation for radial pulsation \eqref{pulsation-equation} together with its boundary conditions (\ref{boundary-condition1}--\ref{boundary-condition2}) constitutes a Sturm--Liouville eigenvalue problem (SL-EVP) for a discrete set of scalar-valued eigenfunctions of radial displacement $\{X_{0}(r),\, X_{1}(r),\, \ldots,\, X_{j}(r),\, \ldots\}$ with their respective eigenvalues $\{\omega^{2}_{0},\, \omega^{2}_{1},\, \ldots,\, \omega^{2}_{j},\, \ldots\}$. Successive eigenvalues correspond to squared complex-valued frequencies of the oscillation modes, arranged in ascending order, where the index $j = 0,\, 1,\, \ldots$ represents the number of nodes of each respective mode inside the star. The smallest eigenvalue $\omega_{0}^{2}$ is associated with the fundamental-mode frequency of radial oscillations which has no nodes between the center and the stellar surface, whereas the first excited mode $(j=1)$ has a node, the second one $(j=2)$ has two, and so forth. The normalized eigenfunctions functions of radial displacement are presented in Fig. \ref{fig:Eigenfunctions}. In terms of a superposition of sinusoids, the $\omega_{0}$ is the lowest-frequency sinusoidal in the sum (see dynamical stability in Sec. \ref{sec:dynamical-stability-fundamental-mode}). Owing to the fact that the general solution to eq. \eqref{pulsation-equation} is always a superposition of both growing and damped oscillation modes, the definitions \eqref{damping-freq} provide the means to separate the real and imaginary parts of the complex frequency:
	\begin{equation}
	\omega^2 = \omega_{n}^2(1-2\zeta^2) + 2i\omega_{n}^2\zeta\sqrt{1-\zeta^2},
	\end{equation}
	thus the imaginary part of inhomogenity of the eigenvalue problem appearing on the right-hand side of eq. \eqref{pulsation-equation} reduces to
	\begin{equation} 
	i(\omega \mathcal{S}_{1} - \omega^{-1}\mathcal{S}_{2}) = i\sqrt{1-\zeta^2}\frac{\omega_{n}\mathcal{S}_{1} - \omega_{n}^{-1}\mathcal{S}_{2}}{p + \epsilon},
	\end{equation}
	where the functions $\mathcal{S}_{1}$ and $\mathcal{S}_{2}$ are expressed by eq. \eqref{S-def}, in terms of transport coefficients $\eta,\, \zeta$, and $\kappa$. The derivation of eq. \eqref{pulsation-equation}, which has been extended to involve transport coefficients in neutron-star matter through $\mathcal{S}_{1}$ and $\mathcal{S}_{2}$, has been outlined in detail in our preceding paper \cite{Barta2019} and shall not be repeated here.
	
	The homogeneous SL-EVP \eqref{pulsation-equation} is associated with harmonic oscillations and can be converted to a self-adjoint form which poses a boundary-value problem. \cite{Barta2019} The eigenvalues and corresponding eigenfunctions of the given boundary-value problem are then uniquely determined in Sec. \ref{sec:method} by the finiteness of $X(r)$ and $dX(r)/dr$ everywhere and by satisfying the following pair of boundary conditions of the third type\footnote{The \emph{Robin boundary condition} or \emph{third-type boundary condition} is a specification of a linear combination of the values of a function and its derivative on the boundary. It is commonly used in solving SL-EVPs.}: \cite{Glass1983}
	\begin{enumerate}
		\item The fluid at the center of the star is assumed to remain at rest; thus,
		\begin{equation} \label{boundary-condition1}
		X = 0 \quad \text{at} \quad r = 0.
		\end{equation}
		\item The perturbed pressure is required to vanish on the perturbed boundary of the star; thus the Lagrangian change in the pressure $\delta p = e^{\nu_{0}/2}r^{-2}\Gamma p_{0} \frac{d}{dr}\left(e^{-\nu_{0}/2}r^{2}X\right)$ has to vanish at the surface; thus, 
		\begin{equation} \label{boundary-condition2}
		\delta p = 0 \quad \text{at} \quad r = R.
		\end{equation}	
	\end{enumerate}
	Let us keep in mind that even the homogeneous SL-EVP cannot be solved analytically in general. However, a rough order-of-magnitude estimate can be made. Provided that $X_{0}(r)$ is nearly linear in $r$ (cf. blue line in Fig. \ref{fig:Eigenfunctions}), $\omega_{0} \propto (G\hat{\rho})^{-1/2}$ is a universal estimate for the f-mode, irrespective of the particular EoS, where $\hat{\rho}$ is the mean density of a star undergoing homologous contraction (see eq. \eqref{time-scale} in Appendix \ref{sec:dynamical-stability}). Detweiler \& Lindblom \cite{Detweiler1983} has found that for most stellar models, the oscillation periods of the fundamental mode
	\begin{equation} \label{oscillation-period}
	T = 1/\nu_{0} = 2\pi/\omega_{0},
	\end{equation}	
	typically range from about 0.2 to 0.9 milliseconds, depending on the particular EoS and the central density. The `ordinary' or `temporal' frequency $\nu_{0}$ and angular frequency $\omega_{0}$ are typically $\sim 10^{4}$ $\text{s}^{-1}$ for neutron stars, but two quantities differ by a factor of $2\pi$. Accordingly, $\tau_{d}$ in eq. \eqref{time-scale}, which is in the range of 0.1 -- 0.3 seconds, can also serve as an estimate of a hydrodynamical timescale for a given star. \cite[p.~291]{Haensel2007} In order to compute the three lowest-frequency oscillation modes with high accuracy for each particular EoS in Table \ref{tab:EoS}, a finite-difference method will be used in the subsequent Sec. \ref{sec:method}.

	\subsection{Dynamical stability criteria based on the fundamental mode} \label{sec:dynamical-stability-fundamental-mode}
	\begin{figure*}
	\begin{minipage}{.48\linewidth}
		\centering
		\subfloat[Schematic mass--radius relation illustrating the spiral structure of the unstable branch.]{\label{fig:MR-unstable-schematic}
			\includegraphics[width=0.81\linewidth]{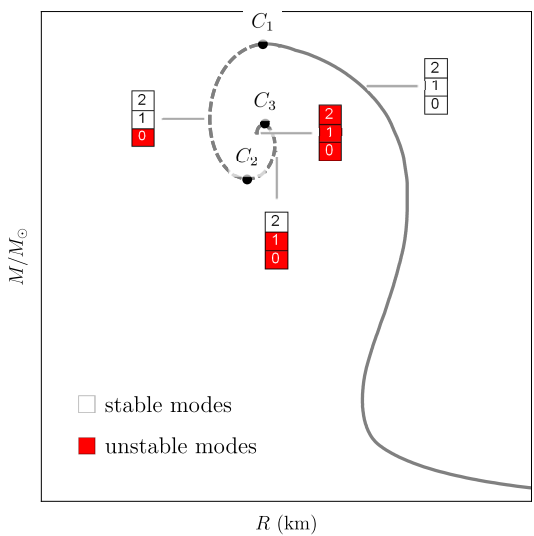}}		
	\end{minipage}
	\begin{minipage}{.48\linewidth}
		\centering
		\subfloat[Numerical mass--radius relation for three selected EoSs extended beyond the stable branch which was depicted in Fig. \ref{fig:MR}.]{\label{fig:MR-unstable-EoS}
			\includegraphics[width=1.06\linewidth]{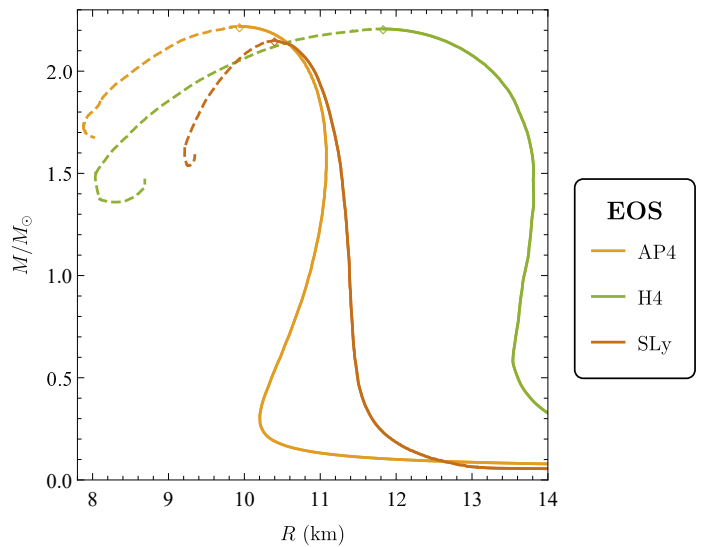}}
	\end{minipage}\par\medskip
	\caption{The left panel schematically illustrates whereas the right panel displays for three selected EoS (APR4, H4, SLy4) the structure of $M(R)$ curves for dynamically stable (solid lines) and unstable states (dashed lines). Critical points are denoted by $C_{1},\, C_{2},\, C_{3}$. The fundamental mode and two higher-order excited modes of radial oscillation are represented on a given segment by a column of numbers $n = 0,\, 1,\, 2$ in cells, with stable oscillation modes in unfilled and unstable modes filled in red, respectively.}
	\label{fig:MR-unstable}
	\end{figure*}
	In Sec. \ref{sec:equilibrium-stellar-model} the hydrostatic stability of equilibrium stellar configurations was shown to be firmly ensured by the TOV equation \eqref{Einstein-eq:b}. On the other hand, the dynamical stability of stellar configurations with respect to small radial perturbations is determined by the fundamental mode:
	\begin{enumerate}
		\item If $\omega_{0}^{2} > 0$, then all the higher eigenfrequencies of the spectrum are real, which indicates that the equilibrium stellar model is dynamically stable with respect to small perturbations, and they form a lower-bounded infinite sequence $\omega_{1} \leq \omega_{2} \leq \omega_{3} \ldots$. In this case the small radial displacements \eqref{time-dependence} exhibit purely harmonic oscillations in the f-mode. \cite{Harrison1965}
		\item If $\omega_{0}^{2} < 0$, then $\omega_{0}$ becomes purely imaginary, which indicates that the equilibrium stellar model comes to be dynamically unstable (at least in the f-mode). In this case the small radial displacements \eqref{time-dependence} exponentially grow or damp with time (depending on the value of the damping ratio $\zeta_{d}$). As discussed in detail by our preceding study \cite{Barta2019}, the highest instability increment, associated with the most rapid exponential growth or dampening, is provided by the f-mode. Instability will indeed occur at a specific central energy density $\epsilon_{0} > \epsilon_{\text{crit}}$ in the crust of those neutron stars with total mass $M$ that exceeds the maximal-mass configuration of a family of non-rotating NS models for the particular EoS. Such NSs would eventually collapse into a black hole. \cite{Zeldovich1971}		
		\item Finally, at $\omega_{0} = 0$ the equilibrium stellar model is at neutral stability; neither stable nor unstable. In this case $\epsilon_{0} = \epsilon_{\text{crit}}$ as marked in Fig. \ref{fig:ME} by the symbols $\diamond$ for states at neutral stability.
	\end{enumerate}
	The dynamical stability is also inherently related to the shape of the $M(R)$ curve. \cite[p.~293]{Haensel2007} As we already stressed in Sec. \ref{sec:equilibrium-stellar-model}, $dM/d\epsilon_{c} > 0$ is necessary but not sufficient condition for dynamical stability. At each critical point where $dR/d\epsilon_{c} < 0$ holds an even number $n$ of radial oscillation mode changes its stability while for $dR/d\epsilon_{c} > 0$ an odd-numbered mode changes its stability. \cite{Harrison1965} The dynamically stable and unstable branches of $M(R)$ curves are presented in Fig. \ref{fig:MR-unstable} by solid lines and dashed lines, respectively. The left panel schematically illustrates the spiral structure of a general $M(R)$ curve for dynamically unstable states whereas the right panel depicts the unstable branches for three selected EoS (APR4, H4, SLy4) with particularly small radii ($R \simeq 8-9$ km) at their respective final critical points. The letters $C_{1},\, C_{2},\, C_{3}$ refer to critical points (turning points) which divide the $M(R)$ curve into four segments, with an increasing number of unstable oscillation modes. As we have seen in Fig. \ref{fig:MR}, the central energy density $\epsilon_{c}$ is increasing from large radii and small masses until a certain critical central energy density $\epsilon_{c}^{\text{crit}}$ is reached at the maximal mass and the minimal radius. All radial oscillation modes are stable in the $\epsilon_{c} < \epsilon_{c}^{\text{crit}}$ regime, but once the central energy density is increased beyond the critical point $C_{1}$, it does not restore the stability. \cite{Schertler2000} At $C_{1}$ the fundamental mode becomes unstable. At $C_{2}$, on account of $(dR/d\epsilon_{c})_{C_{2}} > 0$ the fundamental mode remains unstable and an additional $n = 1$ mode becomes unstable.  At $C_{3}$, all three oscillation modes become unstable.

	\section{Finite-difference method for computing eigenfrequencies} \label{sec:method}
	Various numerical methods have been investigated to accurately compute the eigenfrequencies of radial pulsation for particular nuclear-theory-based EoS models. The scholarly literature \cite{Glass1983,Vath1992,Kokkotas,Glass1983b} refers to two most widespread approach to solve the SL-EVP by Runge--Kutta integration as `shooting method' and `finite-difference method', respectively. We shall adopt the later method by converting the linear ODE \eqref{pulsation-equation} into a system of algebraic equations by replacing the derivatives with finite-difference approximations in the form of a tridiagonal matrix that can be solved by matrix algebra techniques, ideally suited to modern numerical analysis.
	
\begin{figure}
		\centering
		\includegraphics[scale=0.60]{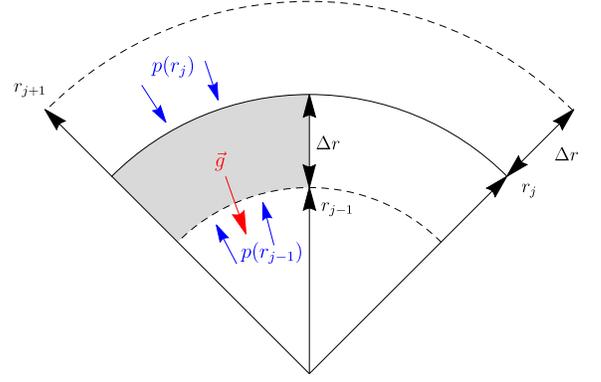}
		\caption{Successive spherical shells of matter with thickness $\Delta r$, on which the fundamental equation for radial pulsation is evaluated, are labelled $r_{j-1},\, r_{j},\, r_{j+1}$ are illustrated by concentric arcs. The pressure at the upper and lower boundary of the $j$th shell, and the vector of gravitational acceleration $|\vec{g}| = -Gm/r^{2}$ acting at on a shell element (highlighted by gray color) are indicated by blue and red arrows, respectively.}
		\label{fig:shells}
	\end{figure}

	Let some grid functions $\{X_{0},\, X_{1},\, \dots,\, X_{N},\, X_{N+1}\}$ be defined on a grid with uniformly-spaced points $r_{j} = j\Delta r$ with a separation $\Delta r = R/(N + 1)$ between adjacent grid points. An illustration of thin spherical shells of matter with thickness $\Delta r$ at successive grid points $r_{j-1},\, r_{j},\, r_{j+1}$ is shown in Fig. \ref{fig:shells}. We consider the finite-difference approximation $X_{j}$ of some eigenfunction $X(r_{j})$ for any positive integer $j = 0,\, \ldots,\, N$. The discretization and evaluation of eq. \eqref{pulsation-equation} on the $j$th shell is formally expressed by
	\begin{equation} \label{pulsation-matrix-eq}
	a_{j}[D^{2}_{r}X]_{j} + b_{j}[D_{r}X]_{j} + c_{j} = \omega_{j}^{2} X_{j}
	\end{equation}
	where $\omega^2_{j}$ is the squared frequency of the $j$th oscillation mode, $[D_{r}X]_{j}$ and $[D^{2}_{r}X]_{j}$ are the first- and second-order discrete derivatives of the grid function $X(r)$ at a corresponding grid point $r = r_{j}$ for $j = 1,\, \ldots,\, N$, and the coefficients $a_{j},\, b_{j},\, c_{j}$ are approximated by
	\begin{equation}
	\begin{array}{l}
	a_{j} \approx \displaystyle -\left[e^{\nu-\lambda}c_{s}^{2}\right]\big|_{r=r_{j}} \\[15pt]
	b_{j} \approx \displaystyle e^{\nu-\lambda}\left[Z + \nu'/2 - (c_{s}^{2})' - 4\pi r \Gamma pe^{\lambda} \right]\big|_{r=r_{j}} \\[15pt]
	c_{j} \approx \displaystyle e^{\nu-\lambda}\left[Z' + 4\pi(\rho + p)Zre^{\lambda} + \frac{2me^{\lambda}}{r^{3}} - \frac{(\nu^{2})'}{2}\right]\Bigg|_{r=r_{j}}.
	\end{array}
	\end{equation}
	Let us consider approximation of the derivatives of grid functions in \eqref{pulsation-matrix-eq} by the following finite central differences:
	\begin{subequations} \label{finite-difference}
		\begin{align}
		& \displaystyle [D_{r}X]_{j} = \frac{X_{j+1} - X_{j-1}}{2\Delta r} = X'(r_{j}) + \epsilon_{t,D},   \label{finite-difference:a} \\
		& \displaystyle [D^{2}_{r}X]_{j} = \frac{X_{j+1} - 2X_{j} + X_{j-1}}{(\Delta r)^2} = X''(r_{j}) + \epsilon_{t,D^{2}}  \label{finite-difference:b}
		\end{align}
	\end{subequations}
	with the truncation error $\epsilon_{t,D} = \epsilon_{t,D^{2}} = \mathcal{O}(\Delta r^{2})$ is to be assessed in Sec. \ref{sec:error}. By re-expressing the finite differences \eqref{finite-difference}, eq. \eqref{pulsation-matrix-eq} reduces to a set of algebraic equations
	\begin{equation} \label{pulsation-matrix-eq2}
	\Sigma^{+}_{j}X_{j+1} + \Sigma^{0}_{j}X_{j} + \Sigma^{-}_{j}X_{j-1} = \omega_{j}^2 X_{j}
	\end{equation}
	for $j = 1,\, \ldots,\, N$ which represent the three successive non-zero entries of the $j$th row of a non-singular tridiagonal matrix $[\Sigma_{ij}] \in \mathbb{R}^{N \times N}$, defined by
	\begin{eqnarray} \label{sigma-matrix}
	\hspace{-5em} \Sigma & = &
	\begin{bmatrix}
	\Sigma^{0}_{1} & \Sigma^{+}_{1} &   &   &   \\
	\Sigma^{-}_{2} & \Sigma^{0}_{2} & \Sigma^{+}_{2} &   &   \\
	& \ddots & \ddots & \ddots &    \\
	&   & \Sigma^{-}_{N-1} & \Sigma^{0}_{N-1} & \Sigma^{+}_{N-1}  \\	
	&   &   & \Sigma^{-}_{N} & \Sigma^{0}_{N}
	\end{bmatrix}
	\end{eqnarray}
	where the superdiagonal, diagonal and subdiagonal entries are as follows:
	\begin{equation}
	\begin{array}{l}
	\displaystyle \Sigma^{+}_{j} = \frac{a_{j}}{(\Delta r)^2}+\frac{b_{j}}{2\Delta r}, \\[10pt]
	\displaystyle \Sigma^{0}_{j} = c_{j}-\frac{2a_{j}}{(\Delta r)^2}, \\[10pt]
	\displaystyle \Sigma^{-}_{j} = \frac{a_{j}}{(\Delta r)^2}-\frac{b_{j}}{2\Delta r}.
	\end{array}
	\end{equation}
	Although $b_{j}$ prevents the matrix \eqref{sigma-matrix} from being symmeric, but due to its tridiagonal structure (inhereted from the central differencing scheme which involved only the closest pair of points), zero entries can be discarded and a significant amount of computational cost is saved. Consequently, a matrix equation of the form
	\begin{equation} \label{matrix-eq}
	(\Sigma - \omega^{2}I)\vec{X} = 0
	\end{equation}
	is constructed from the linear system of $N$ equations \eqref{pulsation-matrix-eq2} for the central differencing scheme where $I$ denotes the $N$-by-$N$ identity matrix, and 
	\begin{eqnarray} \label{grid-functions}
	\hspace{-5em} \vec{X} &\!\! = & \!\! 
	\begin{bmatrix}
	X_{1} + \Sigma^{+}_{0}X_{0} \\
	X_{2} \\
	\vdots \\
	X_{N-1} \\
	X_{N} + \Sigma^{-}_{N}X_{N+1}
	\end{bmatrix}
	\end{eqnarray}
	is the vector which consists of the unknown values $\{X_{1},\, \dots,\, X_{N}\}$ of the eigenfunction of a specific oscillation mode at corresponding grid points $\{r_{1},\, \dots,\, r_{N}\}$. The first and last entry of $\vec{X}$ involve the values $X_{0}$ and $X_{N+1}$, respectively, as a result of the pair of boundary conditions (\ref{boundary-condition1}--\ref{boundary-condition2}) which imply $X_{0} = X_{N+1} = 0$.  
	
	In linear algebra, Cramer's rule states that the homogeneous linear system of equations \eqref{matrix-eq} has a non-trivial solution if and only if the determinant of the corresponding characteristic matrix is zero, expressly
	\begin{equation} \label{characteristic-eq}
	\det(\Sigma - \omega^{2}I) = 0.
	\end{equation}
	The characteristic polynomial of the tridiagonal matrix $\Sigma$, denoted by $p_{\Sigma}(\omega^{2})$ and defined by
	\begin{equation} \label{characteristic-polynomial}
	p_{\Sigma}(\omega^{2}) = \det(\Sigma - \omega^{2}I),
	\end{equation}
	is a monic polynomial of degree $N$ which is invariant under similarity transformation and has the eigenvalues as roots. In other words, eq. \eqref{characteristic-eq} is referred to as the characteristic equation, considering that the roots of $p_{\Sigma}(\omega^{2}) = 0$ are exactly the eigenvalues of $\Sigma$. Harrison et al. \cite{Harrison1965} have argued that a sufficiently accurate numerical algorithm always yield a zero-frequency mode at that specific central density that corresponds to the maximal-mass configuration of a family of stellar models for the particular EoS. In fact, by computing $p_{\Sigma}(\omega^{2})$ for each particular EoS at various values of central density, all the eigenfrequencies and corresponding eigenmodes are simultaneously determined in a single run. Kokkotas \& Ruoff \cite{Kokkotas2001} pointed out that although this method is more time consuming than searching for each eigenvalue separately by shooting method, but one may smoothly refine its accuracy by setting up grids with optimal size. In the next section the truncation error in the eigenvalues due to the finite-difference approximations \eqref{pulsation-matrix-eq} will be estimated. We demonstrate that it does not exceed ca. $10^{-4}$ Hz for setting up regular grid of size $N \simeq 3000$.

	\subsection{An error estimate of the finite-difference approximation} \label{sec:error}
	Two potential sources of numerical errors occur in computing our finite-difference approximation to the solution of the SL-EPV. Errors due to inexactness in the floating-point representation of numbers and the arithmetic operations are called round-off errors and they are determined by the machine precision (roughly $10^{-15}$). Moreover, local truncation errors (LTE) incur due to the inaccuracy in the approximation \eqref{finite-difference} for derivatives of grid functions. The later one, however, can be controlled by varying the spacing of the radial grid on which equilibrium stellar models and the eigenfunctions of radial displacement are determined. \cite{LeVeque2007} 
		
	As stated in the previous section, the discrete derivative $[D_{r}X]_{j}$ is not exactly equal to the derivative $X'(r_{j})$. The truncation error in the first-order centered finite-difference approximation \eqref{finite-difference:a}, expressed by
	\begin{equation} \label{truncation-error}
	\epsilon_{t,D} = [D_{r}X]_{j} - X'(r_{j}),
	\end{equation}
	can be computed in the following manner by expanding $X_{j+1}$ and $X_{j-1}$ in a Taylor series around the point $r_{j}$:
	\begin{subequations} \label{finite-difference2}
		\begin{align}
		X_{j+1} & = \displaystyle X(r + \Delta r) = \sum_{j=1}^{\infty}\frac{1}{j!}\frac{d^{j}X(r_{j})}{dr^{j}}\Delta r^{j} \nonumber \\
		& = \displaystyle X(r_{j}) + X'(r_{j})\Delta r + \frac{1}{2}X''(r_{j})(\Delta r)^{2} + \mathcal{O}(\Delta r^{3}),   \label{finite-difference2:a} \\
		X_{j-1} & = X(r_{j}) - X'(r_{j})\Delta t + \frac{1}{2}X''(r_{j})(\Delta r)^{2} - \mathcal{O}(\Delta r^{3}).   \label{finite-difference2:b}
		\end{align}
	\end{subequations}
	Now, by inserting the quadratic approximations \eqref{finite-difference2} into the finite difference \eqref{finite-difference:a} and collecting terms in $[D_{r}X]_{j} - X'(r_{j})$, the result for the truncation error is
	\begin{equation} \label{truncation-error-1st-order}
	\epsilon_{t,D} = \displaystyle \frac{X_{j+1} - X_{j-1}}{2\Delta r} - X'(r_{j}) = \displaystyle \frac{1}{6}X'''(r_{j})\Delta r^{2} + \mathcal{O}(\Delta r^{4}),
	\end{equation}
	an expression given in terms of a power-series in $\Delta r$ where the lowest power is proportional to $\Delta r^{2}$. \cite[p.~419]{Langtangen2017} The truncation error for the approximant of the second-order derivative \eqref{finite-difference:b} is
	\begin{equation}
	\epsilon_{t,D^{2}} = \displaystyle \frac{1}{12}X''''(r_{j})\Delta r^{2} + \mathcal{O}(\Delta r^{4}).
	\end{equation}
	The truncation error, therefore, in both expressions is said to be of second order in $\Delta r$, as we stated after eq. \eqref{finite-difference}.	The accuracy of our computations was tested by varying the spacing of grid points, defined by $\Delta r \equiv R/(N + 1)$. Our grids were set up over $N \simeq 3000$ points for each stellar models where the circumferential radius ranged from about $8$ to $14$ km, therefore the spacing varied between $2.666$ and $4.665$ metres (or equivalently between $8.892$ and $15.561$ nanoseconds of oscillation period $T$). Subsequently, our tests indicated that the truncation error \eqref{truncation-error-1st-order} in the computation of eigenfrequencies did not exceed $\epsilon_{t} \sim 10^{-4}$ Hz.

	\subsection{Numerical results and interpretation}
	This subsection discusses the numerical results obtained by the implementation of the finite-difference method for the homogeneous SL-EVP \eqref{pulsation-equation}.	We present the first three radial-mode frequencies as a function of central energy density	in the form of tables and accompanying figures for each realistic EoS. In each panel of Table \ref{tab:frequency} we list the central energy density ($\epsilon_{c}$), the circumferential radius ($R$), the gravitational mass ($M$), the frequencies of the fundamental mode ($\nu_{0}$) and the first two lowest-frequency excited modes ($\nu_{1}$ and $\nu_{2}$) of the respective stellar model.\footnote{Let us note that the eigenfrequencies in Sec. \ref{sec:method} were in fact angular frequencies $(\omega)$, whereas the use of `ordinary' or `temporal' frequencies $(\nu)$, which differ only by a factor of $2\pi$ in eq. \eqref{oscillation-period}, are more conventional in asteroseismology.} In addition to the stable model, we also included a stellar configuration just beyond the limit of dynamical stability where, as stated above, the total mass $M$ as function of the central energy density $\epsilon_{c}$ exhibits a local maximum. The f-mode frequency of such configurations is marked by an asterisk. Considering that the f-mode oscillation period \eqref{oscillation-period} relies extremely sensitively on the particular equilibrium stellar model (and thus inherently on the choice of interpolants used to fill the gaps in the associated discontinuous EoS, the smoothness of $\Gamma(\epsilon)$, etc.), slight deviances between each numerical predictions of the scholarly literature \cite{Chanmugam1977,Glass1983,Vath1992,Kokkotas2001} and also our results for the eigenfrequencies are anticipated. Ref. \cite{Vath1992} pointed out that the frequencies obtained by different authors are not substantially different at low central densities, where effects of GR remain substantially negligible, but at higher central densities where the redshift of the star $z = (1-2GM/R c^{2})^{-1/2} - 1$ exceeds $z \simeq 0.1$, discrepancies occur which may be assigned to the slight deviations in the equilibrium stellar models.\footnote{As we started in Sec. \ref{sec:equilibrium-stellar-model}, equilibrium configurations are generally parameterized by $\rho_{c}$, $\epsilon_{c}$ or equivalently by $M$. However, the oscillation spectra of stellar models constructed from different EOSs are more directly comparable with a parametrization by the redshift parameter $z$ than by either of the earlier parameters (see \cite{Glass1983,Chirenti2015}).}.

\begin{figure*}
		\subfloat[APR4]{\includegraphics[scale=0.60]{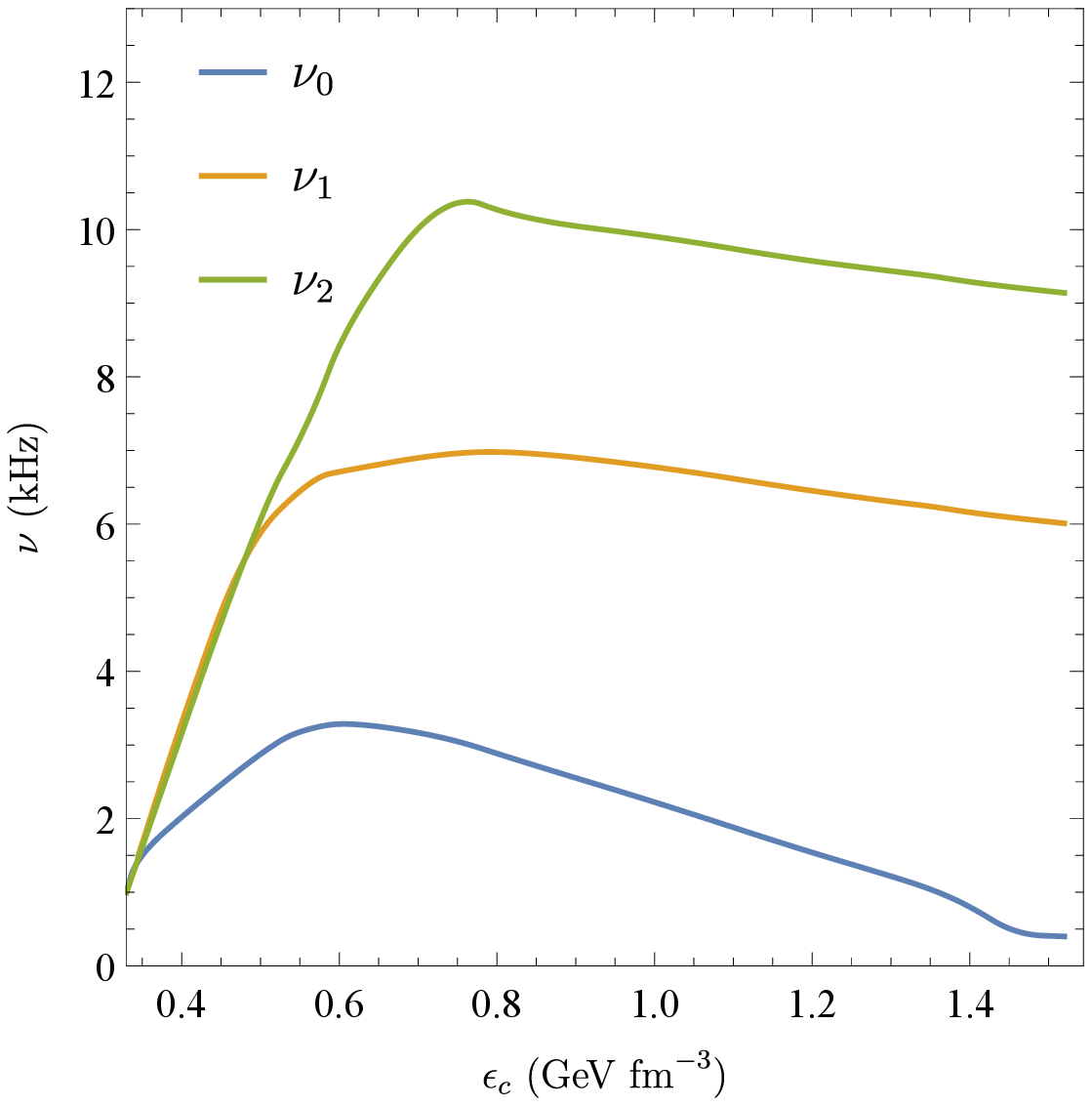}} \hspace{40pt}
		\subfloat[H4]{\includegraphics[scale=0.60]{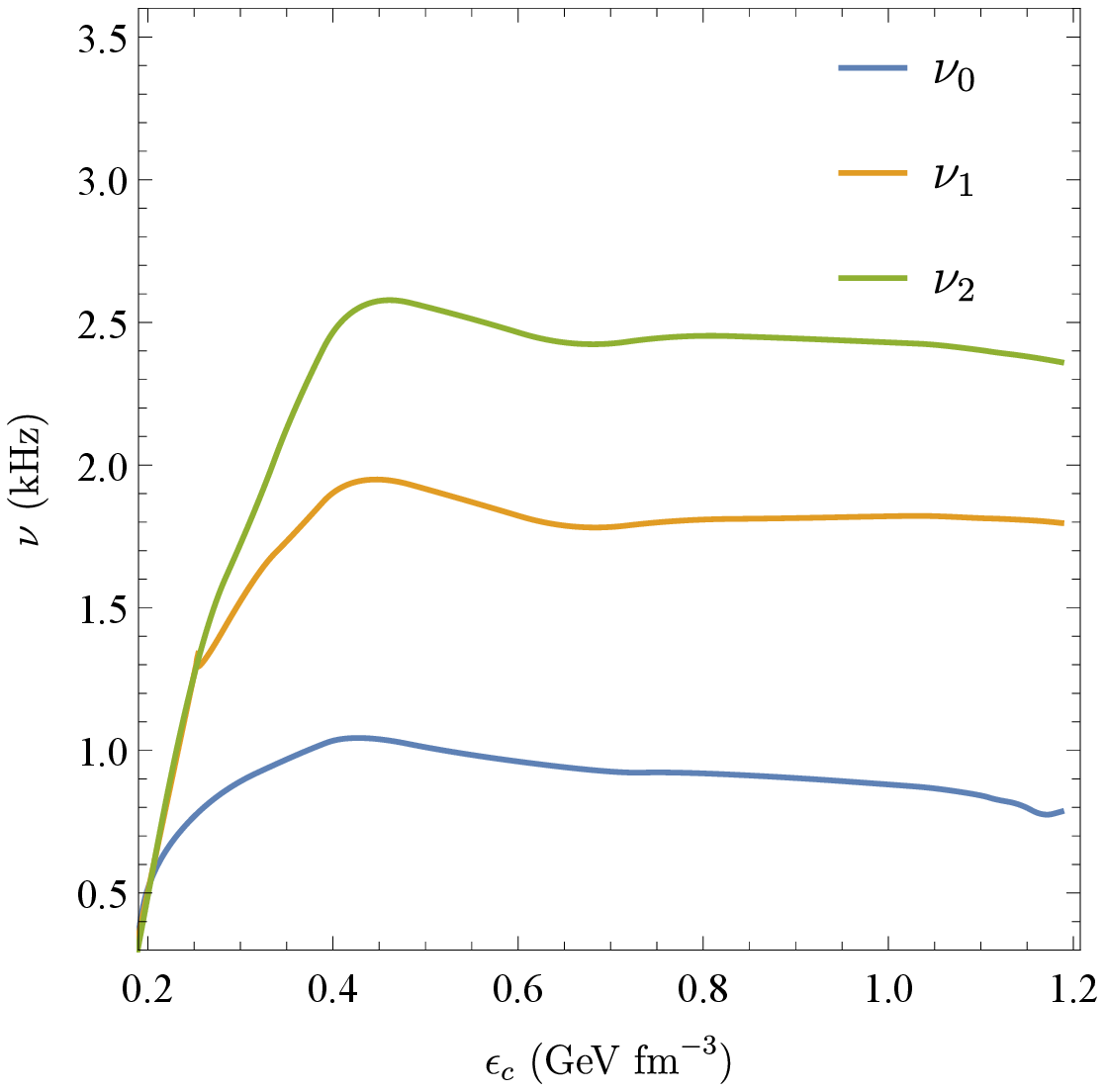}}
		\hfill
		\subfloat[MPA1]{\includegraphics[scale=0.60]{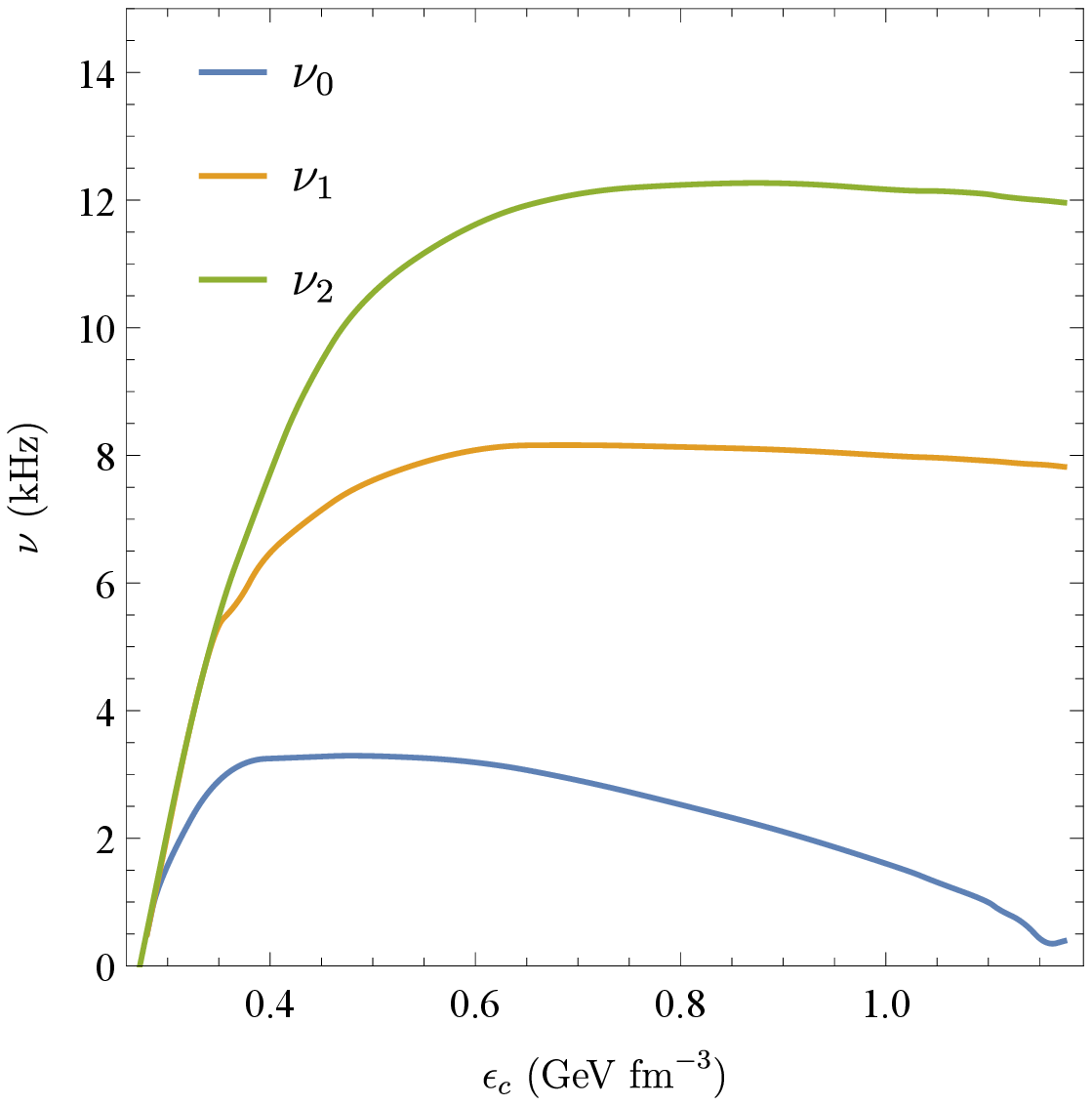}} \hspace{40pt}
		\subfloat[MS1]{\includegraphics[scale=0.60]{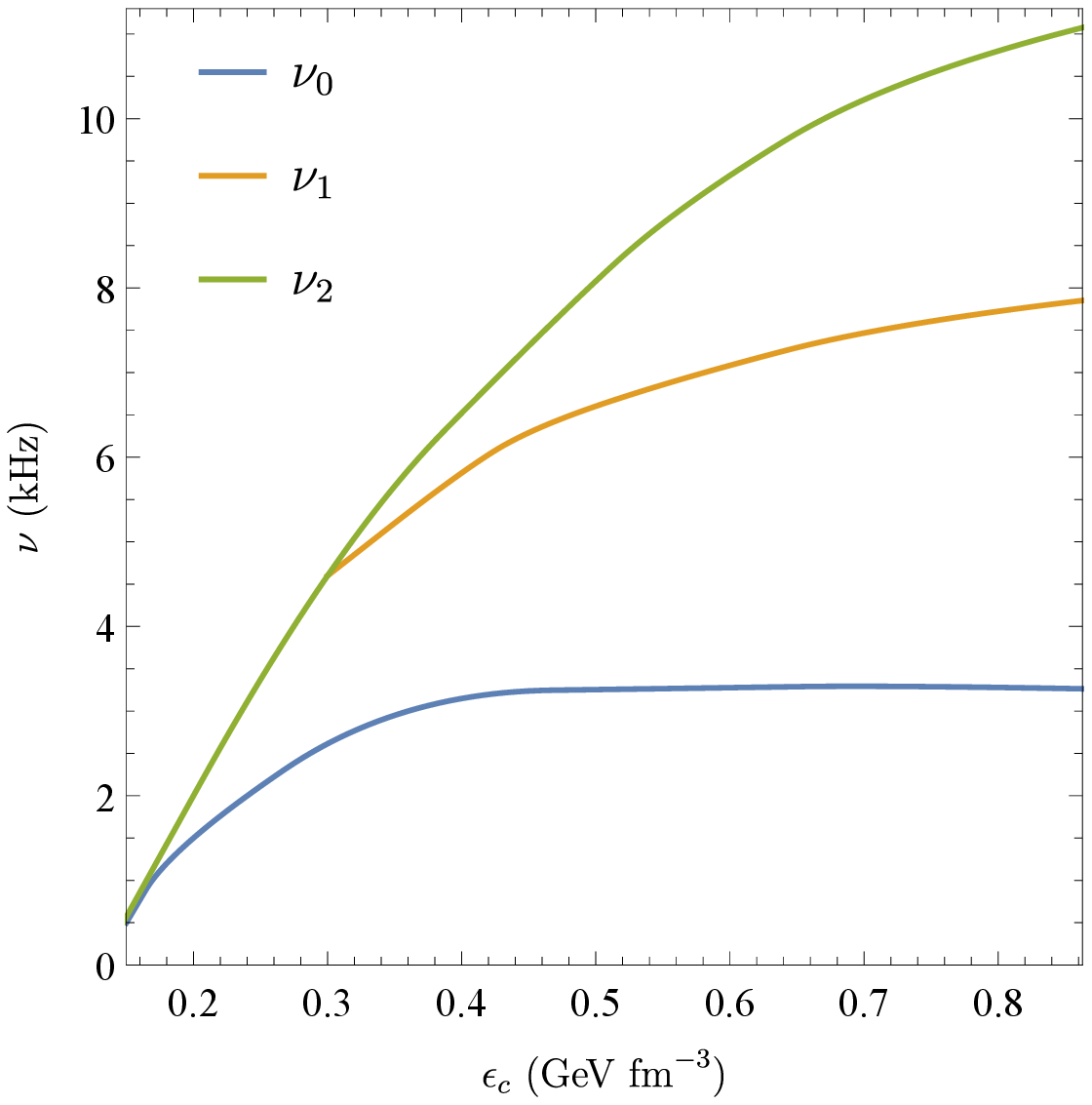}}
		\hfill
		\subfloat[Sly4]{\includegraphics[scale=0.60]{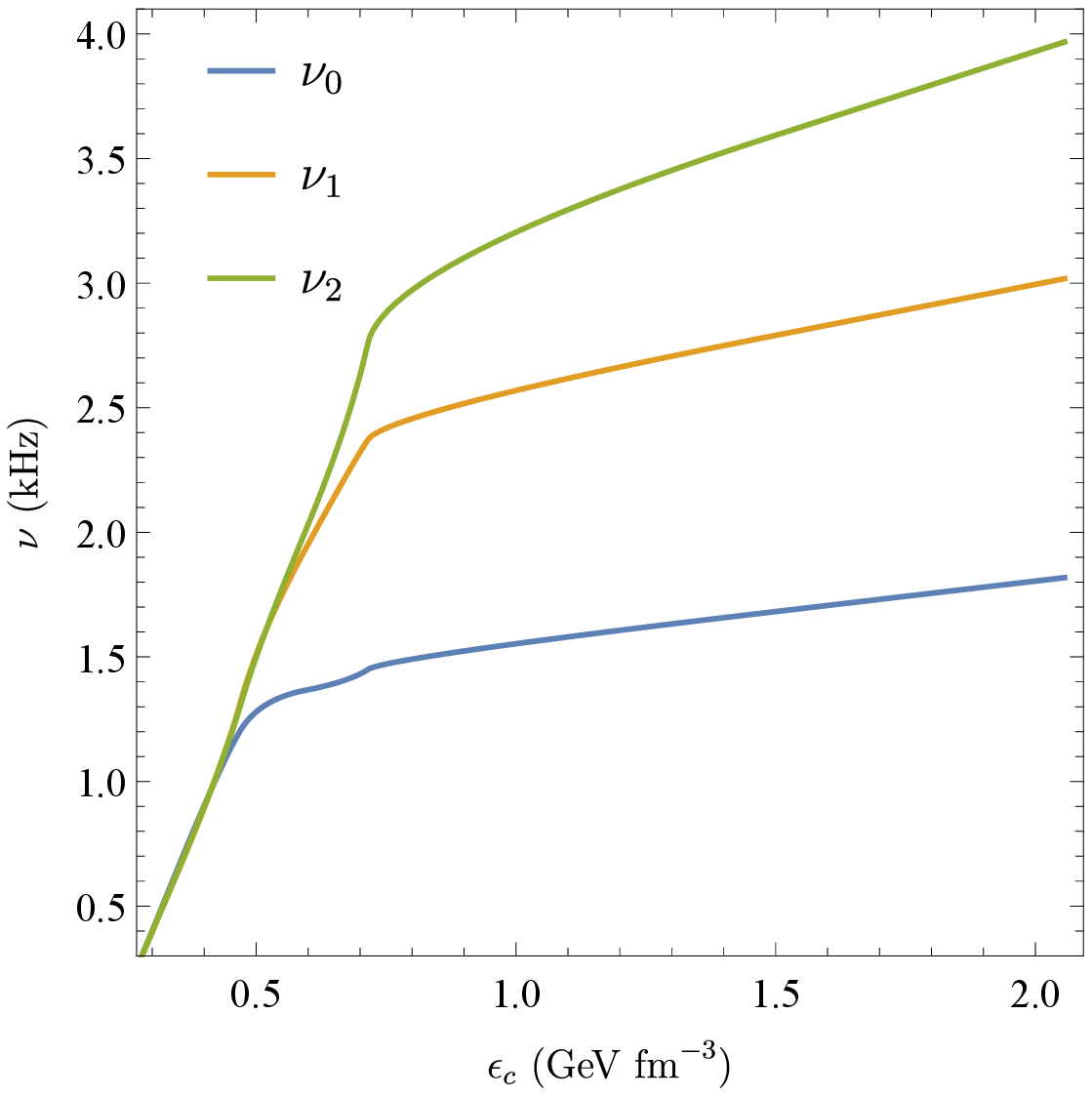}} \hspace{40pt}
		\subfloat[SQM1]{\includegraphics[scale=0.60]{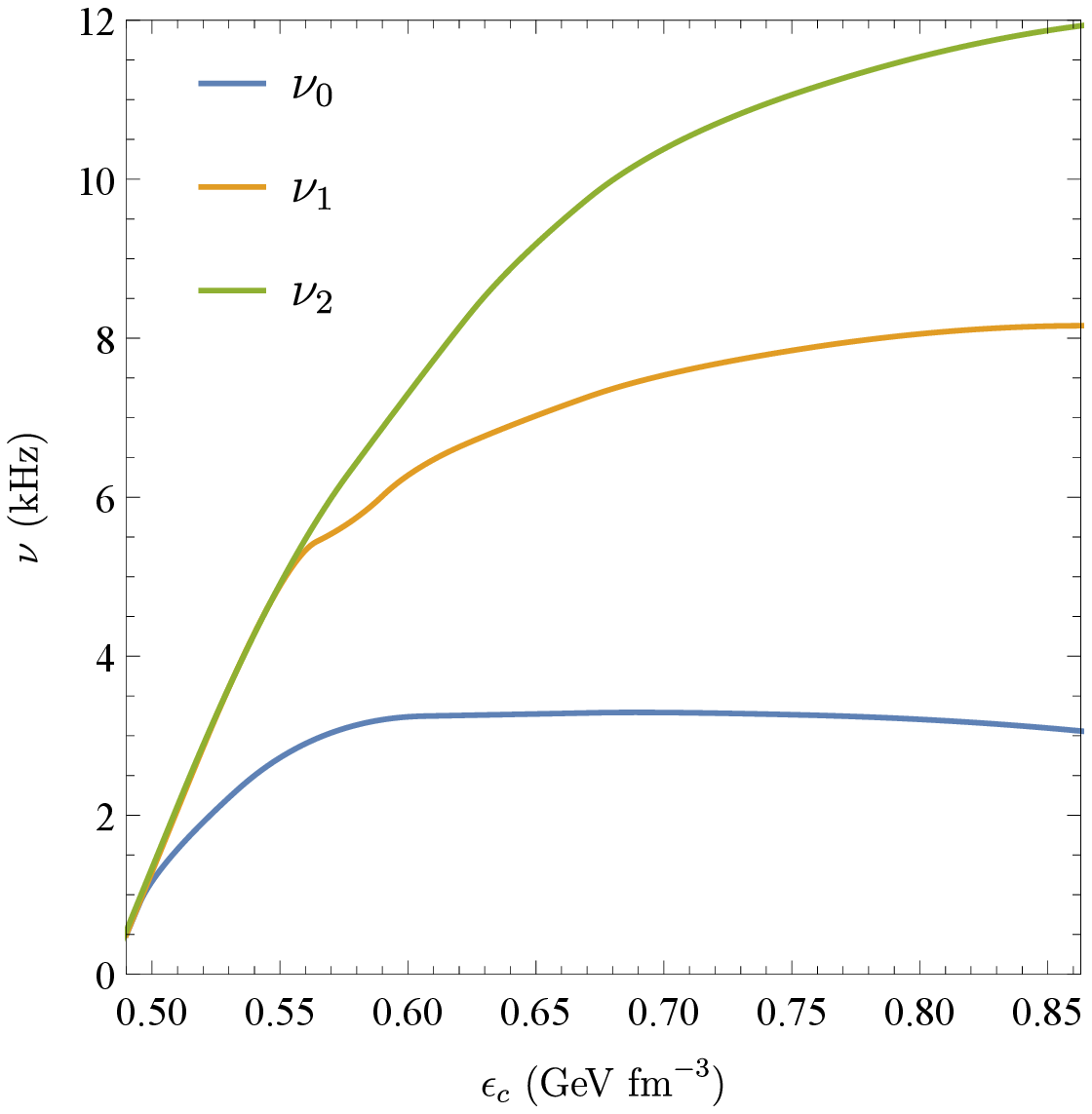}}	
		\caption{The frequencies of the fundamental mode ($\nu_{0}$) and the first two lowest-frequency excited modes ($\nu_{1}$ and $\nu_{2}$) of radial oscillation as functions of central density ($\epsilon_{c}$) for each EoS of nucleonic state (APR4, MPA1, MS1, SLy4) and hybrid nucleon--hyperon--quark state (H4, SQM1), listed in Table \ref{tab:EoS}.}
		\label{fig:frequency}
	\end{figure*}

	On each panel of Fig. \ref{fig:frequency}, the three lines represent the three lowest-frequency eigenmodes of radial oscillation as functions of central energy density. On each panel, the f-mode frequency drops towards zero as the particular stellar model approaches its dynamical stability limit which, indeed, is indicated by the presence of an eigenmode with zero frequency (cf. \cite[p.~21]{Ruoff2000}).	The dynamical instability in stars with MS1 and APR4 EoSs is exposed by the presence of a very low frequency of the f-mode, which has dropped to less than $5\%$ of that of the first excited mode, at central energy densities associated with the maximal-mass stable configurations. The decay of the lowest-frequency eigenmodes appears to be a general feature of Fig. \ref{fig:frequency}, irrespective of the particular EoS. The oscillation frequency of higher modes is always larger than that of a lower stable mode and for all modes it appears to decrease as the central energy density approches the smallest possible value $\epsilon_{\text{min}}$ of the particular stellar model. The explanation for our observations can thus be traced back to the fact that NSs at high densities can be approximated as being homogeneous \cite{Arnett1977} and the $\omega_{0}^2$ can be given by the relation
	\begin{equation} \label{omega-approximation}
	\omega_{0}^{2} \simeq G\bar{\rho}(4 - 3\Gamma).
	\end{equation}	
	Ref. \cite[p.~253]{Vath1992} confirms that when the central energy density of NSs is approaching $\epsilon_{\text{min}}$, such compact objects become approximately homogeneous and due to their small mass, non-relativistic \cite{Alcock1988}. The relation \ref{omega-approximation}, which shall be explicitly explained by eq. \ref{omega-approximation2}, expresses that the eigenvalue $\omega_{0}^2$ for the fundamental mode is proportional to the squared average density, as well as that the value of $\omega_{0}^2$ scales with the adiabatic index $\Gamma$. 
	
	It is also apparent from Fig. \ref{fig:frequency} that stellar models of softer EoSs have higher frequencies in the f-mode than the stiffer ones for the same central density. As reported by Ref. \cite{Lindblom1992}, stellar models of softer EoSs are generally associated with more centrally condensed stars with larger average densities. Having the highest pressure among the considered types of EoSs at a high-density region $\epsilon \lesssim 0.57 \text{ GeV fm}^{-3}$, MS1 and H4 are the stiffest EoSs and producing NSs with comparatively low-frequency oscillations. When the central energy density exceeds $\epsilon \lesssim 0.42 \text{ GeV fm}^{-3}$, the frequency for stars with H4 EoS decreases by about $15\%$ in the fundamental mode compared to the frequency of the MS1 EoS. This is due to the fact that H4 being a stiffer mixture of nuclear matter at low densities and hyperons at high densities (cf. Sec. \ref{sec:realistic-EoS}), a phase transition to hyperon-admixed matter takes place in the core of NSs. As reflected in the decreasing frequency, the phase transition causes a drop of adiabatic index with respect to stars with MS1 EoS, which do not contain hyperons but are otherwise mostly similar.
	
	\begin{table*}
		\centering
		\subfloat[APR4]{%
			\vspace{0.5cm}
			\begin{tabular}[t]{|c c c c c c|}
				\hwidth{1pt}
				$\epsilon_{c}$ & $R$ & $M$ & $\nu_{0}$ & $\nu_{1}$ & $\nu_{2}$ \\
				$(\text{GeV fm}^{-3})$ & (km) & $(M_{\odot})$ & (kHz) & (kHz) & (kHz) \\ 
				\hwidth{.4pt}
				1.518 & 9.940 & 2.219 & $0.398^{*}$ & 6.012 & 9.142 \\
				1.420 & 10.051 & 2.217 & 0.684 & 6.131 & 9.261 \\		
				1.320 & 10.172 & 2.210 & 1.145 & 6.279 & 9.409 \\
				1.220 & 10.304 & 2.195 & 1.470 & 6.418 & 9.542 \\
				1.120 & 10.441 & 2.171 & 1.799 & 6.576 & 9.698 \\
				1.030 & 10.589 & 2.131 & 2.143 & 6.738 & 9.868 \\
				0.930 & 10.739 & 2.068 & 2.471 & 6.883 & 10.005 \\
				0.830 & 10.886 & 1.974 & 2.797 & 6.965 & 10.196 \\
				0.730 & 11.008 & 1.832 & 3.108 & 6.937 & 10.254 \\
				0.630 & 11.080 & 1.625 & 3.278 & 6.769 & 8.984 \\
				0.530 & 11.038 & 1.336 & 3.063 & 6.254 & 6.645 \\
				0.430 & 10.836 & 0.977 & 2.269 & 4.191 & 4.156 \\
				1.330 & 10.474 & 0.600 & 1.001 & 0.984 & 0.998 \\
				\hwidth{1pt}
			\end{tabular}
			\vspace{0.5cm}
		} \hspace{40pt}
		\subfloat[H4]{%
			\vspace{0.5cm}
			\begin{tabular}[t]{|c c c c c c|}
				\hwidth{1pt}
				$\epsilon_{c}$ & $R$ & $M$ & $\nu_{0}$ & $\nu_{1}$ & $\nu_{2}$ \\
				$(\text{GeV fm}^{-3})$ & (km) & $(M_{\odot})$ & (kHz) & (kHz) & (kHz) \\ 
				\hwidth{.4pt}
				1.179 & 11.811 & 2.039 & 0.777* & 1.800 & 2.365 \\
				1.000 & 12.15 & 2.031 & 0.880 & 1.821 & 2.430 \\
				0.950 & 12.255 & 2.025 & 0.892 & 1.818 & 2.437 \\
				0.900 & 12.375 & 2.016 & 0.903 & 1.814 & 2.444 \\
				0.800 & 12.488 & 2.004 & 0.912 & 1.812 & 2.450 \\
				0.700 & 12.625 & 1.988 & 0.919 & 1.809 & 2.453 \\
				0.600 & 12.900 & 1.940 & 0.927 & 1.783 & 2.427 \\
				0.500 & 13.206 & 1.867 & 0.956 & 1.820 & 2.460 \\
				0.450 & 13.495 & 1.753 & 1.018 & 1.927 & 2.572 \\
				0.350 & 13.608 & 1.673 & 1.045 & 1.960 & 2.600 \\
				0.550 & 13.809 & 1.430 & 0.970 & 1.733 & 2.130 \\
				0.250 & 13.704 & 0.980 & 0.793 & 1.257 & 1.262 \\
				0.190 & 13.563 & 0.624 & 0.352 & 0.374 & 0.341 \\
				\hwidth{1pt}
			\end{tabular} 
			\vspace{0.5cm}
		} \hspace{40pt}
		\hfill
		\subfloat[MPA1]{%
			\vspace{0.5cm}
			\begin{tabular}[t]{|c c c c c c|}
				\hwidth{1pt}
				$\epsilon_{c}$ & $R$ & $M$ & $\nu_{0}$ & $\nu_{1}$ & $\nu_{2}$ \\
				$(\text{GeV fm}^{-3})$ & (km) & $(M_{\odot})$ & (kHz) & (kHz) & (kHz) \\ 
				\hwidth{.4pt}
				1.174 & 11.250 & 2.470 & 0.390* & 7.820 & 11.960 \\
				1.100 & 11.353 & 2.469 & 0.985 & 7.914 & 12.088 \\
				1.000 & 11.507 & 2.458 & 1.608 & 7.997 & 12.166 \\
				0.950 & 11.593 & 2.448 & 1.868 & 8.043 & 12.221 \\
				0.850 & 11.772 & 2.411 & 2.320 & 8.111 & 12.263 \\
				0.800 & 11.864 & 2.382 & 2.526 & 8.132 & 12.237 \\
				0.750 & 11.955 & 2.343 & 2.725 & 8.151 & 12.194 \\
				0.700 & 12.042 & 2.291 & 2.907 & 8.160 & 12.094 \\
				0.600 & 12.195 & 2.129 & 3.188 & 8.082 & 11.617 \\
				0.500 & 12.246 & 1.846 & 3.290 & 7.612 & 10.550 \\
				0.400 & 12.034 & 1.392 & 3.250 & 6.479 & 7.720 \\
				0.350 & 11.765 & 1.106 & 2.907 & 5.181 & 5.447 \\
				0.280 & 11.189 & 0.703 & 0.501 & 0.512 & 0.608 \\
				\hwidth{1pt}
			\end{tabular}
			\vspace{0.5cm}
		} \hspace{40pt}
		\subfloat[MS1]{%
			\vspace{0.5cm}
			\begin{tabular}[t]{|c c c c c c|}
				\hwidth{1pt}
				$\epsilon_{c}$ & $R$ & $M$ & $\nu_{0}$ & $\nu_{1}$ & $\nu_{2}$ \\
				$(\text{GeV fm}^{-3})$ & (km) & $(M_{\odot})$ & (kHz) & (kHz) & (kHz) \\ 
				\hwidth{.4pt}
				0.863 & 13.265 & 2.782 & $3.270^{*}$ & 7.873 & 11.117 \\
				0.800 & 13.410 & 2.780 & 3.284 & 7.741 & 10.830 \\		
				0.750 & 13.535 & 2.773 & 3.290 & 7.609 & 10.544 \\
				0.680 & 13.710 & 2.752 & 3.288 & 7.385 & 10.054 \\
				0.630 & 13.863 & 2.725 & 3.280 & 7.198 & 9.635 \\
				0.560 & 14.070 & 2.663 & 3.268 & 6.906 & 8.880 \\
				0.500 & 14.252 & 2.573 & 3.257 & 6.625 & 8.070 \\
				0.450 & 14.394 & 2.459 & 3.243 & 6.280 & 7.315 \\
				0.390 & 14.523 & 2.247 & 3.118 & 5.702 & 6.337 \\
				0.330 & 14.555 & 1.923 & 2.844 & 4.996 & 5.232 \\
				0.270 & 14.392 & 1.484 & 2.416 & 4.175 & 3.914 \\
				0.210 & 13.965 & 1.012 & 1.665 & 3.237 & 2.380 \\
				0.150 & 13.184 & 0.560 & 0.502 & 2.182 & 0.631 \\
				\hwidth{1pt}
			\end{tabular}
			\vspace{0.5cm}
		} \hspace{40pt}
		\hfill
		\subfloat[Sly4]{%
			\vspace{0.5cm}
			\begin{tabular}[t]{|c c c c c c|}
				\hwidth{1pt}
				$\epsilon_{c}$ & $R$ & $M$ & $\nu_{0}$ & $\nu_{1}$ & $\nu_{2}$ \\
				$(\text{GeV fm}^{-3})$ & (km) & $(M_{\odot})$ & (kHz) & (kHz) & (kHz) \\ 
				\hwidth{.4pt}
				2.056 & 9.238 & 2.025 & 1.818* & 3.018 & 3.968 \\
				1.900 & 9.496 & 2.042 & 1.814 & 2.983 & 4.029 \\
				1.800 & 9.641 & 2.050 & 1.805 & 2.956 & 4.036 \\
				1.600 & 9.887 & 2.056 & 1.773 & 2.887 & 3.986 \\
				1.450 & 10.065 & 2.053 & 1.738 & 2.825 & 3.886 \\
				1.300 & 10.262 & 2.040 & 1.691 & 2.753 & 3.734 \\
				1.150 & 10.486 & 2.011 & 1.635 & 2.671 & 3.530 \\
				1.000 & 10.727 & 1.956 & 1.568 & 2.579 & 3.274 \\
				0.850 & 10.982 & 1.857 & 1.490 & 2.478 & 2.965 \\
				0.700 & 11.218 & 1.688 & 1.415 & 2.291 & 2.541 \\
				0.550 & 11.360 & 1.408 & 1.358 & 1.750 & 1.791 \\
				0.400 & 11.251 & 0.979 & 0.916 & 0.886 & 0.891 \\
				0.280 & 10.831 & 0.553 & 0.296 & 0.299 & 0.300 \\
				\hwidth{1pt}
			\end{tabular}
			\vspace{0.5cm}
		} \hspace{40pt}
		\subfloat[SQM1]{%
			\vspace{0.5cm}
			\begin{tabular}[t]{|c c c c c c|}
				\hwidth{1pt}
				$\epsilon_{c}$ & $R$ & $M$ & $\nu_{0}$ & $\nu_{1}$ & $\nu_{2}$ \\
				$(\text{GeV fm}^{-3})$ & (km) & $(M_{\odot})$ & (kHz) & (kHz) & (kHz) \\ 
				\hwidth{.4pt}
				0.823 & 13.357 &2.782 & $3.165^{*}$ & 8.109 & 11.703 \\
				0.800 & 13.410 & 2.780 & 3.207 & 8.057 & 11.544 \\		
				0.770 & 13.484 & 2.776 & 3.251 & 7.956 & 11.297 \\
				0.740 & 13.560 & 2.770 & 3.278 & 7.815 & 10.990 \\
				0.710 & 13.637 & 2.762 & 3.290 & 7.612 & 10.550 \\
				0.680 & 13.720 & 2.752 & 3.286 & 7.345 & 9.967 \\
				0.660 & 13.775 & 2.742 & 3.278 & 7.134 & 9.481 \\
				0.630 & 13.863 & 2.725 & 3.262 & 6.778 & 8.511 \\
				0.600 & 13.951 & 2.703 & 3.240 & 6.272 & 7.301 \\
				0.570 & 14.040 & 2.674 & 3.035 & 5.474 & 5.954 \\
				0.550 & 14.100 & 2.651 & 2.750 & 4.820 & 4.894 \\
				0.520 & 14.190 & 2.608 & 2.005 & 3.120 & 2.958 \\
				0.490 & 14.280 & 2.554 & 0.512 & 0.524 & 0.608 \\
				\hwidth{1pt}
			\end{tabular}
			\vspace{0.5cm}
		}
		\caption{The frequencies of the fundamental mode ($\nu_{0}$) and the first two lowest-frequency excited modes ($\nu_{1}$ and $\nu_{2}$) of radial oscillation as functions of central density ($\epsilon_{c}$) for each EoS of nucleonic state (APR4, MPA1, MS1, SLy4) and hybrid nucleon--hyperon--quark state (H4, SQM1), listed in Table \ref{tab:EoS}. The asterisk indicates $\nu_{0}$ corresponds to the maximal-mass stable configuration for the particular EoS (marked by the symbols $\diamond$ in Fig. \ref{fig:MR}). The gravitational mass and circumferential radius of each corresponding stellar model are in units of $M_{\odot}$ and km, respectively.} \label{tab:frequency}
	\end{table*}

	\appendix
	
	\section{Components of the stress--energy tensor} \label{sec:stress-energy-components}
	The non-vanishing components of the viscosity and heat flux stress--energy tensors defined by \eqref{energy-momentum-tensor2} and quantities involving them shall be explicitly evaluated below. Here, we only repeat those quantities from our preceding paper \cite{Barta2019} that appear in the present paper. The non-vanishing components of the unperturbed and perturbed states are written in the notation we used in this paper as
	\begin{equation} \label{t1components}
	\begin{array}{l}
	\displaystyle (T_{\text{visc}} + T_{\text{heat}})^{1}_{\;0} = \frac{1}{4}e^{-\nu_{0}/2}\left[8\kappa T'+\left(12\eta-\kappa T\right)\nu'_{0}\right], \\[10pt]
	\displaystyle (\delta T_{\text{visc}} + \delta T_{\text{heat}})^{1}_{\;0} = \displaystyle e^{-\nu_{0}/2}\left[3\eta\nu'_{0}\delta\nu-\left(\eta-\frac{1}{4}\kappa T\right)\delta\nu'\right],
	\end{array}
	\end{equation}
	respectively. The inhomogenity of the Sturm--Liouville equation \eqref{pulsation-equation} that determines the damped solution is expressed by
	\begin{equation} \label{S-def}
	\begin{array}{ll}
	\mkern-20mu \mathcal{S}_{1} = & \!\!\displaystyle \frac{e^{-\nu_{0}/2}}{r}\left(\eta\delta\lambda + A\delta u_{r} - A\frac{(\delta T_{\text{visc}} + \delta T_{\text{heat}})^{1}_{\;0}}{\bar{p}_{0} + \bar{\epsilon}_{0}}\right) \\[10pt]
	& \!\!\displaystyle - \frac{1}{2}(T_{\text{visc}} + T_{\text{heat}})^{1}_{\;0}(\delta\lambda + \delta\nu), \\[10pt]
	\mkern-20mu \mathcal{S}_{2} = & \!\!\displaystyle e^{-(\lambda_{0}+2\nu_{0})/2}\frac{\mathrm{d}}{\mathrm{d}r}\bigg[\frac{e^{\lambda_{0}+2\nu_{0}}}{r^2}\left(\frac{\partial \bar{N}_{0}}{\partial\bar{p}_{0}}\right)^{-1} \times \\[10pt]
	& \!\!\displaystyle \frac{\mathrm{d}}{\mathrm{d}r}\left(r^2 e^{-(\lambda_{0}+2\nu_{0})/2}\bar{N}_{0}\frac{(\delta T_{\text{visc}} + \delta T_{\text{heat}})^{1}_{\;0}}{\bar{p}_{0} + \bar{\epsilon}_{0}}\right)\bigg],
	\end{array}
	\end{equation}
	where $\delta u_{r}$ is the radial velocity defined by eq. \eqref{radial-velocity} and
	\begin{equation}
	A = \eta\left[\left(1+3e^{\nu_{0}-\lambda_{0}}\right)\nu'_{0} - 9\lambda'_{0} - \frac{2}{r}e^{\lambda_{0}}\right] + \kappa\left[T\nu'_{0}-8T'\right].
	\end{equation}

	\section{Dynamical stability of non-rotating spherical stars in Newtonian gravity} \label{sec:dynamical-stability}
	This section provides a Lagrangian description of radial oscillations of stars in the framework of classical hydrodynamics in order for one to find a linear approximation for the dynamical times cale of these oscillations. Let us consider a mass element $dm$ at a radius $r$ inside the star at a given moment $t = t_{0}$ contained in a thin spherical shell with a thickness $dr$ (see Fig. \ref{fig:shells}). Considering that the spatial coordinate of the given mass element does not vary in time, it is more convenient to take a Lagrangian coordinate, which is connected to the mass element
	\begin{equation} 
	dm  = \rho dr,
	\end{equation}
	instead of the radius $r$. \cite{Kippenhahn1990} The spatial behavior of the radial function $r(m,t)$ is then described by differential equation
	\begin{equation} \label{mass-conservation}
	\frac{\partial r}{\partial m}  = \frac{1}{4\pi r^{2}\rho},
	\end{equation}
	in accordance with the definition of gravitational mass \eqref{mass-func2}. The Navier--Stokes equation of hydrodynamics with spherical symmetry provide the greatly simplified expression 
	\begin{equation} \label{equation-of-motion0}
	\frac{\partial^{2} r}{\partial t^{2}} = F_{g} - \frac{1}{\rho}\frac{\partial p}{\partial r}
	\end{equation}
	for the equation of motion of mass elements, where 
	\begin{equation} \label{forces}
	F_{g} = -\frac{Gm}{r^{2}}, \quad \frac{\partial p}{\partial r} = 4\pi r^{2}\frac{\partial p}{\partial m}
	\end{equation}
	are the gravitational force gravity and pressure gradient, respectively. \cite{Maeder2009} In order to prevent the mass elements of the shell from being accelerated by the gravitational force towards the center of the star (as indicated by the minus sign), a net force of the same absolute value due to pressure acts on the shell in an outward direction (see Fig. \ref{fig:shells}). In the state of hydrostatic equilibrium $\partial^{2} r/\partial t^{2} = 0$ and the sum of the forces arising from the gravity and pressure has to be zero as well, which yields the ODE of stellar structure
	\begin{equation} \label{hydrostatic-equilibrium-condition}
	\frac{\partial p }{\partial m} = -\frac{Gm}{4\pi r^{4}}\rho,
	\end{equation}
	where the pressure, defined by
	\begin{equation} \label{pressure-force}
	p_{j-1} - p_{j}  = \frac{\partial p}{\partial r}dr,
	\end{equation}
	is consequently always greater at the lower boundary than at the outer boundary of the shell, expressively, $p_{j} > p_{j-1}$ (see Fig. \ref{fig:shells}). The pair of basic equations \eqref{mass-conservation} and \eqref{hydrostatic-equilibrium-condition} in the Lagrangian description corresponds to the conservation of mass and of momentum, respectively.
	
	Let us now recall the small, adiabatic perturbations of the equilibrium stellar configuration described in Sec. \ref{sec:perturbations}. We have seen that the change in the radius of every mass shell is $r = r_{0}(1 + \delta)$, where the small parameter $\delta \ll 1$ may vary only with time (see eq. \eqref{time-dependence3}). Mass conservation expressed by eq. \eqref{mass-conservation} entails that the change in the rest-mass density is $\rho = \rho_{0}(1 + 3\delta)$. Considering that the perturbation is adiabatic, the definition \eqref{adiabatic-index} for the adiabatic index $\Gamma_{1}$ implies that
	\begin{equation}
	\frac{p - p_{0}}{p_{0}} = - \Gamma_{1} \frac{\rho - \rho_{0}}{\rho_{0}},
	\end{equation}
	so the change in the pressure is $p = p_{0}(1 - 3\delta)$, where the pressure and density perturbations are assumed to be very small: $|(p - p_{0})/p_{0}| \ll 1, \quad |(\rho - \rho_{0})/\rho_{0}| \ll 1$. \cite{Phillips1994}
	
	The perturbed quantities no longer satisfy the equation of hydrostatic equilibrium \eqref{hydrostatic-equilibrium-condition}, but rather the equation of motion \eqref{equation-of-motion0}. As the only time variable quantity is $\delta$ we may write the equation of motion in the form
	\begin{equation} \label{equation-of-motion}
	\frac{d^{2}\delta r_{0}}{dt^{2}} = - \frac{Gm}{r_{0}^{2}}(1 - 2\delta) - 4\pi r_{0}^{2}\frac{\partial p_{0}}{\partial m}(1 - 2\delta - 3\Gamma \delta)
	\end{equation}
	where any quantity with the subscript `0' satisfies the equation of hydrostatic equilibrium \eqref{hydrostatic-equilibrium-condition}. Therefore, combining equations \eqref{mass-conservation} and \eqref{equation-of-motion} we obtain
	\begin{equation}
	\frac{d^{2}\delta}{dt^{2}} = - \frac{Gm}{r_{0}^{3}}(4 - 3\Gamma)\delta.
	\end{equation}
	Now we make another crude approximation, $M/r_{0}^{3} \approx \bar{\rho} = \text{const.}$ where $M$ is the total mass of the star, and we obtain
	\begin{equation} \label{omega-approximation2}
	\frac{d^{2}\delta}{dt^{2}} = G\bar{\rho}(4 - 3\Gamma)\delta = \omega^{2}\delta, \quad \omega^{2} \equiv G\bar{\rho}(4 - 3\Gamma).
	\end{equation}
	This equation has a solution
	\begin{equation}
	\delta = \delta_{1}e^{\omega t} + \delta_{2}e^{-\omega t}.
	\end{equation}
	If $\omega^{2} < 0$, then the motion is oscillatory, i.e. the star is dynamically stable. If $\omega^{2} > 0$, then there is a solution that increases exponentially, i.e. the star is dynamically unstable. The proper analysis leads to the requirement, which was already pointed out in Sec. \ref{sec:polytropic-EoS}, that the average value of $\Gamma$ within the star has to be less than $4/3$ for the star to be dynamically unstable. The linear instability of an oscillating dynamical system occurs on a dynamical timescale
	\begin{equation} \label{time-scale}
	\tau_{d} \approx \omega^{-1} \approx \left(\frac{R^{3}}{GM}\right)^{1/2} \approx \frac{1}{2}(G\bar{\rho})^{-1/2}
	\end{equation}
	where $\hat{\rho}$ is the mean density of a star. \cite{Haensel2007} $\tau_{d}$ is often referred as the hydrostatic timescale, considering that it represents the typical time after which a slightly perturbed star can return to equilibrium. The critical value for the adiabatic exponent, $\Gamma_{cr} = 4/3$, is a consequence of the gravitational force varying as $r^{-2}$. If the gravitational acceleration was given as $-GM/r^{2+\alpha}$, then the critical value would be $\Gamma_{cr} = (4 + \alpha)/3$. One of the effects of GR is to make gravity stronger than Newtonian near a black hole, but the effect is there even when the field is weak. For $r \gg r_{s} = 2GM/c^{2}$ we have roughly $\alpha \approx r_{s}/r$. This means that dynamical instability may occur when $\Gamma$ is just very close to, but somewhat larger than $4/3$.

	\begin{acknowledgments}
	This work was supported by the National Research, Development and Innovation Office of Hungary (NKFIH, by Hungarian abbreviation) via grants NKFIH K--124366 and K--124508. I also acknowledge the partial support of networking activities by the ``PHAROS'' EU COST Action CA16214. I wish to express my gratitude and appreciation to Prof. Kostas Kokkotas, director of Institute for Astronomy and Astrophysics at the University of T\"{u}bingen (IAAT), for his helpful discussions and expert guidance in dealing with numerical issues in computing accurate eigenfrequencies of the fundamental and higher-order radial modes of neutron-star oscillations. I would also like to extend my gratitude to my former doctoral advisor, Dr. M\'{a}ty\'{a}s Vas\'{u}th, head of Gravitational Physics Research Group at the Wigner Research Centre for Physics, for his critical comments and for reviewing some of the most important calculations.
	\end{acknowledgments}

	\bibliography{cikk3}

\begin{thebibliography}{72}%
\makeatletter
\providecommand \@ifxundefined [1]{%
 \@ifx{#1\undefined}
}%
\providecommand \@ifnum [1]{%
 \ifnum #1\expandafter \@firstoftwo
 \else \expandafter \@secondoftwo
 \fi
}%
\providecommand \@ifx [1]{%
 \ifx #1\expandafter \@firstoftwo
 \else \expandafter \@secondoftwo
 \fi
}%
\providecommand \natexlab [1]{#1}%
\providecommand \enquote  [1]{``#1''}%
\providecommand \bibnamefont  [1]{#1}%
\providecommand \bibfnamefont [1]{#1}%
\providecommand \citenamefont [1]{#1}%
\providecommand \href@noop [0]{\@secondoftwo}%
\providecommand \href [0]{\begingroup \@sanitize@url \@href}%
\providecommand \@href[1]{\@@startlink{#1}\@@href}%
\providecommand \@@href[1]{\endgroup#1\@@endlink}%
\providecommand \@sanitize@url [0]{\catcode `\\12\catcode `\$12\catcode
  `\&12\catcode `\#12\catcode `\^12\catcode `\_12\catcode `\%12\relax}%
\providecommand \@@startlink[1]{}%
\providecommand \@@endlink[0]{}%
\providecommand \url  [0]{\begingroup\@sanitize@url \@url }%
\providecommand \@url [1]{\endgroup\@href {#1}{\urlprefix }}%
\providecommand \urlprefix  [0]{URL }%
\providecommand \Eprint [0]{\href }%
\providecommand \doibase [0]{https://doi.org/}%
\providecommand \selectlanguage [0]{\@gobble}%
\providecommand \bibinfo  [0]{\@secondoftwo}%
\providecommand \bibfield  [0]{\@secondoftwo}%
\providecommand \translation [1]{[#1]}%
\providecommand \BibitemOpen [0]{}%
\providecommand \bibitemStop [0]{}%
\providecommand \bibitemNoStop [0]{.\EOS\space}%
\providecommand \EOS [0]{\spacefactor3000\relax}%
\providecommand \BibitemShut  [1]{\csname bibitem#1\endcsname}%
\let\auto@bib@innerbib\@empty
\bibitem [{\citenamefont {Chirenti}\ \emph {et~al.}(2015)\citenamefont
  {Chirenti}, \citenamefont {de~Souza},\ and\ \citenamefont
  {Kastaun}}]{Chirenti2015}%
  \BibitemOpen
  \bibfield  {author} {\bibinfo {author} {\bibfnamefont {C.}~\bibnamefont
  {Chirenti}}, \bibinfo {author} {\bibfnamefont {G.~H.}\ \bibnamefont
  {de~Souza}},\ and\ \bibinfo {author} {\bibfnamefont {W.}~\bibnamefont
  {Kastaun}},\ }\bibfield  {title} {\bibinfo {title} {Fundamental oscillation
  modes of neutron stars: validity of universal relations},\ }\href
  {https://doi.org/10.1103/PhysRevD.91.044034} {\bibfield  {journal} {\bibinfo
  {journal} {Phys. Rev. D}\ }\textbf {\bibinfo {volume} {91}},\ \bibinfo
  {pages} {044034} (\bibinfo {year} {2015})}\BibitemShut {NoStop}%
\bibitem [{\citenamefont {\"{O}zel}\ and\ \citenamefont
  {Freire}(2016)}]{Ozel2016}%
  \BibitemOpen
  \bibfield  {author} {\bibinfo {author} {\bibfnamefont {F.}~\bibnamefont
  {\"{O}zel}}\ and\ \bibinfo {author} {\bibfnamefont {P.}~\bibnamefont
  {Freire}},\ }\bibfield  {title} {\bibinfo {title} {Masses, radii, and the
  equation of state of neutron stars},\ }\href
  {https://doi.org/10.1146/annurev-astro-081915-023322} {\bibfield  {journal}
  {\bibinfo  {journal} {Annu. Rev. Astron. Astrophys.}\ }\textbf {\bibinfo
  {volume} {54}},\ \bibinfo {pages} {401} (\bibinfo {year} {2016})}\BibitemShut
  {NoStop}%
\bibitem [{\citenamefont {{Lattimer}}\ and\ \citenamefont
  {{Prakash}}(2001)}]{Lattimer2001}%
  \BibitemOpen
  \bibfield  {author} {\bibinfo {author} {\bibfnamefont {J.~M.}\ \bibnamefont
  {{Lattimer}}}\ and\ \bibinfo {author} {\bibfnamefont {M.}~\bibnamefont
  {{Prakash}}},\ }\bibfield  {title} {\bibinfo {title} {{Neutron star structure
  and the equation of state}},\ }\href {https://doi.org/10.1086/319702}
  {\bibfield  {journal} {\bibinfo  {journal} {\apj}\ }\textbf {\bibinfo
  {volume} {550}},\ \bibinfo {pages} {426} (\bibinfo {year}
  {2001})}\BibitemShut {NoStop}%
\bibitem [{\citenamefont {Read}\ \emph {et~al.}(2009)\citenamefont {Read},
  \citenamefont {Lackey}, \citenamefont {Owen},\ and\ \citenamefont
  {Friedman}}]{Read2009}%
  \BibitemOpen
  \bibfield  {author} {\bibinfo {author} {\bibfnamefont {J.~S.}\ \bibnamefont
  {Read}}, \bibinfo {author} {\bibfnamefont {B.~D.}\ \bibnamefont {Lackey}},
  \bibinfo {author} {\bibfnamefont {B.~J.}\ \bibnamefont {Owen}},\ and\
  \bibinfo {author} {\bibfnamefont {J.~L.}\ \bibnamefont {Friedman}},\
  }\bibfield  {title} {\bibinfo {title} {Constraints on a phenomenologically
  parametrized neutron-star equation of state},\ }\href
  {https://doi.org/10.1103/PhysRevD.79.124032} {\bibfield  {journal} {\bibinfo
  {journal} {Phys. Rev. D}\ }\textbf {\bibinfo {volume} {79}},\ \bibinfo
  {pages} {124032} (\bibinfo {year} {2009})}\BibitemShut {NoStop}%
\bibitem [{\citenamefont {Fonseca}\ \emph {et~al.}(2016)\citenamefont {Fonseca}
  \emph {et~al.}}]{Fonseca2016}%
  \BibitemOpen
  \bibfield  {author} {\bibinfo {author} {\bibfnamefont {E.}~\bibnamefont
  {Fonseca}} \emph {et~al.},\ }\bibfield  {title} {\bibinfo {title} {{The
  NANOGrav nine-year data set: mass and geometric measurements of binary
  millisecond pulsars}},\ }\href {https://doi.org/10.3847/0004-637X/832/2/167}
  {\bibfield  {journal} {\bibinfo  {journal} {\apj}\ }\textbf {\bibinfo
  {volume} {832}},\ \bibinfo {eid} {167} (\bibinfo {year} {2016})}\BibitemShut
  {NoStop}%
\bibitem [{\citenamefont {Antoniadis}\ \emph {et~al.}(2013)\citenamefont
  {Antoniadis} \emph {et~al.}}]{Antoniadis2013}%
  \BibitemOpen
  \bibfield  {author} {\bibinfo {author} {\bibfnamefont {J.}~\bibnamefont
  {Antoniadis}} \emph {et~al.},\ }\bibfield  {title} {\bibinfo {title} {{A
  massive pulsar in a compact relativistic binary}},\ }\href
  {https://doi.org/10.1126/science.1233232} {\bibfield  {journal} {\bibinfo
  {journal} {Science}\ }\textbf {\bibinfo {volume} {340}},\ \bibinfo {pages}
  {448} (\bibinfo {year} {2013})}\BibitemShut {NoStop}%
\bibitem [{\citenamefont {Miller}\ and\ \citenamefont
  {Lamb}(2016)}]{Miller2016}%
  \BibitemOpen
  \bibfield  {author} {\bibinfo {author} {\bibfnamefont {M.~C.}\ \bibnamefont
  {Miller}}\ and\ \bibinfo {author} {\bibfnamefont {F.~K.}\ \bibnamefont
  {Lamb}},\ }\bibfield  {title} {\bibinfo {title} {{Observational constraints
  on neutron star masses and radii}},\ }\href
  {https://doi.org/10.1140/epja/i2016-16063-8} {\bibfield  {journal} {\bibinfo
  {journal} {Eur. Phys. J.}\ }\textbf {\bibinfo {volume} {A52}},\ \bibinfo
  {pages} {63} (\bibinfo {year} {2016})}\BibitemShut {NoStop}%
\bibitem [{\citenamefont {Gendreau}\ \emph {et~al.}(2012)\citenamefont
  {Gendreau}, \citenamefont {Arzoumanian},\ and\ \citenamefont
  {Okajima}}]{Gendreau2012}%
  \BibitemOpen
  \bibfield  {author} {\bibinfo {author} {\bibfnamefont {K.~C.}\ \bibnamefont
  {Gendreau}}, \bibinfo {author} {\bibfnamefont {Z.}~\bibnamefont
  {Arzoumanian}},\ and\ \bibinfo {author} {\bibfnamefont {T.}~\bibnamefont
  {Okajima}},\ }\bibfield  {title} {\bibinfo {title} {The neutron star interior
  composition explorer (nicer): an explorer mission of opportunity for soft
  {X}-ray timing spectroscopy},\ }\href {https://doi.org/10.1117/12.926396}
  {\bibfield  {journal} {\bibinfo  {journal} {Proc. SPIE}\ }\textbf {\bibinfo
  {volume} {8443}},\ \bibinfo {pages} {844313} (\bibinfo {year}
  {2012})}\BibitemShut {NoStop}%
\bibitem [{\citenamefont {Watts}\ \emph {et~al.}(2016)\citenamefont {Watts}
  \emph {et~al.}}]{Watts2016}%
  \BibitemOpen
  \bibfield  {author} {\bibinfo {author} {\bibfnamefont {A.~L.}\ \bibnamefont
  {Watts}} \emph {et~al.},\ }\bibfield  {title} {\bibinfo {title} {Colloquium:
  Measuring the neutron star equation of state using {X}-ray timing},\ }\href
  {https://doi.org/10.1103/RevModPhys.88.021001} {\bibfield  {journal}
  {\bibinfo  {journal} {Rev. Mod. Phys.}\ }\textbf {\bibinfo {volume} {88}},\
  \bibinfo {pages} {021001} (\bibinfo {year} {2016})}\BibitemShut {NoStop}%
\bibitem [{\citenamefont {Abbott}\ \emph {et~al.}(2018)\citenamefont {Abbott}
  \emph {et~al.}}]{Abbott2018}%
  \BibitemOpen
  \bibfield  {author} {\bibinfo {author} {\bibfnamefont {B.~P.}\ \bibnamefont
  {Abbott}} \emph {et~al.} (\bibinfo {collaboration} {The LIGO Scientific
  Collaboration and the Virgo Collaboration}),\ }\bibfield  {title} {\bibinfo
  {title} {{GW}170817: Measurements of neutron star radii and equation of
  state},\ }\href {https://doi.org/10.1103/PhysRevLett.121.161101} {\bibfield
  {journal} {\bibinfo  {journal} {Phys. Rev. Lett.}\ }\textbf {\bibinfo
  {volume} {121}},\ \bibinfo {pages} {161101} (\bibinfo {year}
  {2018})}\BibitemShut {NoStop}%
\bibitem [{\citenamefont {Abbott}\ \emph {et~al.}(2020)\citenamefont {Abbott}
  \emph {et~al.}}]{Abbott2020}%
  \BibitemOpen
  \bibfield  {author} {\bibinfo {author} {\bibfnamefont {B.~P.}\ \bibnamefont
  {Abbott}} \emph {et~al.} (\bibinfo {collaboration} {The LIGO Scientific
  Collaboration and the Virgo Collaboration}),\ }\bibfield  {title} {\bibinfo
  {title} {{GW}190425: Observation of a compact binary coalescence with total
  mass $\sim 3.4 {M}_{\odot}$},\ }\href
  {https://doi.org/10.3847/2041-8213/ab75f5} {\bibfield  {journal} {\bibinfo
  {journal} {Astrophys. J.}\ }\textbf {\bibinfo {volume} {892}},\ \bibinfo
  {pages} {L3} (\bibinfo {year} {2020})}\BibitemShut {NoStop}%
\bibitem [{\citenamefont {Lindblom}(1992)}]{Lindblom1992}%
  \BibitemOpen
  \bibfield  {author} {\bibinfo {author} {\bibfnamefont {L.}~\bibnamefont
  {Lindblom}},\ }\bibfield  {title} {\bibinfo {title} {{Determining the nuclear
  equation of state from neutron-star masses and radii}},\ }\href
  {https://doi.org/10.1086/171882} {\bibfield  {journal} {\bibinfo  {journal}
  {\apj}\ }\textbf {\bibinfo {volume} {398}},\ \bibinfo {pages} {569} (\bibinfo
  {year} {1992})}\BibitemShut {NoStop}%
\bibitem [{\citenamefont {Ramaty}\ and\ \citenamefont
  {Lingenfelter}(1981)}]{Ramaty1981}%
  \BibitemOpen
  \bibfield  {author} {\bibinfo {author} {\bibfnamefont {R.}~\bibnamefont
  {Ramaty}}\ and\ \bibinfo {author} {\bibfnamefont {R.~E.}\ \bibnamefont
  {Lingenfelter}},\ }\bibfield  {title} {\bibinfo {title} {Interpretations and
  implications of ${\beta}$-ray lines from solar flares, the galactic centre
  and ${\beta}$-ray transients},\ }\href
  {https://doi.org/10.1098/rsta.1981.0151} {\bibfield  {journal} {\bibinfo
  {journal} {Phil. Trans. Roy. Soc.}\ }\textbf {\bibinfo {volume} {301}},\
  \bibinfo {pages} {671} (\bibinfo {year} {1981})}\BibitemShut {NoStop}%
\bibitem [{\citenamefont {Sagun}\ \emph {et~al.}(2020)\citenamefont {Sagun},
  \citenamefont {Panotopoulos},\ and\ \citenamefont {Lopes}}]{Sagun2020}%
  \BibitemOpen
  \bibfield  {author} {\bibinfo {author} {\bibfnamefont {V.}~\bibnamefont
  {Sagun}}, \bibinfo {author} {\bibfnamefont {G.}~\bibnamefont
  {Panotopoulos}},\ and\ \bibinfo {author} {\bibfnamefont {I.}~\bibnamefont
  {Lopes}},\ }\bibfield  {title} {\bibinfo {title} {Asteroseismology: Radial
  oscillations of neutron stars with realistic equation of state},\ }\href
  {https://doi.org/10.1103/PhysRevD.101.063025} {\bibfield  {journal} {\bibinfo
   {journal} {Phys. Rev. D}\ }\textbf {\bibinfo {volume} {101}},\ \bibinfo
  {pages} {063025} (\bibinfo {year} {2020})}\BibitemShut {NoStop}%
\bibitem [{\citenamefont {{Detweiler}}\ and\ \citenamefont
  {{Lindblom}}(1985)}]{Detweiler1983}%
  \BibitemOpen
  \bibfield  {author} {\bibinfo {author} {\bibfnamefont {S.}~\bibnamefont
  {{Detweiler}}}\ and\ \bibinfo {author} {\bibfnamefont {L.}~\bibnamefont
  {{Lindblom}}},\ }\bibfield  {title} {\bibinfo {title} {{On the nonradial
  pulsations of general relativistic stellar models}},\ }\href
  {https://doi.org/10.1086/163127} {\bibfield  {journal} {\bibinfo  {journal}
  {\apj}\ }\textbf {\bibinfo {volume} {292}},\ \bibinfo {pages} {12} (\bibinfo
  {year} {1985})}\BibitemShut {NoStop}%
\bibitem [{\citenamefont {Phelan}\ \emph {et~al.}(2008)\citenamefont {Phelan},
  \citenamefont {Ryan},\ and\ \citenamefont {Shearer}}]{Phelan2008}%
  \BibitemOpen
  \bibinfo {editor} {\bibfnamefont {D.}~\bibnamefont {Phelan}}, \bibinfo
  {editor} {\bibfnamefont {O.}~\bibnamefont {Ryan}},\ and\ \bibinfo {editor}
  {\bibfnamefont {A.}~\bibnamefont {Shearer}},\ eds.,\ \href
  {https://doi.org/10.1007/978-1-4020-6518-7_1} {\emph {\bibinfo {title} {High
  Time Resolution Astrophysics}}},\ \bibinfo {series} {Astrophysics and Space
  Science Library}, Vol.\ \bibinfo {volume} {351}\ (\bibinfo  {publisher}
  {Springer Publishing},\ \bibinfo {address} {Dordrecht, The Netherlands},\
  \bibinfo {year} {2008})\BibitemShut {NoStop}%
\bibitem [{\citenamefont {Chandrasekhar}(1964)}]{Chandrasekhar1964}%
  \BibitemOpen
  \bibfield  {author} {\bibinfo {author} {\bibfnamefont {S.}~\bibnamefont
  {Chandrasekhar}},\ }\bibfield  {title} {\bibinfo {title} {{The dynamical
  instability of gaseous masses approaching the Schwarzschild limit in General
  Relativity}},\ }\href {https://doi.org/10.1086/147938} {\bibfield  {journal}
  {\bibinfo  {journal} {\apj}\ }\textbf {\bibinfo {volume} {140}},\ \bibinfo
  {pages} {417} (\bibinfo {year} {1964})},\ \bibinfo {note} {erratum:
  Astrophys. J. \textbf{140}, 1342 (1964)}\BibitemShut {NoStop}%
\bibitem [{\citenamefont {Harrison}\ \emph {et~al.}(1965)\citenamefont
  {Harrison}, \citenamefont {Thorne}, \citenamefont {Wakano},\ and\
  \citenamefont {Wheeler}}]{Harrison1965}%
  \BibitemOpen
  \bibfield  {author} {\bibinfo {author} {\bibfnamefont {B.~K.}\ \bibnamefont
  {Harrison}}, \bibinfo {author} {\bibfnamefont {K.~S.}\ \bibnamefont
  {Thorne}}, \bibinfo {author} {\bibfnamefont {M.}~\bibnamefont {Wakano}},\
  and\ \bibinfo {author} {\bibfnamefont {J.~A.}\ \bibnamefont {Wheeler}},\
  }\href {https://doi.org/10.1007/978-0-226-31802-8} {\emph {\bibinfo {title}
  {Gravitation Theory and Gravitational Collapse, Chicago: University of
  Chicago Press, 1965}}},\ \bibinfo {edition} {1st}\ ed.,\ Vol.~\bibinfo
  {volume} {1}\ (\bibinfo  {publisher} {University of Chicago Press},\ \bibinfo
  {address} {Chicago, IL},\ \bibinfo {year} {1965})\BibitemShut {NoStop}%
\bibitem [{\citenamefont {{Zel'dovich}}\ and\ \citenamefont
  {{Novikov}}(1971)}]{Zeldovich1971}%
  \BibitemOpen
  \bibfield  {author} {\bibinfo {author} {\bibfnamefont {{\relax Ya}.~B.}\
  \bibnamefont {{Zel'dovich}}}\ and\ \bibinfo {author} {\bibfnamefont {I.~D.}\
  \bibnamefont {{Novikov}}},\ }\href@noop {} {\emph {\bibinfo {title}
  {Relativistic Astrophysics: Stars and Relativity}}},\ \bibinfo {edition}
  {1st}\ ed.,\ Vol.~\bibinfo {volume} {1}\ (\bibinfo  {publisher} {University
  of Chicago Press},\ \bibinfo {year} {1971})\BibitemShut {NoStop}%
\bibitem [{\citenamefont {{Chanmugam}}(1977)}]{Chanmugam1977}%
  \BibitemOpen
  \bibfield  {author} {\bibinfo {author} {\bibfnamefont {G.}~\bibnamefont
  {{Chanmugam}}},\ }\bibfield  {title} {\bibinfo {title} {{Radial oscillations
  of zero-temperature white dwarfs and neutron stars below nuclear
  densities}},\ }\href {https://doi.org/10.1086/155627} {\bibfield  {journal}
  {\bibinfo  {journal} {\apj}\ }\textbf {\bibinfo {volume} {217}},\ \bibinfo
  {pages} {799} (\bibinfo {year} {1977})}\BibitemShut {NoStop}%
\bibitem [{\citenamefont {{Glass}}\ and\ \citenamefont
  {{Lindblom}}(1983)}]{Glass1983}%
  \BibitemOpen
  \bibfield  {author} {\bibinfo {author} {\bibfnamefont {E.~N.}\ \bibnamefont
  {{Glass}}}\ and\ \bibinfo {author} {\bibfnamefont {L.}~\bibnamefont
  {{Lindblom}}},\ }\bibfield  {title} {\bibinfo {title} {{The radial
  oscillations of neutron stars}},\ }\href {https://doi.org/10.1086/190885}
  {\bibfield  {journal} {\bibinfo  {journal} {Astrophys. J. Sup.}\ }\textbf
  {\bibinfo {volume} {53}},\ \bibinfo {pages} {93} (\bibinfo {year}
  {1983})}\BibitemShut {NoStop}%
\bibitem [{\citenamefont {{V{\"a}th}}\ and\ \citenamefont
  {{Chanmugam}}(1992)}]{Vath1992}%
  \BibitemOpen
  \bibfield  {author} {\bibinfo {author} {\bibfnamefont {H.~M.}\ \bibnamefont
  {{V{\"a}th}}}\ and\ \bibinfo {author} {\bibfnamefont {G.}~\bibnamefont
  {{Chanmugam}}},\ }\bibfield  {title} {\bibinfo {title} {{Radial oscillations
  of neutron stars and strange stars}},\ }\href@noop {} {\bibfield  {journal}
  {\bibinfo  {journal} {Astron. Astrophys.}\ }\textbf {\bibinfo {volume}
  {260}},\ \bibinfo {pages} {250} (\bibinfo {year} {1992})}\BibitemShut
  {NoStop}%
\bibitem [{\citenamefont {Kokkotas}\ and\ \citenamefont
  {Ruoff}(2001)}]{Kokkotas2001}%
  \BibitemOpen
  \bibfield  {author} {\bibinfo {author} {\bibfnamefont {K.~D.}\ \bibnamefont
  {Kokkotas}}\ and\ \bibinfo {author} {\bibfnamefont {J.}~\bibnamefont
  {Ruoff}},\ }\bibfield  {title} {\bibinfo {title} {Radial oscillations of
  relativistic stars},\ }\href {https://doi.org/10.1051/0004-6361:20000216}
  {\bibfield  {journal} {\bibinfo  {journal} {Astron. Astrophys.}\ }\textbf
  {\bibinfo {volume} {366}},\ \bibinfo {pages} {565} (\bibinfo {year}
  {2001})}\BibitemShut {NoStop}%
\bibitem [{\citenamefont {Barta}(2019)}]{Barta2019}%
  \BibitemOpen
  \bibfield  {author} {\bibinfo {author} {\bibfnamefont {D.}~\bibnamefont
  {Barta}},\ }\bibfield  {title} {\bibinfo {title} {Effect of viscosity and
  thermal conductivity on the radial oscillation and relaxation of relativistic
  stars},\ }\href {https://doi.org/10.1088/1361-6382/ab472e} {\bibfield
  {journal} {\bibinfo  {journal} {Class. Quant. Grav.}\ }\textbf {\bibinfo
  {volume} {36}},\ \bibinfo {pages} {215012} (\bibinfo {year}
  {2019})}\BibitemShut {NoStop}%
\bibitem [{\citenamefont {Gondek}\ \emph {et~al.}(1997)\citenamefont {Gondek},
  \citenamefont {Haensel},\ and\ \citenamefont {Zdunik}}]{Gondek1997}%
  \BibitemOpen
  \bibfield  {author} {\bibinfo {author} {\bibfnamefont {D.}~\bibnamefont
  {Gondek}}, \bibinfo {author} {\bibfnamefont {P.}~\bibnamefont {Haensel}},\
  and\ \bibinfo {author} {\bibfnamefont {J.~L.}\ \bibnamefont {Zdunik}},\
  }\bibfield  {title} {\bibinfo {title} {{Radial pulsations and stability of
  protoneutron stars}},\ }\href@noop {} {\bibfield  {journal} {\bibinfo
  {journal} {Astron. Astrophys.}\ }\textbf {\bibinfo {volume} {325}},\ \bibinfo
  {pages} {217} (\bibinfo {year} {1997})}\BibitemShut {NoStop}%
\bibitem [{\citenamefont {Gondek}\ and\ \citenamefont
  {Zdunik}(1999)}]{Gondek1999}%
  \BibitemOpen
  \bibfield  {author} {\bibinfo {author} {\bibfnamefont {D.}~\bibnamefont
  {Gondek}}\ and\ \bibinfo {author} {\bibfnamefont {J.}~\bibnamefont
  {Zdunik}},\ }\bibfield  {title} {\bibinfo {title} {{Avoided crossings in
  radial pulsations of neutron and strange stars}},\ }\href@noop {} {\bibfield
  {journal} {\bibinfo  {journal} {Astron. Astrophys.}\ }\textbf {\bibinfo
  {volume} {344}},\ \bibinfo {pages} {117} (\bibinfo {year}
  {1999})}\BibitemShut {NoStop}%
\bibitem [{\citenamefont {Shternin}\ and\ \citenamefont
  {Yakovlev}(2008)}]{Shternin2008}%
  \BibitemOpen
  \bibfield  {author} {\bibinfo {author} {\bibfnamefont {P.~S.}\ \bibnamefont
  {Shternin}}\ and\ \bibinfo {author} {\bibfnamefont {D.~G.}\ \bibnamefont
  {Yakovlev}},\ }\bibfield  {title} {\bibinfo {title} {Shear viscosity in
  neutron star cores},\ }\href {https://doi.org/10.1103/PhysRevD.78.063006}
  {\bibfield  {journal} {\bibinfo  {journal} {Phys. Rev. D}\ }\textbf {\bibinfo
  {volume} {78}},\ \bibinfo {pages} {063006} (\bibinfo {year}
  {2008})}\BibitemShut {NoStop}%
\bibitem [{\citenamefont {Shternin}\ \emph {et~al.}(2013)\citenamefont
  {Shternin}, \citenamefont {Baldo},\ and\ \citenamefont
  {Haensel}}]{Shternin2013}%
  \BibitemOpen
  \bibfield  {author} {\bibinfo {author} {\bibfnamefont {P.~S.}\ \bibnamefont
  {Shternin}}, \bibinfo {author} {\bibfnamefont {M.}~\bibnamefont {Baldo}},\
  and\ \bibinfo {author} {\bibfnamefont {P.}~\bibnamefont {Haensel}},\
  }\bibfield  {title} {\bibinfo {title} {Transport coefficients of nuclear
  matter in neutron star cores},\ }\href
  {https://doi.org/10.1103/PhysRevC.88.065803} {\bibfield  {journal} {\bibinfo
  {journal} {Phys. Rev. C}\ }\textbf {\bibinfo {volume} {88}},\ \bibinfo
  {pages} {065803} (\bibinfo {year} {2013})}\BibitemShut {NoStop}%
\bibitem [{\citenamefont {Shternin}\ \emph {et~al.}(2017)\citenamefont
  {Shternin}, \citenamefont {Baldo},\ and\ \citenamefont
  {Schulze}}]{Shternin2017}%
  \BibitemOpen
  \bibfield  {author} {\bibinfo {author} {\bibfnamefont {P.~S.}\ \bibnamefont
  {Shternin}}, \bibinfo {author} {\bibfnamefont {M.}~\bibnamefont {Baldo}},\
  and\ \bibinfo {author} {\bibfnamefont {H.~J.}\ \bibnamefont {Schulze}},\
  }\bibfield  {title} {\bibinfo {title} {Transport coefficients in neutron star
  cores in {BHF} approach. comparison of different nucleon potentials},\ }\href
  {https://doi.org/10.1088/1742-6596/932/1/012042} {\bibfield  {journal}
  {\bibinfo  {journal} {J. Phys. Conf. Ser.}\ }\textbf {\bibinfo {volume}
  {932}},\ \bibinfo {pages} {012042} (\bibinfo {year} {2017})}\BibitemShut
  {NoStop}%
\bibitem [{\citenamefont {Tolos}\ \emph {et~al.}(2016)\citenamefont {Tolos},
  \citenamefont {Manuel}, \citenamefont {Sarkar},\ and\ \citenamefont
  {Tarrus}}]{Tolos2016}%
  \BibitemOpen
  \bibfield  {author} {\bibinfo {author} {\bibfnamefont {L.}~\bibnamefont
  {Tolos}}, \bibinfo {author} {\bibfnamefont {C.}~\bibnamefont {Manuel}},
  \bibinfo {author} {\bibfnamefont {S.}~\bibnamefont {Sarkar}},\ and\ \bibinfo
  {author} {\bibfnamefont {J.}~\bibnamefont {Tarrus}},\ }\bibfield  {title}
  {\bibinfo {title} {Transport coefficients in superfluid neutron stars},\
  }\href {https://doi.org/10.1063/1.4938690} {\bibfield  {journal} {\bibinfo
  {journal} {AIP Conference Proceedings}\ }\textbf {\bibinfo {volume} {1701}},\
  \bibinfo {pages} {080001} (\bibinfo {year} {2016})}\BibitemShut {NoStop}%
\bibitem [{\citenamefont {Kopal}(1964)}]{Kopal1964}%
  \BibitemOpen
  \bibfield  {author} {\bibinfo {author} {\bibfnamefont {Z.}~\bibnamefont
  {Kopal}},\ }\bibfield  {title} {\bibinfo {title} {{The effects of viscosity
  and radiative braking on stellar pulsations}},\ }\href@noop {} {\bibfield
  {journal} {\bibinfo  {journal} {Astrophys. Nor.}\ }\textbf {\bibinfo {volume}
  {9}},\ \bibinfo {pages} {239} (\bibinfo {year} {1964})}\BibitemShut {NoStop}%
\bibitem [{\citenamefont {Higgins}\ and\ \citenamefont
  {Kopal}(1968)}]{Higgins1968}%
  \BibitemOpen
  \bibfield  {author} {\bibinfo {author} {\bibfnamefont {T.~P.}\ \bibnamefont
  {Higgins}}\ and\ \bibinfo {author} {\bibfnamefont {Z.}~\bibnamefont
  {Kopal}},\ }\bibfield  {title} {\bibinfo {title} {Volume integrals of the
  products of spherical harmonics and their application to viscous dissipation
  phenomena in fluids},\ }\href {https://doi.org/10.1007/BF00650913} {\bibfield
   {journal} {\bibinfo  {journal} {Astrophys. Space Sci.}\ }\textbf {\bibinfo
  {volume} {2}},\ \bibinfo {pages} {352} (\bibinfo {year} {1968})}\BibitemShut
  {NoStop}%
\bibitem [{\citenamefont {{Mihalas}}(1983)}]{Mihalas1983}%
  \BibitemOpen
  \bibfield  {author} {\bibinfo {author} {\bibfnamefont {D.}~\bibnamefont
  {{Mihalas}}},\ }\bibfield  {title} {\bibinfo {title} {{Comments on the
  dynamical effects of radiative viscosity}},\ }\href
  {https://doi.org/10.1086/160772} {\bibfield  {journal} {\bibinfo  {journal}
  {Astrophys. J.}\ }\textbf {\bibinfo {volume} {266}},\ \bibinfo {pages} {242}
  (\bibinfo {year} {1983})}\BibitemShut {NoStop}%
\bibitem [{\citenamefont {Herrera}\ \emph {et~al.}(1989)\citenamefont
  {Herrera}, \citenamefont {Le~Denmat},\ and\ \citenamefont
  {Santos}}]{Herrera1989}%
  \BibitemOpen
  \bibfield  {author} {\bibinfo {author} {\bibfnamefont {L.}~\bibnamefont
  {Herrera}}, \bibinfo {author} {\bibfnamefont {G.}~\bibnamefont {Le~Denmat}},\
  and\ \bibinfo {author} {\bibfnamefont {N.~O.}\ \bibnamefont {Santos}},\
  }\bibfield  {title} {\bibinfo {title} {{Dynamical instability for
  non-adiabatic spherical collapse}},\ }\href
  {https://doi.org/10.1093/mnras/237.1.257} {\bibfield  {journal} {\bibinfo
  {journal} {Mon. Not. R. Astron. Soc.}\ }\textbf {\bibinfo {volume} {237}},\
  \bibinfo {pages} {257} (\bibinfo {year} {1989})}\BibitemShut {NoStop}%
\bibitem [{\citenamefont {Chan}\ \emph {et~al.}(1993)\citenamefont {Chan},
  \citenamefont {Herrera},\ and\ \citenamefont {Santos}}]{Chan1993}%
  \BibitemOpen
  \bibfield  {author} {\bibinfo {author} {\bibfnamefont {R.}~\bibnamefont
  {Chan}}, \bibinfo {author} {\bibfnamefont {L.}~\bibnamefont {Herrera}},\ and\
  \bibinfo {author} {\bibfnamefont {N.~O.}\ \bibnamefont {Santos}},\ }\bibfield
   {title} {\bibinfo {title} {{Dynamical Instability for Radiating Anisotropic
  Collapse}},\ }\href {https://doi.org/10.1093/mnras/265.3.533} {\bibfield
  {journal} {\bibinfo  {journal} {Mon. Not. R. Astron. Soc.}\ }\textbf
  {\bibinfo {volume} {265}},\ \bibinfo {pages} {533} (\bibinfo {year}
  {1993})}\BibitemShut {NoStop}%
\bibitem [{\citenamefont {Herrera}\ and\ \citenamefont
  {Santos}(1997)}]{Herrera1997}%
  \BibitemOpen
  \bibfield  {author} {\bibinfo {author} {\bibfnamefont {L.}~\bibnamefont
  {Herrera}}\ and\ \bibinfo {author} {\bibfnamefont {N.~O.}\ \bibnamefont
  {Santos}},\ }\bibfield  {title} {\bibinfo {title} {{Thermal evolution of
  compact objects and relaxation time}},\ }\href
  {https://doi.org/10.1093/mnras/287.1.161} {\bibfield  {journal} {\bibinfo
  {journal} {Mon. Not. R. Astron. Soc.}\ }\textbf {\bibinfo {volume} {287}},\
  \bibinfo {pages} {161} (\bibinfo {year} {1997})}\BibitemShut {NoStop}%
\bibitem [{\citenamefont {Lackey}\ \emph {et~al.}(2014)\citenamefont {Lackey},
  \citenamefont {Kyutoku}, \citenamefont {Shibata}, \citenamefont {Brady},\
  and\ \citenamefont {Friedman}}]{Lackey2014}%
  \BibitemOpen
  \bibfield  {author} {\bibinfo {author} {\bibfnamefont {B.~D.}\ \bibnamefont
  {Lackey}}, \bibinfo {author} {\bibfnamefont {K.}~\bibnamefont {Kyutoku}},
  \bibinfo {author} {\bibfnamefont {M.}~\bibnamefont {Shibata}}, \bibinfo
  {author} {\bibfnamefont {P.~R.}\ \bibnamefont {Brady}},\ and\ \bibinfo
  {author} {\bibfnamefont {J.~L.}\ \bibnamefont {Friedman}},\ }\bibfield
  {title} {\bibinfo {title} {Extracting equation of state parameters from black
  hole-neutron star mergers: Aligned-spin black holes and a preliminary
  waveform model},\ }\href {https://doi.org/10.1103/PhysRevD.89.043009}
  {\bibfield  {journal} {\bibinfo  {journal} {Phys. Rev. D}\ }\textbf {\bibinfo
  {volume} {89}},\ \bibinfo {pages} {043009} (\bibinfo {year}
  {2014})}\BibitemShut {NoStop}%
\bibitem [{\citenamefont {Lackey}\ \emph {et~al.}(2006)\citenamefont {Lackey},
  \citenamefont {Nayyar},\ and\ \citenamefont {Owen}}]{Lackey2006}%
  \BibitemOpen
  \bibfield  {author} {\bibinfo {author} {\bibfnamefont {B.~D.}\ \bibnamefont
  {Lackey}}, \bibinfo {author} {\bibfnamefont {M.}~\bibnamefont {Nayyar}},\
  and\ \bibinfo {author} {\bibfnamefont {B.~J.}\ \bibnamefont {Owen}},\
  }\bibfield  {title} {\bibinfo {title} {Observational constraints on hyperons
  in neutron stars},\ }\href {https://doi.org/10.1103/PhysRevD.73.024021}
  {\bibfield  {journal} {\bibinfo  {journal} {Phys. Rev. D}\ }\textbf {\bibinfo
  {volume} {73}},\ \bibinfo {pages} {024021} (\bibinfo {year}
  {2006})}\BibitemShut {NoStop}%
\bibitem [{\citenamefont {Alford}\ \emph {et~al.}(2005)\citenamefont {Alford},
  \citenamefont {Braby}, \citenamefont {Paris},\ and\ \citenamefont
  {Reddy}}]{Alford2005}%
  \BibitemOpen
  \bibfield  {author} {\bibinfo {author} {\bibfnamefont {M.}~\bibnamefont
  {Alford}}, \bibinfo {author} {\bibfnamefont {M.}~\bibnamefont {Braby}},
  \bibinfo {author} {\bibfnamefont {M.}~\bibnamefont {Paris}},\ and\ \bibinfo
  {author} {\bibfnamefont {S.}~\bibnamefont {Reddy}},\ }\bibfield  {title}
  {\bibinfo {title} {Hybrid stars that masquerade as neutron stars},\ }\href
  {https://doi.org/10.1086/430902} {\bibfield  {journal} {\bibinfo  {journal}
  {\apj}\ }\textbf {\bibinfo {volume} {629}},\ \bibinfo {pages} {969} (\bibinfo
  {year} {2005})}\BibitemShut {NoStop}%
\bibitem [{\citenamefont {Akmal}\ \emph {et~al.}(1998)\citenamefont {Akmal},
  \citenamefont {Pandharipande},\ and\ \citenamefont {Ravenhall}}]{Akmal1998}%
  \BibitemOpen
  \bibfield  {author} {\bibinfo {author} {\bibfnamefont {A.}~\bibnamefont
  {Akmal}}, \bibinfo {author} {\bibfnamefont {V.~R.}\ \bibnamefont
  {Pandharipande}},\ and\ \bibinfo {author} {\bibfnamefont {D.~G.}\
  \bibnamefont {Ravenhall}},\ }\bibfield  {title} {\bibinfo {title} {Equation
  of state of nucleon matter and neutron star structure},\ }\href
  {https://doi.org/10.1103/PhysRevC.58.1804} {\bibfield  {journal} {\bibinfo
  {journal} {Phys. Rev. C}\ }\textbf {\bibinfo {volume} {58}},\ \bibinfo
  {pages} {1804} (\bibinfo {year} {1998})}\BibitemShut {NoStop}%
\bibitem [{\citenamefont {M\"{u}ther}\ \emph {et~al.}(1987)\citenamefont
  {M\"{u}ther}, \citenamefont {Prakash},\ and\ \citenamefont
  {Ainsworth}}]{Muther1987}%
  \BibitemOpen
  \bibfield  {author} {\bibinfo {author} {\bibfnamefont {H.}~\bibnamefont
  {M\"{u}ther}}, \bibinfo {author} {\bibfnamefont {M.}~\bibnamefont
  {Prakash}},\ and\ \bibinfo {author} {\bibfnamefont {T.}~\bibnamefont
  {Ainsworth}},\ }\bibfield  {title} {\bibinfo {title} {The nuclear symmetry
  energy in relativistic brueckner-hartree-fock calculations},\ }\href@noop {}
  {\bibfield  {journal} {\bibinfo  {journal} {Physics Letters B}\ }\textbf
  {\bibinfo {volume} {199}},\ \bibinfo {pages} {469} (\bibinfo {year}
  {1987})}\BibitemShut {NoStop}%
\bibitem [{\citenamefont {{M\"{u}ller}}\ and\ \citenamefont
  {Serot}(1996)}]{Mueller1996}%
  \BibitemOpen
  \bibfield  {author} {\bibinfo {author} {\bibfnamefont {H.}~\bibnamefont
  {{M\"{u}ller}}}\ and\ \bibinfo {author} {\bibfnamefont {B.~D.}\ \bibnamefont
  {Serot}},\ }\bibfield  {title} {\bibinfo {title} {{Relativistic mean-field
  theory and the high-density nuclear equation of state}},\ }\href
  {https://doi.org/10.1016/0375-9474(96)00187-X} {\bibfield  {journal}
  {\bibinfo  {journal} {Nucl. Phys.}\ }\textbf {\bibinfo {volume} {A606}},\
  \bibinfo {pages} {508} (\bibinfo {year} {1996})}\BibitemShut {NoStop}%
\bibitem [{\citenamefont {Douchin}\ and\ \citenamefont
  {Haensel}(2001)}]{Douchin2001}%
  \BibitemOpen
  \bibfield  {author} {\bibinfo {author} {\bibfnamefont {F.}~\bibnamefont
  {Douchin}}\ and\ \bibinfo {author} {\bibfnamefont {P.}~\bibnamefont
  {Haensel}},\ }\bibfield  {title} {\bibinfo {title} {A unified equation of
  state of dense matter and neutron star structure},\ }\href
  {https://doi.org/10.1051/0004-6361:20011402} {\bibfield  {journal} {\bibinfo
  {journal} {Astron. Astrophys.}\ }\textbf {\bibinfo {volume} {380}},\ \bibinfo
  {pages} {151} (\bibinfo {year} {2001})}\BibitemShut {NoStop}%
\bibitem [{\citenamefont {Prakash}\ \emph {et~al.}(1995)\citenamefont
  {Prakash}, \citenamefont {Cooke},\ and\ \citenamefont
  {Lattimer}}]{Prakash1995}%
  \BibitemOpen
  \bibfield  {author} {\bibinfo {author} {\bibfnamefont {M.}~\bibnamefont
  {Prakash}}, \bibinfo {author} {\bibfnamefont {J.~R.}\ \bibnamefont {Cooke}},\
  and\ \bibinfo {author} {\bibfnamefont {J.~M.}\ \bibnamefont {Lattimer}},\
  }\bibfield  {title} {\bibinfo {title} {Quark-hadron phase transition in
  protoneutron stars},\ }\href {https://doi.org/10.1103/PhysRevD.52.661}
  {\bibfield  {journal} {\bibinfo  {journal} {Phys. Rev. D}\ }\textbf {\bibinfo
  {volume} {52}},\ \bibinfo {pages} {661} (\bibinfo {year} {1995})}\BibitemShut
  {NoStop}%
\bibitem [{\citenamefont {Ruoff}(2000)}]{Ruoff2000}%
  \BibitemOpen
  \bibfield  {author} {\bibinfo {author} {\bibfnamefont {J.}~\bibnamefont
  {Ruoff}},\ }\emph {\bibinfo {title} {The numerical evolution of neutron star
  oscillations}},\ \href@noop {} {Ph.D. thesis},\ \bibinfo  {school}
  {University of T\"{u}bingen} (\bibinfo {year} {2000}),\ \bibinfo {note}
  {\href{https://arxiv.org/abs/gr-qc/0010041}{arXiv:gr-qc/0010041}}\BibitemShut
  {NoStop}%
\bibitem [{\citenamefont {Tooper}(1964)}]{Tooper1964}%
  \BibitemOpen
  \bibfield  {author} {\bibinfo {author} {\bibfnamefont {R.~F.}\ \bibnamefont
  {Tooper}},\ }\bibfield  {title} {\bibinfo {title} {Stability of massive stars
  in general relativity},\ }\href {https://doi.org/10.1086/147980} {\bibfield
  {journal} {\bibinfo  {journal} {Astrophys. J.}\ }\textbf {\bibinfo {volume}
  {140}},\ \bibinfo {pages} {811} (\bibinfo {year} {1964})}\BibitemShut
  {NoStop}%
\bibitem [{\citenamefont {Bona}\ \emph {et~al.}(2009)\citenamefont {Bona},
  \citenamefont {Palenzuela-Luque},\ and\ \citenamefont
  {Bona-Casas}}]{Bona2009}%
  \BibitemOpen
  \bibfield  {author} {\bibinfo {author} {\bibfnamefont {C.}~\bibnamefont
  {Bona}}, \bibinfo {author} {\bibfnamefont {C.}~\bibnamefont
  {Palenzuela-Luque}},\ and\ \bibinfo {author} {\bibfnamefont {C.}~\bibnamefont
  {Bona-Casas}},\ }\href {https://doi.org/10.1007/978-3-319-97616-7} {\emph
  {\bibinfo {title} {Elements of numerical relativity and relativistic
  hydrodynamics: From Einstein's equations to astrophysical simulations}}},\
  \bibinfo {edition} {2nd}\ ed.,\ \bibinfo {series} {Lecture Notes in Physics},
  Vol.\ \bibinfo {volume} {783}\ (\bibinfo  {publisher} {Springer-Verlag,
  Berlin; Heidelberg},\ \bibinfo {year} {2009})\ p.\ \bibinfo {pages}
  {214}\BibitemShut {NoStop}%
\bibitem [{\citenamefont {Ib\'{a}\~{n}ez}\ \emph {et~al.}(2018)\citenamefont
  {Ib\'{a}\~{n}ez}, \citenamefont {Marquina}, \citenamefont {Serna},\ and\
  \citenamefont {Aloy}}]{Ibanez2018}%
  \BibitemOpen
  \bibfield  {author} {\bibinfo {author} {\bibfnamefont {J.~M.}\ \bibnamefont
  {Ib\'{a}\~{n}ez}}, \bibinfo {author} {\bibfnamefont {A.}~\bibnamefont
  {Marquina}}, \bibinfo {author} {\bibfnamefont {S.}~\bibnamefont {Serna}},\
  and\ \bibinfo {author} {\bibfnamefont {M.~A.}\ \bibnamefont {Aloy}},\
  }\bibfield  {title} {\bibinfo {title} {{Anomalous dynamics triggered by a
  non-convex equation of state in relativistic flows}},\ }\href
  {https://doi.org/10.1093/mnras/sty137} {\bibfield  {journal} {\bibinfo
  {journal} {Mon. Not. R. Astron. Soc.}\ }\textbf {\bibinfo {volume} {476}},\
  \bibinfo {pages} {1100} (\bibinfo {year} {2018})}\BibitemShut {NoStop}%
\bibitem [{\citenamefont {Ivanov}(2017)}]{Ivanov2017}%
  \BibitemOpen
  \bibfield  {author} {\bibinfo {author} {\bibfnamefont {B.~V.}\ \bibnamefont
  {Ivanov}},\ }\bibfield  {title} {\bibinfo {title} {Analytical study of
  anisotropic compact star models},\ }\href
  {https://doi.org/10.1140/epjc/s10052-017-5322-7} {\bibfield  {journal}
  {\bibinfo  {journal} {Eur. Phys. J. C}\ }\textbf {\bibinfo {volume} {77}},\
  \bibinfo {pages} {738} (\bibinfo {year} {2017})}\BibitemShut {NoStop}%
\bibitem [{\citenamefont {Hotokezaka}\ \emph {et~al.}(2013)\citenamefont
  {Hotokezaka} \emph {et~al.}}]{Hotokezaka2012}%
  \BibitemOpen
  \bibfield  {author} {\bibinfo {author} {\bibfnamefont {K.}~\bibnamefont
  {Hotokezaka}} \emph {et~al.},\ }\bibfield  {title} {\bibinfo {title} {Mass
  ejection from the merger of binary neutron stars},\ }\href
  {https://doi.org/10.1103/PhysRevD.87.024001} {\bibfield  {journal} {\bibinfo
  {journal} {Phys. Rev. D}\ }\textbf {\bibinfo {volume} {87}},\ \bibinfo
  {pages} {024001} (\bibinfo {year} {2013})}\BibitemShut {NoStop}%
\bibitem [{\citenamefont {Lim}\ \emph {et~al.}(2017)\citenamefont {Lim},
  \citenamefont {Hyun},\ and\ \citenamefont {Lee}}]{Lim2017}%
  \BibitemOpen
  \bibfield  {author} {\bibinfo {author} {\bibfnamefont {Y.}~\bibnamefont
  {Lim}}, \bibinfo {author} {\bibfnamefont {C.~H.}\ \bibnamefont {Hyun}},\ and\
  \bibinfo {author} {\bibfnamefont {C.~H.}\ \bibnamefont {Lee}},\ }\bibfield
  {title} {\bibinfo {title} {Nuclear equation of state and neutron star
  cooling},\ }\href {https://doi.org/10.1142/S021830131750015X} {\bibfield
  {journal} {\bibinfo  {journal} {Int. J. Mod. Phys.}\ }\textbf {\bibinfo
  {volume} {E26}},\ \bibinfo {pages} {1750015} (\bibinfo {year}
  {2017})}\BibitemShut {NoStop}%
\bibitem [{\citenamefont {Gusakov}\ \emph {et~al.}(2005)\citenamefont
  {Gusakov}, \citenamefont {Yakovlev},\ and\ \citenamefont
  {Gnedin}}]{Gusakov2005}%
  \BibitemOpen
  \bibfield  {author} {\bibinfo {author} {\bibfnamefont {M.~E.}\ \bibnamefont
  {Gusakov}}, \bibinfo {author} {\bibfnamefont {D.~G.}\ \bibnamefont
  {Yakovlev}},\ and\ \bibinfo {author} {\bibfnamefont {O.~Y.}\ \bibnamefont
  {Gnedin}},\ }\bibfield  {title} {\bibinfo {title} {Thermal evolution of a
  pulsating neutron star},\ }\href
  {https://doi.org/10.1111/j.1365-2966.2005.09295.x} {\bibfield  {journal}
  {\bibinfo  {journal} {Mon. Not. R. Astron. Soc.}\ }\textbf {\bibinfo {volume}
  {361}},\ \bibinfo {pages} {1415} (\bibinfo {year} {2005})}\BibitemShut
  {NoStop}%
\bibitem [{\citenamefont {Eckart}(1940)}]{Eckart1940}%
  \BibitemOpen
  \bibfield  {author} {\bibinfo {author} {\bibfnamefont {C.}~\bibnamefont
  {Eckart}},\ }\bibfield  {title} {\bibinfo {title} {The thermodynamics of
  irreversible processes. iii. relativistic theory of the simple fluid},\
  }\href {https://doi.org/10.1103/PhysRev.58.919} {\bibfield  {journal}
  {\bibinfo  {journal} {Phys. Rev.}\ }\textbf {\bibinfo {volume} {58}},\
  \bibinfo {pages} {919} (\bibinfo {year} {1940})}\BibitemShut {NoStop}%
\bibitem [{\citenamefont {Rezzolla}\ and\ \citenamefont
  {Zanotti}(2013)}]{Rezzola2013}%
  \BibitemOpen
  \bibfield  {author} {\bibinfo {author} {\bibfnamefont {L.}~\bibnamefont
  {Rezzolla}}\ and\ \bibinfo {author} {\bibfnamefont {O.}~\bibnamefont
  {Zanotti}},\ }\href
  {https://doi.org/10.1093/acprof:oso/9780198528906.001.0001} {\emph {\bibinfo
  {title} {Relativistic hydrodynamics}}},\ \bibinfo {edition} {1st}\ ed.,\
  Vol.~\bibinfo {volume} {1}\ (\bibinfo  {publisher} {Oxford University
  Press},\ \bibinfo {year} {2013})\BibitemShut {NoStop}%
\bibitem [{\citenamefont {Chamel}\ and\ \citenamefont
  {Haensel}(2008)}]{Chamel2008}%
  \BibitemOpen
  \bibfield  {author} {\bibinfo {author} {\bibfnamefont {N.}~\bibnamefont
  {Chamel}}\ and\ \bibinfo {author} {\bibfnamefont {P.}~\bibnamefont
  {Haensel}},\ }\bibfield  {title} {\bibinfo {title} {{Physics of neutron star
  crusts}},\ }\href {https://doi.org/10.12942/lrr-2008-10} {\bibfield
  {journal} {\bibinfo  {journal} {Living Rev. Relativ.}\ }\textbf {\bibinfo
  {volume} {11}},\ \bibinfo {pages} {10} (\bibinfo {year} {2008})}\BibitemShut
  {NoStop}%
\bibitem [{\citenamefont {Camenzind}(2007)}]{Camenzind2007}%
  \BibitemOpen
  \bibfield  {author} {\bibinfo {author} {\bibfnamefont {M.}~\bibnamefont
  {Camenzind}},\ }\bibfield  {title} {\bibinfo {title} {Relativistic stellar
  structure},\ }in\ \href {https://doi.org/10.1007/978-3-540-49912-1_4} {\emph
  {\bibinfo {booktitle} {Compact objects in Astrophysics: White dwarfs, neutron
  stars and black holes}}}\ (\bibinfo  {publisher} {Springer},\ \bibinfo
  {address} {Berlin, Heidelberg, Germany},\ \bibinfo {year} {2007})\ pp.\
  \bibinfo {pages} {123--135}\BibitemShut {NoStop}%
\bibitem [{\citenamefont {Suwa}\ \emph {et~al.}(2018)\citenamefont {Suwa},
  \citenamefont {Yoshida}, \citenamefont {Shibata}, \citenamefont {Umeda},\
  and\ \citenamefont {Takahashi}}]{Suwa2018}%
  \BibitemOpen
  \bibfield  {author} {\bibinfo {author} {\bibfnamefont {Y.}~\bibnamefont
  {Suwa}}, \bibinfo {author} {\bibfnamefont {T.}~\bibnamefont {Yoshida}},
  \bibinfo {author} {\bibfnamefont {M.}~\bibnamefont {Shibata}}, \bibinfo
  {author} {\bibfnamefont {H.}~\bibnamefont {Umeda}},\ and\ \bibinfo {author}
  {\bibfnamefont {K.}~\bibnamefont {Takahashi}},\ }\bibfield  {title} {\bibinfo
  {title} {{On the minimum mass of neutron stars}},\ }\href
  {https://doi.org/10.1093/mnras/sty2460} {\bibfield  {journal} {\bibinfo
  {journal} {MNRAS}\ }\textbf {\bibinfo {volume} {481}},\ \bibinfo {pages}
  {3305} (\bibinfo {year} {2018})}\BibitemShut {NoStop}%
\bibitem [{\citenamefont {Haensel}\ \emph {et~al.}(2009)\citenamefont
  {Haensel}, \citenamefont {Zdunik}, \citenamefont {Bejger},\ and\
  \citenamefont {Lattimer}}]{Haensel2009}%
  \BibitemOpen
  \bibfield  {author} {\bibinfo {author} {\bibfnamefont {P.}~\bibnamefont
  {Haensel}}, \bibinfo {author} {\bibfnamefont {J.~L.}\ \bibnamefont {Zdunik}},
  \bibinfo {author} {\bibfnamefont {M.}~\bibnamefont {Bejger}},\ and\ \bibinfo
  {author} {\bibfnamefont {J.~M.}\ \bibnamefont {Lattimer}},\ }\bibfield
  {title} {\bibinfo {title} {Keplerian frequency of uniformly rotating neutron
  stars and strange stars},\ }\href
  {https://doi.org/10.1051/0004-6361/200811605} {\bibfield  {journal} {\bibinfo
   {journal} {Astron. Astrophys.}\ }\textbf {\bibinfo {volume} {502}},\
  \bibinfo {pages} {605} (\bibinfo {year} {2009})}\BibitemShut {NoStop}%
\bibitem [{\citenamefont {Lattimer}(2012)}]{Lattimer2012}%
  \BibitemOpen
  \bibfield  {author} {\bibinfo {author} {\bibfnamefont {J.~M.}\ \bibnamefont
  {Lattimer}},\ }\bibfield  {title} {\bibinfo {title} {The nuclear equation of
  state and neutron star masses},\ }\href
  {https://doi.org/10.1146/annurev-nucl-102711-095018} {\bibfield  {journal}
  {\bibinfo  {journal} {Annu. Rev. Nucl. Part. Sci.}\ }\textbf {\bibinfo
  {volume} {62}},\ \bibinfo {pages} {485} (\bibinfo {year} {2012})}\BibitemShut
  {NoStop}%
\bibitem [{\citenamefont {Glass}\ and\ \citenamefont
  {Harpaz}(1983)}]{Glass1983b}%
  \BibitemOpen
  \bibfield  {author} {\bibinfo {author} {\bibfnamefont {E.~N.}\ \bibnamefont
  {Glass}}\ and\ \bibinfo {author} {\bibfnamefont {A.}~\bibnamefont {Harpaz}},\
  }\bibfield  {title} {\bibinfo {title} {{The stability of relativistic gas
  spheres}},\ }\href {https://doi.org/10.1093/mnras/202.1.159} {\bibfield
  {journal} {\bibinfo  {journal} {Mon. Not. R. Astron. Soc.}\ }\textbf
  {\bibinfo {volume} {202}},\ \bibinfo {pages} {159} (\bibinfo {year}
  {1983})}\BibitemShut {NoStop}%
\bibitem [{\citenamefont {{Whitham}}(1974)}]{Whitham1974}%
  \BibitemOpen
  \bibfield  {author} {\bibinfo {author} {\bibfnamefont {G.~B.}\ \bibnamefont
  {{Whitham}}},\ }\href {https://doi.org/10.1002/9781118032954.ch11} {\emph
  {\bibinfo {title} {Linear Dispersive Waves}}},\ \bibinfo {edition} {1st}\
  ed.,\ Vol.~\bibinfo {volume} {1}\ (\bibinfo  {publisher} {New York: John
  Wiley \& Sons},\ \bibinfo {year} {1974})\BibitemShut {NoStop}%
\bibitem [{\citenamefont {Cutler}\ and\ \citenamefont
  {Lindblom}(1987)}]{Cutler1987}%
  \BibitemOpen
  \bibfield  {author} {\bibinfo {author} {\bibfnamefont {C.}~\bibnamefont
  {Cutler}}\ and\ \bibinfo {author} {\bibfnamefont {L.}~\bibnamefont
  {Lindblom}},\ }\bibfield  {title} {\bibinfo {title} {The effect of viscosity
  on neutron star oscillations},\ }\href {https://doi.org/10.1086/165052}
  {\bibfield  {journal} {\bibinfo  {journal} {\apj}\ }\textbf {\bibinfo
  {volume} {314}},\ \bibinfo {pages} {234} (\bibinfo {year}
  {1987})}\BibitemShut {NoStop}%
\bibitem [{\citenamefont {Chandrasekhar}(1984)}]{Chandrasekhar1984}%
  \BibitemOpen
  \bibfield  {author} {\bibinfo {author} {\bibfnamefont {S.}~\bibnamefont
  {Chandrasekhar}},\ }\bibfield  {title} {\bibinfo {title} {On stars, their
  evolution and their stability},\ }\href
  {https://doi.org/10.1103/RevModPhys.56.137} {\bibfield  {journal} {\bibinfo
  {journal} {Rev. Mod. Phys.}\ }\textbf {\bibinfo {volume} {56}},\ \bibinfo
  {pages} {137} (\bibinfo {year} {1984})}\BibitemShut {NoStop}%
\bibitem [{\citenamefont {Haensel}\ \emph {et~al.}(2007)\citenamefont
  {Haensel}, \citenamefont {Potekhin},\ and\ \citenamefont
  {Yakovlev}}]{Haensel2007}%
  \BibitemOpen
  \bibinfo {editor} {\bibfnamefont {P.}~\bibnamefont {Haensel}}, \bibinfo
  {editor} {\bibfnamefont {A.~Y.}\ \bibnamefont {Potekhin}},\ and\ \bibinfo
  {editor} {\bibfnamefont {D.~G.}\ \bibnamefont {Yakovlev}},\ eds.,\ \href
  {https://doi.org/10.1007/978-0-387-47301-7_6} {\emph {\bibinfo {title}
  {Neutron Stars 1}}},\ \bibinfo {series} {Astrophysics and Space Science
  Library}, Vol.\ \bibinfo {volume} {326}\ (\bibinfo  {publisher} {Springer
  Publishing},\ \bibinfo {address} {New York, NY},\ \bibinfo {year}
  {2007})\BibitemShut {NoStop}%
\bibitem [{\citenamefont {Schertler}\ \emph {et~al.}(2000)\citenamefont
  {Schertler}, \citenamefont {Greiner}, \citenamefont {J.},\ and\ \citenamefont
  {Thoma}}]{Schertler2000}%
  \BibitemOpen
  \bibfield  {author} {\bibinfo {author} {\bibfnamefont {K.}~\bibnamefont
  {Schertler}}, \bibinfo {author} {\bibfnamefont {C.}~\bibnamefont {Greiner}},
  \bibinfo {author} {\bibfnamefont {S.-B.}\ \bibnamefont {J.}},\ and\ \bibinfo
  {author} {\bibfnamefont {M.~H.}\ \bibnamefont {Thoma}},\ }\bibfield  {title}
  {\bibinfo {title} {Quark phases in neutron stars and a third family of
  compact stars as signature for phase transitions},\ }\href
  {https://doi.org/https://doi.org/10.1016/S0375-9474(00)00305-5} {\bibfield
  {journal} {\bibinfo  {journal} {Nucl. Phys. A}\ }\textbf {\bibinfo {volume}
  {677}},\ \bibinfo {pages} {463 } (\bibinfo {year} {2000})}\BibitemShut
  {NoStop}%
\bibitem [{\citenamefont {LeVeque}(2007)}]{LeVeque2007}%
  \BibitemOpen
  \bibfield  {author} {\bibinfo {author} {\bibfnamefont {R.}~\bibnamefont
  {LeVeque}},\ }\href {https://doi.org/10.1137/1.9780898717839} {\emph
  {\bibinfo {title} {Finite difference methods for ordinary and partial
  differential equations}}},\ Classics in Applied Mathematics\ (\bibinfo
  {publisher} {SIAM},\ \bibinfo {address} {Philadelphia, PA, US},\ \bibinfo
  {year} {2007})\BibitemShut {NoStop}%
\bibitem [{\citenamefont {Langtangen}\ and\ \citenamefont
  {Svein}(2017)}]{Langtangen2017}%
  \BibitemOpen
  \bibfield  {author} {\bibinfo {author} {\bibfnamefont {H.~P.}\ \bibnamefont
  {Langtangen}}\ and\ \bibinfo {author} {\bibfnamefont {L.}~\bibnamefont
  {Svein}},\ }\href {https://doi.org/10.1007/978-3-319-55456-3} {\emph
  {\bibinfo {title} {Finite difference computing with PDEs: A modern software
  approach}}},\ Texts in Computational Science and Engineering\ (\bibinfo
  {publisher} {Springer Nature},\ \bibinfo {address} {Cham, Germany},\ \bibinfo
  {year} {2017})\BibitemShut {NoStop}%
\bibitem [{\citenamefont {Arnett}\ and\ \citenamefont
  {Bowers}(1977)}]{Arnett1977}%
  \BibitemOpen
  \bibfield  {author} {\bibinfo {author} {\bibfnamefont {W.~D.}\ \bibnamefont
  {Arnett}}\ and\ \bibinfo {author} {\bibfnamefont {R.~L.}\ \bibnamefont
  {Bowers}},\ }\bibfield  {title} {\bibinfo {title} {A microscopic
  interpretation of neutron star structure},\ }\href
  {https://doi.org/10.1086/190434} {\bibfield  {journal} {\bibinfo  {journal}
  {ApJS}\ }\textbf {\bibinfo {volume} {33}},\ \bibinfo {pages} {415} (\bibinfo
  {year} {1977})}\BibitemShut {NoStop}%
\bibitem [{\citenamefont {Alcock}\ and\ \citenamefont
  {Olinto}(1988)}]{Alcock1988}%
  \BibitemOpen
  \bibfield  {author} {\bibinfo {author} {\bibfnamefont {C.}~\bibnamefont
  {Alcock}}\ and\ \bibinfo {author} {\bibfnamefont {A.}~\bibnamefont
  {Olinto}},\ }\bibfield  {title} {\bibinfo {title} {Exotic phases of hadronic
  matter and their astrophysical application},\ }\href
  {https://doi.org/10.1146/annurev.ns.38.120188.001113} {\bibfield  {journal}
  {\bibinfo  {journal} {Annu. Rev. Nucl. Part. Sci.}\ }\textbf {\bibinfo
  {volume} {38}},\ \bibinfo {pages} {161} (\bibinfo {year} {1988})}\BibitemShut
  {NoStop}%
\bibitem [{\citenamefont {Kippenhahn}\ and\ \citenamefont
  {Weigert}(1990)}]{Kippenhahn1990}%
  \BibitemOpen
  \bibinfo {editor} {\bibfnamefont {R.}~\bibnamefont {Kippenhahn}}\ and\
  \bibinfo {editor} {\bibfnamefont {A.}~\bibnamefont {Weigert}},\ eds.,\ \href
  {https://doi.org/10.1007/978-3-642-30304-3} {\emph {\bibinfo {title} {Stellar
  structure and evolution}}},\ \bibinfo {series} {Astrophysics and Space
  Science Library}, Vol.~\bibinfo {volume} {16}\ (\bibinfo  {publisher}
  {Springer-Verlag},\ \bibinfo {address} {Berlin, Heidelberg, New York},\
  \bibinfo {year} {1990})\BibitemShut {NoStop}%
\bibitem [{\citenamefont {Maeder}(2009)}]{Maeder2009}%
  \BibitemOpen
  \bibinfo {editor} {\bibfnamefont {A.}~\bibnamefont {Maeder}},\ ed.,\ \href
  {https://doi.org/10.1007/978-3-540-76949-1_1} {\emph {\bibinfo {title}
  {Physics, formation and evolution of rotating stars}}},\ \bibinfo {series}
  {Astrophysics and Space Science Library}, Vol.~\bibinfo {volume} {1}\
  (\bibinfo  {publisher} {Springer-Verlag},\ \bibinfo {address} {Berlin,
  Heidelberg},\ \bibinfo {year} {2009})\BibitemShut {NoStop}%
\bibitem [{\citenamefont {Phillips}(1994)}]{Phillips1994}%
  \BibitemOpen
  \bibinfo {editor} {\bibfnamefont {A.~C.}\ \bibnamefont {Phillips}},\ ed.,\
  \href {https://doi.org/10.1007/978-0-4719-8798-7} {\emph {\bibinfo {title}
  {The physics of stars}}}\ (\bibinfo  {publisher} {Wiley},\ \bibinfo {address}
  {Chichester, UK},\ \bibinfo {year} {1994})\BibitemShut {NoStop}%
\end{thebibliography}%
	
\end{document}